\documentclass[10pt]{article}

\usepackage{color}
\usepackage{graphicx}
\usepackage{amssymb, amsfonts, amsmath}
\usepackage{hyperref, cite}
\usepackage{subfigure}
\usepackage{authblk}
\usepackage[utf8]{inputenc}

\renewcommand{\l}{\left}
\renewcommand{\r}{\right}
\newcommand{\g}[1]{\gamma_{#1}} 
\newcommand{\DB}[1]{\stackrel{\leftarrow}{D}_{#1}} 
\newcommand{\DF}[1]{\stackrel{\rightarrow}{D}_{#1}} 
\newcommand{\DBF}[1]{\stackrel{\leftrightarrow}{D}_{#1}} 
\newcommand{\tr}{\mathrm{tr}}
\newcommand{\bra}[1]{\left< #1 \right|} 
\newcommand{\ket}[1]{\left| #1 \right>} 

\newcommand{\chiral}[1]{\mathring{#1}} 
\newcommand{\gev}{\,\mathrm{GeV}}

\newcommand{\mev}{\,\mathrm{MeV}}
\newcommand{\fm}{\,\mathrm{fm}}

\newcommand{\MSbar}{\overline{\mathrm{MS}}}

\newcommand{\phys}{\mathrm{phys}}
\newcommand{\stat}{\mathrm{stat}}
\newcommand{\sys}{\mathrm{sys}}
\newcommand{\syserr}[1]{(#1)_\mathrm{sys}}
\newcommand{\staterr}[1]{(#1)_\mathrm{stat}}
\newcommand{\chierr}[1]{(#1)_\chi}
\newcommand{\conterr}[1]{(#1)_\mathrm{cont}}
\newcommand{\FSerr}[1]{(#1)_\mathrm{FS}}

\newcommand{\tsep}{t_\mathrm{sep}}

\newcommand{\Mpi}{M_\pi}

\newcommand{\avgx}[2]{\langle x \rangle_{#2 u #1 #2 d}}

\newcommand{\Ctwopt}[2]{C^\mathrm{2pt}(\vec{#2}, #1)}
\newcommand{\Cthreept}[5]{C^{#1}_{#2}(\vec{#5},#3,#4)}

\title{\textbf{Nucleon isovector charges and twist-2 matrix elements with $N_f=2+1$ dynamical Wilson quarks} \\[1cm]
}

\author[(a,b)]{Tim~Harris}
\author[(c)]{Georg~von~Hippel}
\author[(a,d)]{Parikshit~Junnarkar}
\author[(a,c)]{Harvey~B.~Meyer}
\author[(c)]{Konstantin~Ottnad}
\author[(c)]{Jonas~Wilhelm}
\author[(a,c)]{Hartmut~Wittig}
\author[(c,e)]{Linus~Wrang}

\affil[(a)]{Helmholtz~Institute~Mainz, Johannes~Gutenberg-Universität~Mainz, 55099~Mainz, Germany}
\affil[(b)]{Dip.~di~Fisica~G.~Occhialini, Università~degli~Studi~di~Milano-Bicocca and INFN, sezione~di~Milano-Bicocca, Piazza~della~Scienza~3, 20126~Milano, Italy}
\affil[(c)]{PRISMA$^+$~Cluster~of~Excellence and Institut~f\"ur~Kernphysik, Johannes~Gutenberg-Universität~Mainz, 55099~Mainz, Germany}
\affil[(d)]{Tata~Institute~of~Fundamental~Research~(TIFR), Homi~Bhabha~Road, Mumbai~400005, India}
\affil[(e)]{Department of Earth Sciences, University~of~Uppsala, Villavägen~16, 752~36~Uppsala, Sweden}

\begin{document}
 \date{}
 \maketitle
 \begin{abstract}
  \noindent We present results from a lattice QCD study of nucleon matrix elements at vanishing momentum transfer for local and twist-2 isovector operator insertions. Computations are performed on gauge ensembles with non-perturbatively improved $N_f=2+1$ Wilson fermions, covering four values of the lattice spacing and pion masses down to $M_\pi \approx 200\,\mev$. Several source-sink separations (typically $\sim 1.0\,\fm$ to $\sim 1.5\,\fm$) allow us to assess excited-state contamination. Results on individual ensembles are obtained from simultaneous two-state fits across all observables and all available source-sink separations with the energy gap as a common fit parameter. Renormalization has been performed non-perturbatively using the Rome-Southampton method for all but the finest lattice spacing for which an extrapolation has been used. Physical results are quoted in the $\MSbar$ scheme at a scale of $\mu=2\gev$ and are obtained from a combined chiral, continuum  and finite-size extrapolation. For the nucleon isovector axial, scalar and tensor charges we find physical values of
$g_A^{u-d} = 1.242\staterr{25}\syserr{\genfrac{}{}{0pt}{2}{+00}{-31}}$, 
$g_S^{u-d} = 1.13 \staterr{11}\syserr{\genfrac{}{}{0pt}{2}{+07}{-06}}$ and 
$g_T^{u-d} = 0.965\staterr{38}\syserr{\genfrac{}{}{0pt}{2}{+13}{-41}}$, respectively, where individual systematic errors in each direction from the chiral, continuum and finite-size extrapolation have been added in quadrature.
Our final results for the isovector average quark momentum fraction and the isovector helicity and transversity moments are given by
$\avgx{-}{} = 0.180\staterr{25}\syserr{\genfrac{}{}{0pt}{2}{+14}{-06}}$,
$\avgx{-}{\Delta} = 0.221\staterr{25}\syserr{\genfrac{}{}{0pt}{2}{+10}{-00}}$ and
$\avgx{-}{\delta} = 0.212\staterr{32}\syserr{\genfrac{}{}{0pt}{2}{+20}{-10}}$, respectively.
 \end{abstract}

\clearpage

\section{Introduction} \label{sec:intro}

Nucleon matrix elements carry information on the internal structure and properties of nucleons, which can be related to a large variety of physical processes. Using calculations within the framework of lattice QCD, these matrix elements can be studied from first principles. Considering local, isovector operator insertions and vanishing momentum transfer, the corresponding matrix elements give access to isovector nucleon charges. These can be obtained from lattice QCD without the need to consider contributions from quark-disconnected diagrams, which are computationally particularly difficult.

For the isovector axial charge the experimental value is precisely known, i.e. $g_A^{u-d}=1.2724(23)$ \cite{Tanabashi:2018oca}, as it can be measured from the $\beta$-decay of a neutron into a proton, hence providing a crucial test for lattice QCD. This has led to considerable interest in computing the axial charge on the lattice \cite{Chang:2018uxx,Gupta:2018qil,Ishikawa:2018rew,Liang:2018pis,Yamanaka:2018uud,Alexandrou:2017hac,Berkowitz:2017gql,Capitani:2017qpc,Rajan:2017lxk,Bhattacharya:2016zcn,Bouchard:2016heu,Abdel-Rehim:2015owa,Bali:2014nma,Horsley:2013ayv,Capitani:2012gj,Green:2012ud}. Recently, $g_A^{u-d}$ has been included in the FLAG review (Ref.~\cite{Aoki:2019cca}) together with the isovector scalar and tensor charges of the nucleon. \par

Unlike the axial charge, the scalar and tensor charges, which can contribute to the $\beta$ decay of the nucleon through non-standard couplings outside the Standard Model (SM) \cite{Bhattacharya:2011qm} and are important for interpreting the results of dark matter searches \cite{DelNobile:2013sia}, are much less well-determined from phenomenology. Therefore, in this case lattice QCD can provide crucial input to searches for Beyond the Standard Model (BSM) physics. The tensor charge also enters in searches for BSM sources of $CP$-violation as it governs the contribution of quark electric dipole moments to the neutron electric dipole moment \cite{Bhattacharya:2015esa}. Future experimental results will likely improve the precision of phenomenological determinations of the tensor charge \cite{Ye:2016prn}, which should allow for future tests of predictions from lattice QCD. \par

Beyond nucleon matrix elements of local operators, there are observables related to higher-twist operators, such as parton distribution functions (PDFs). In particular, the average quark momentum fraction of the nucleon is of considerable phenomenological interest, as it contributes in the gauge-invariant decomposition of the nucleon spin given by Ji's sum rule \cite{Ji:1996ek}. For twist-2 operator insertions, as required for e.g. the isovector average quark momentum fraction and the second moments of helicity and transversity PDFs, again lattice QCD can be used to compute the relevant matrix elements, which are typically less well determined than the ones related to local operators. Even higher moments of PDFs would involve also higher-twist operators rendering lattice calculations infeasible due to operator mixing and further decreasing signal-to-noise ratios. \par

Excited-state contamination is one of the dominant sources of systematic uncertainty in contemporary lattice QCD calculations of nucleon matrix elements \cite{Capitani:2010sg,Capitani:2012uca,Capitani:2012ef,vonHippel:2014hla,Green:2018vxw}. This is caused by an exponentially decreasing signal-to-noise ratio making sufficiently large Euclidean time separations to suppress such unwanted contributions from excited states unaffordable in terms of computational cost. Several approaches have been used in the past in an attempt to tame these effects. They mostly rely on either explicitly fitting excited states in two- and three-point functions \cite{Bhattacharya:2013ehc,Bhattacharya:2016zcn} for a given nucleon matrix element, or on performing a summation over the operator insertion \cite{Maiani:1987by,Bulava:2011yz,Bouchard:2016heu} to achieve additional suppression of excited-state contaminations. Here we investigate an approach to simultaneously fit data for nucleon charges and second moments of PDFs at multiple source-sink separations with a common, fitted energy gap. While nucleon charges and moments of PDFs are typically studied separately on the lattice, we find that such a combined analysis has several advantages. First of all, the spectrum and thus the energy gaps depend only on the quantum numbers of the interpolating operators chosen for the nucleon state, but not on the operator insertion itself. Therefore, fitting a common energy gap allows us to fully exploit correlations between different matrix elements. We find that such simultaneous fits are much more stable compared to fitting single observables with an energy gap left free. Moreover, assuming sufficient statistics, the convergence of the fitted gap towards its theoretical expectation can be tracked as a function of the fit range in our approach, and no additional assumption is required with respect to the energy gap. \par

In this paper we present physical results for the isovector axial, scalar and tensor charge of the nucleon, its average quark momentum fraction, and the isovector moments for the helicity and transversity PDFs from isovector twist-2 operator insertions. Some preliminary results have been published in Refs.~\cite{Ottnad:2017mzd,Ottnad:2018fri}. Since we consider only isovector quantities, there are no contributions from quark-disconnected diagrams. However, we are planning to add isoscalar observables in a future publication; a first account of related work can be found in Refs.~\cite{Djukanovic:2018iir,Djukanovic:2019jtp}. \par

This paper is organized as follows: In section~\ref{sec:setup} we present the setup for our lattice calculations, including an overview of the ensembles, operators and matrix elements, technical details on the calculation of two-point and three-point functions as well as a discussion of the renormalization required to obtain physical results. Section~\ref{sec:ground state dominance} deals with the methods we employ to ensure ground-state dominance in the desired nucleon matrix elements, which is required for the extraction of physical observables from lattice data. Physical results from chiral, continuum and finite size (CCF) extrapolations are discussed in section~\ref{sec:CCF extrapolation}, and some concluding remarks are contained in section~\ref{sec:summary}. Additional technical details related to renormalization have been moved to an appendix. \par

\section{Lattice setup} \label{sec:setup}

\subsection{Ensembles} \label{subsec:ensembles}

Calculations have been performed on eleven gauge ensembles provided by the Coordinated Lattice Simulations (CLS) initiative \cite{Bruno:2014jqa}. These ensembles have been generated with $N_f=2+1$ flavors of non-perturbatively $\mathcal{O}(a)$-improved dynamical Wilson fermions and the tree-level Symanzik gauge action. A twisted-mass regulator has been introduced in the simulations to suppress exceptional configurations \cite{Luscher:2012av} and open boundary conditions in time direction are employed to alleviate the issue of long autocorrelations in the topological charge~\cite{Luscher:2011kk}. For further details on the simulations we refer to Ref.~\cite{Bruno:2014jqa}. \par

\begin{table}[!t]
 \centering
  \begin{tabular}{llllllllllll}
   \hline\hline
   ID  & $\beta$ & $T/a$ & $L/a$ & $M_\pi / \mev$ & $M_\pi L$ & $M_N / \gev$ & $N_\mathrm{HP}$ & $N_\mathrm{LP}$ & twist-2 & $\tsep / \fm$ \\
   \hline\hline
   H102 & 3.40 &  96 & 32 & 352(4) & 4.93 & 1.078(15) & 7988 &     0 & no  & 1.0, 1.2, 1.4           \\ 
   H105 & 3.40 &  96 & 32 & 278(4) & 3.90 & 1.020(18) & 4076 & 48912 & yes & 1.0, 1.2, 1.4           \\ 
   C101 & 3.40 &  96 & 48 & 223(3) & 4.68 & 0.984(12) & 2000 & 64000 & yes & 1.0, 1.2, 1.4           \\ 
   \hline
   S400 & 3.46 & 128 & 32 & 350(4) & 4.34 & 1.123(15) & 1725 & 27600 & yes & 1.1, 1.2, 1.4, 1.5, 1.7 \\ 
   N401 & 3.46 & 128 & 48 & 287(4) & 5.33 & 1.058(15) &  701 & 11216 & yes & 1.1, 1.2, 1.4, 1.5, 1.7 \\ 
   \hline
   N203 & 3.55 & 128 & 48 & 347(4) & 5.42 & 1.105(13) & 1540 & 24640 & yes & 1.0, 1.2, 1.3, 1.4, 1.5 \\ 
   S201 & 3.55 & 128 & 32 & 293(4) & 3.05 & 1.097(21) & 2092 & 66944 & yes & 1.0, 1.2, 1.3, 1.4      \\ 
   N200 & 3.55 & 128 & 48 & 283(3) & 4.42 & 1.053(14) & 1697 & 20364 & yes & 1.0, 1.2, 1.3, 1.4      \\ 
   D200 & 3.55 & 128 & 64 & 203(3) & 4.23 & 0.960(13) & 1021 & 32672 & yes & 1.0, 1.2, 1.3, 1.4      \\ 
   \hline
   N302 & 3.70 & 128 & 48 & 353(4) & 4.28 & 1.117(15) & 1177 & 18832 & yes & 1.0, 1.1, 1.2, 1.3, 1.4 \\ 
   J303 & 3.70 & 192 & 64 & 262(3) & 4.24 & 1.052(17) &  531 &  8496 & yes & 1.0, 1.1, 1.2, 1.3      \\ 
   \hline\hline
   \vspace*{0.1cm}
  \end{tabular}
  \caption{Overview of ensembles used in this study. The error on the pion and nucleon masses include the error from the scale setting. $N_\mathrm{HP}$ and $N_\mathrm{LP}$ denote to the number of high-precision (HP) and low-precision (LP) measurements on each value of $\tsep$, respectively. The column labelled ``twist-2'' indicates whether twist-2 operator insertions are available on a given ensemble. The statistics for the two-point function is always the same as for the three-point functions.}
 \label{tab:ensembles}
\end{table}

An overview of the ensembles used in the present study is shown in Tab.~\ref{tab:ensembles}. The ensembles cover four values of the lattice spacing $a$ and pion masses in a range of $\sim200\mev$ to $\sim350\mev$. Lattice volumes are chosen such that $M_\pi L \gtrsim 4$, with the exception of the S201 ensemble which has been included to enable a direct test of finite-size effects.  Values for the pion mass have been (re-)measured for most ensembles on the same set of gauge configurations that has been used in the calculation of nucleon matrix elements, hence they may slightly differ from the values originally published in Ref.~\cite{Bruno:2014jqa}. The only exception are ensembles H102 and H105 at the coarsest lattice spacing, for which we employ the values from Ref.~\cite{Bruno:2014jqa}. However, the precision of the values on the pion mass is in any case not yet relevant to the present study. \par

In Table~\ref{tab:beta_a_t0} we list the values of the lattice spacing, corresponding to the four values of $\beta$ in Tab.~\ref{tab:ensembles}, together with values for the gradient flow scale $t_0/a^2$ introduced in Ref.~\cite{Luscher:2010iy}. All results in Table~\ref{tab:beta_a_t0} are taken from Ref.~\cite{Bruno:2016plf} and we refer to this publication for further details on the scale-setting procedure. In order to set the scale in our study the physical value of $t_0$ is required, which has also been determined in Ref.~\cite{Bruno:2016plf}
\begin{equation}
 \sqrt{8 t_{0,\phys}} = 0.415(4)_\stat(2)_\sys \fm \,,
 \label{eq:t_0_phys}
\end{equation}
through the physical quantity $f_{\pi K} = \frac{2}{3}(f_K + \frac{1}{2} f_\pi)$ employing Particle Data Group values for the pion and kaon decay constant $f_\pi=130.4(2)\mev$ and $f_K=156.2(7)\mev$ \cite{Agashe:2014kda}. \par

\begin{table}[!t]
 \centering
  \begin{tabular}{lll}
   \hline\hline
   $\beta$ & $a/\mathrm{fm}$ & $t_0/a^2$ \\
   \hline\hline
   3.40 & 0.08636(98)(40) & 2.860(11)(03) \\
   3.46 & 0.07634(92)(31) & 3.659(16)(03) \\
   3.55 & 0.06426(74)(17) & 5.164(18)(03) \\
   3.70 & 0.04981(56)(10) & 8.595(29)(02) \\
   \hline\hline
  \end{tabular}
  \caption{Values of the lattice spacing $a$ and $t_0/a^2$ for each value of $\beta$ used in this study. Values are taken from Ref.~\cite{Bruno:2016plf}. The first error is statistical, second one systematic.}
 \label{tab:beta_a_t0}
\end{table}

\subsection{Operators and matrix elements} \label{subsec:operators}

In this study we aim at computing isovector axial, scalar and tensor charges that are related to the following local dimension-three operators
\begin{equation}
  \mathcal{O}^{A}_{\mu}(x)=\bar{q}(x) \g{\mu}\g{5} q(x) \,, \qquad \mathcal{O}^{S}_{}(x)=\bar{q}(x) q(x) \,, \qquad \mathcal{O}^{T}_{\mu\nu}(x)=\bar{q}(x) \sigma_{\mu\nu} q(x) \,.
 \label{eq:local_operators}
\end{equation}
Additionally, we are interested in forward matrix elements of twist-2, dimension-four operators
\begin{equation}
  \mathcal{O}^{vD}_{\mu\nu}=\bar{q} \g{\l\{\mu\r.} \DBF{\l.\nu\r\}} q \,, \qquad \mathcal{O}^{aD}_{\mu\nu}=\bar{q} \g{\l\{\mu\r.} \g{5} \DBF{\l.\nu\r\}} q \,, \qquad \mathcal{O}^{tD}_{\mu\nu\rho}=\bar{q} \sigma_{\l[\mu\l\{\nu\r.\r]} \DBF{\l.\rho\r\}} q \,,
 \label{eq:twist2_operators}
\end{equation}
where $\{...\}$ indicates symmetrization over indices with subtraction of the trace and $[...]$ denotes anti-symmetrization. Dirac matrices are labelled by $\g{\mu,5}$, $\sigma_{\mu\nu}=\frac{1}{2}\l[\g{\mu},\g{\nu}\r]$. The symmetric derivative $\DBF{}$ is defined as $\DBF{\mu}=\frac{1}{2} (\DF{\mu}-\DB{\mu})$. \par

Throughout this study we will work in Euclidean spacetime. Besides, we introduce a compact notation for which the matrix element of a given operator insertion $\mathcal{O}^X_{\mu_1...\mu_n}$ with $X\in\{A,S,T,vD,aD,tD\}$ and $n$ Lorentz indices reads
\begin{equation}
 \bra{N(p_f,s_f)} \mathcal{O}^X_{\mu_1...\mu_n} \ket{N(p_i,s_i)} = \bar{u}(p_f, s_f) W^X_{\mu_1...\mu_n}(Q^2)  u(p_i,s_i) \,,
 \label{eq:matrix_element}
\end{equation}
where $u(p_i,s_i)$, $\bar{u}(p_f, s_f)$ denote Dirac spinors with initial (final) state momentum $p_i$ ($p_f$) and spin $s_i$ ($s_f$). $W^X_{\mu_1...\mu_n}(Q^2)$ on the right-hand side is an operator-dependent form factor decomposition. For example, for the axial vector current one has
\begin{equation}
 W^A_{\mu}(Q^2) = \g{\mu} \g{5} G_A(Q^2) - i\g{5} \frac{Q_\mu}{2M_N} G_P(Q^2) \,,
 \label{eq:axial_decomposition}
\end{equation}
where $G_A(Q^2)$, $G_P(Q^2)$ are the axial and induced pseudoscalar form factor, $Q_\mu=(iE_f-iE_i, \vec{q})$ is the Euclidean four-momentum transfer with $\vec{q}=\vec{p}_f-\vec{p}_i$ and $M_N$ the nucleon mass. For further details on the relevant form factor decompositions for generalized parton distribution functions (GPDFs) we refer to Ref.~\cite{Hagler:2004yt}. \par

\begin{figure}[t]
 \centering
 \subfigure{\includegraphics[width=0.475\textwidth]{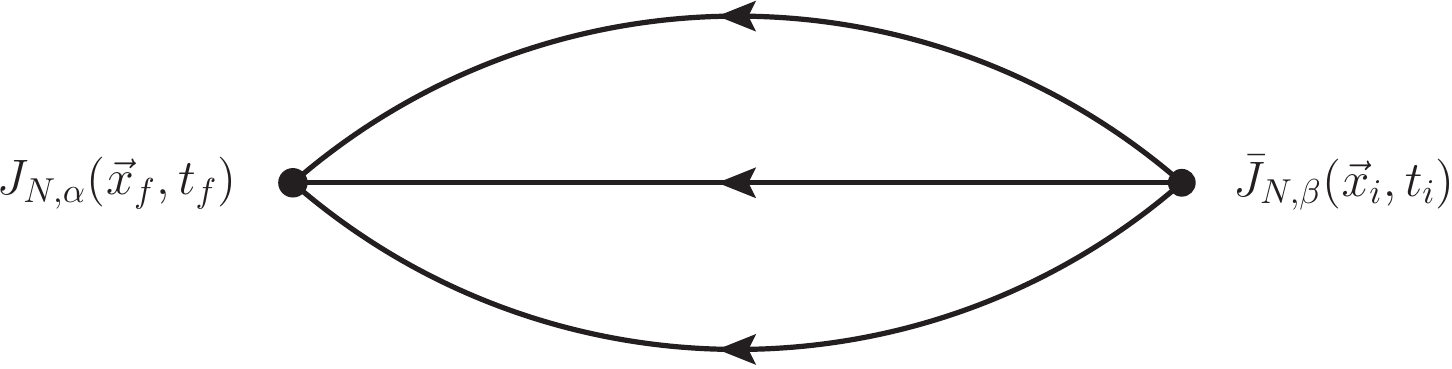}} \hfill
 \subfigure{\includegraphics[width=0.475\textwidth]{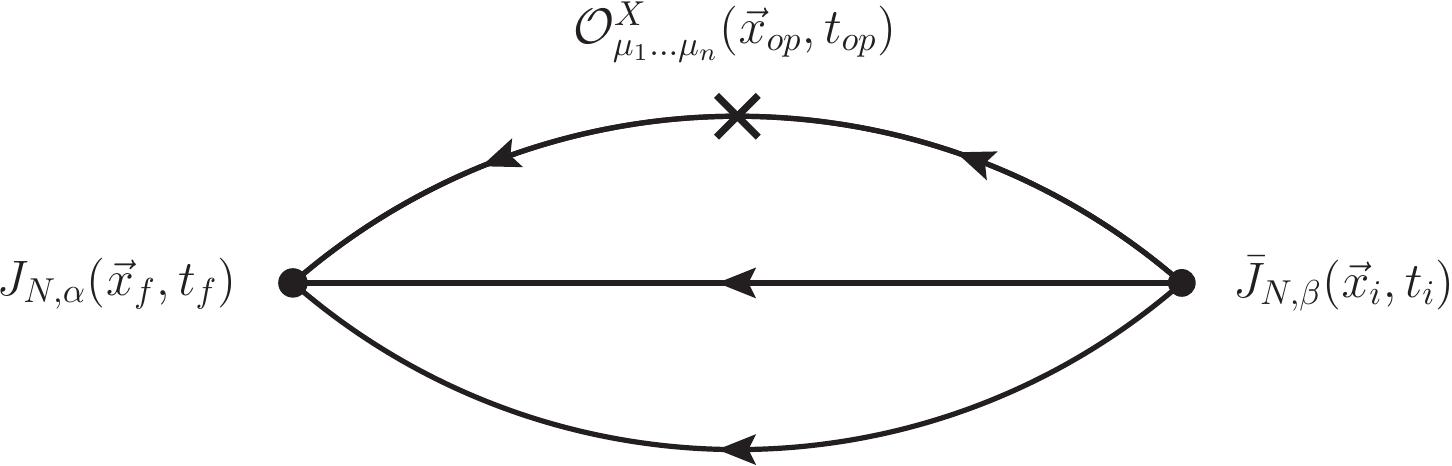}}
 \caption{Left panel: Nucleon two-point function. Right panel: Quark-connected nucleon (isovector) three-point function.}
 \label{fig:2pt_and_3pt}
\end{figure}

Obtaining nucleon matrix elements in lattice QCD requires the computation of spin-projected two- and three-point functions as depicted in Fig.~\ref{fig:2pt_and_3pt}
\begin{align}
 \Ctwopt{t_f-t_i}{p} &= \Gamma_0^{\alpha\beta} \sum_{\vec{x}_f} e^{i \vec{p} \cdot (\vec{x}_f-\vec{x}_i)} \langle J_{N,\alpha}(\vec{x}_f,t_f) \bar{J}_{N,\beta}(\vec{x}_i,t_i)\rangle \,, \label{eq:2pt_xspace} \\
 C^X_{\mu_1...\mu_n}(\vec{q}, t_{op}-t_i, t_f-t_i) &= \Gamma_z^{\alpha\beta} \sum_{\vec{x}_f, \vec{x}_{op}}  e^{i \vec{p}_f \cdot (\vec{x}_{f}-\vec{x}_{op})} e^{i \vec{p} \cdot (\vec{x}_{op}-\vec{x}_{i})} \langle J_{N,\alpha}(\vec{x}_f, t_f) \mathcal{O}^X_{\mu_1...\mu_n}(\vec{x}_{op}, t_{op}) \bar{J}_{N,\beta}(\vec{x}_i, t_i)\rangle \,. \label{eq:3pt_xspace}
\end{align}
In case of the three-point functions we employ polarization in the $z$-direction, i.e. we project with $\Gamma_z = \Gamma_0 (1+i\g{5}\g{3})$, while for the two-point functions $\Gamma_0=\frac{1}{2}(1+\g{0})$ is used, effectively averaging over all three spatial polarizations. For the two-point function we find that the latter yields a slightly better signal-to-noise ratio for e.g. the resulting nucleon masses, than using $\Gamma_z$. The proton interpolating field is given in position space by
\begin{equation}
 J_{N,\alpha}(x) = \epsilon_{abc} \l(\tilde{u}_a^T(x) C \g{5} \tilde{d}_b(x)\r) \tilde{u}_{c,\alpha}(x) \,.
 \label{eq:interpolating_operator}
\end{equation}
where $C$ is the charge conjugation matrix, and we have introduced Gaussian-smeared quark fields
\begin{equation}
 \tilde{q}=\l(1+\kappa_G\Delta\r)^N q \,, \qquad q=u,d \,.
\end{equation}
The values for the parameters $\kappa_G$ and $N$ have been chosen to correspond to a smearing radius of $\sim0.5\fm$ for each value of $\beta$. Furthermore, we apply spatial APE-smearing \cite{Albanese:1987ds} to the gauge links entering the three-dimensional Laplacian $\Delta$, to improve the ground state projection for the relevant matrix elements and to gain additional noise reduction. \par

For the following discussion we define the source-sink separation $\tsep=t_f - t_i$ and introduce the shorthand $t=t_{op}-t_i$. W.l.o.g. we will assume that the source time is zero, i.e. $t_i=0$, corresponding to a index shift in the actual calculation. Moreover, we demand that the final state is produced at rest, i.e. $\vec{p}_f=0$, $\vec{q}=-\vec{p}_i$. In momentum space the two- and three-point functions in Eqs.~(\ref{eq:2pt_xspace},\ref{eq:3pt_xspace}) can then be written as
\begin{align}
 \Ctwopt{\tsep}{q} &= \Gamma_0^{\alpha\beta} \langle J_{N,\alpha}(-\vec{q}, \tsep) \bar{J}_{N,\beta}(-\vec{q}, 0)\rangle \,, \label{eq:2pt} \\
 C^X_{\mu_1...\mu_n}(\vec{q}, t, \tsep) &= \Gamma_z^{\alpha\beta} \langle J_{N,\alpha}(\vec{0}, \tsep) \mathcal{O}^X_{\mu_1...\mu_n}(\vec{q}, t) \bar{J}_{N,\beta}(-\vec{q}, 0)\rangle \,. \label{eq:3pt}
\end{align}
Extracting the physical matrix elements requires the cancelation of unknown overlap factors in the three-point function, which in the case of vanishing momentum transfer $Q^2=0$ can be achieved by forming the ratio
\begin{equation}
 R^X_{\mu_1...\mu_n}(\vec{0},t,\tsep) = \frac{\Cthreept{X}{\mu_1...\mu_n}{t}{\tsep}{0}}{\Ctwopt{\tsep}{0}} \,.
 \label{eq:ratio}
\end{equation}
In the limit of large Euclidean time separations $t$ and $\tsep-t$ the ratio turns into a plateau as it becomes dominated by the ground state, i.e.
\begin{equation}
 \lim_{t\rightarrow\infty} \ \lim_{(\tsep-t)\rightarrow\infty} R^X_{\mu_1...\mu_n}(\vec{0},t,\tsep) = \mathrm{const}
\end{equation}
For the local operators in Eqs.~(\ref{eq:local_operators}) one obtains the following, asymptotic relations at large Euclidean times for the isovector axial-, scalar- and tensor charges $g_A^{u-d}$, $g_S^{u-d}$ and $g_T^{u-d}$
\begin{align}
 R^{A}_{\mu}(\vec{0},t,\tsep)    &\rightarrow i\delta_{3\mu} g_A^{u-d} \,, \\
 R^{S}(\vec{0},t,\tsep)          &\rightarrow g_S^{u-d} \,, \\
 R^{T}_{\mu\nu}(\vec{0},t,\tsep) &\rightarrow \epsilon_{03\mu\nu} g_T^{u-d} \,.
 \label{eq:charges}
\end{align}
The decompositions for the isovector combinations of the dimension-four operators in Eqs.~(\ref{eq:twist2_operators}) lead to
\begin{align}
 R^{vD}_{\mu\nu}(\vec{0},t,\tsep) &\rightarrow m \l( \delta_{0\mu} \delta_{0\nu} - \frac{1}{4} \delta_{\mu\nu} \r) \avgx{-}{} \,, \\ 
 R^{aD}_{\mu\nu}(\vec{0},t,\tsep) &\rightarrow \frac{i m}{2} \l(\delta_{3\mu} \delta_{0\nu} + \delta_{0\mu} \delta_{3\nu}\r) \avgx{-}{\Delta} \,, \\ 
 R^{tD}_{\mu\nu\rho}(\vec{0},t,\tsep) &\rightarrow -\frac{i m}{4} \epsilon_{\mu\nu\rho 3} \l(2 \delta_{0\rho} - \delta_{0\nu} - \delta_{0\mu} \r) \avgx{-}{\delta} \,, 
 \label{eq:twist2}
\end{align}
where in the GPDF notation of Ref.~\cite{Hagler:2004yt} we have defined the isovector average quark momentum fraction $\avgx{-}{}=A_{20}^{u-d}(Q^2=0)$, helicity moment $\avgx{-}{\Delta}=\tilde{A}_{20}^{u-d}(Q^2=0)$ and transversity moment $\avgx{-}{\delta}=A_{T20}^{u-d}(Q^2=0)$. In the actual calculation we always average over all contributing, numerically non-identical index permutations. \par

\subsection{Computation of two- and three-point functions} \label{subsec:npt_functions}

Apart from the actual generation of the gauge ensembles, the computationally most expensive part of this study is the calculation of two- and especially three-point functions in Eqs.~(\ref{eq:2pt},\ref{eq:3pt}). Therefore, we employ the truncated solver method \cite{Bali:2009hu,Blum:2012uh,Shintani:2014vja} on most ensembles to reduce the cost of the required inversions. The method is based on the idea of using a (relatively) large number of low-precision $N_\mathrm{LP}$ inversions to obtain a statistically precise estimate of the actual observable and only a small number $N_\mathrm{HP}$ of high-precision measurements to correct for the resulting bias in the final expectation value
\begin{equation}
 \langle\mathcal{O}\rangle = \langle\frac{1}{N_\mathrm{LP}}\sum_{i=1}^{N_\mathrm{LP}}\mathcal{O}_i^\mathrm{LP}\rangle + \langle \mathcal{O}_\mathrm{bias}\rangle \,, \quad \mathcal{O}_\mathrm{bias} = \frac{1}{N_\mathrm{HP}}\sum_{i=1}^{N_\mathrm{HP}}(\mathcal{O}_i^\mathrm{HP} - \mathcal{O}_i^\mathrm{LP}) \,.
  \label{eq:truncated_solver_method}
\end{equation}
For the ensembles in Table~\ref{tab:ensembles} we typically observe a factor $\sim2$ to $3$ improvement in computer time compared to using only exact solves. The total numbers of low- and high-precision inversions for each ensembles can be found in Tab.~\ref{tab:ensembles}. \par

For the computation of three-point functions we perform sequential inversions through the sink with the final state produced at rest. Depending on the value of the lattice spacing and the available statistics, we compute three-point functions for at least three and up to five values of $t_\mathrm{sep}$. This allows us to check the dependence on the source-sink separation, which is instrumental in dealing with excited-state contamination. The values of $t_\mathrm{sep}$ in physical units are shown in Table~\ref{tab:ensembles}. Note that we do not include values of $t_\mathrm{sep}$ smaller than $1\fm$. For the initial (forward) propagator we use point sources distributed on a single timeslice in the center bulk of the lattice. Typically, the actual position of the source timeslice $t_i$ (before performing the index shift $t_i\rightarrow 0$) on a given ensemble is chosen such that $ t_\mathrm{sep}^\mathrm{min} = T - 2t_i$ holds for the smallest, available value of the source-sink separation $t_\mathrm{sep}^\mathrm{min}$. Since the ensembles used in this study have been generated with open boundary conditions, this choice guarantees that all operators remain sufficiently far away from the boundaries in time, hence preventing further contamination due to boundary effects. Finally, two-point functions are generally computed on the same source timeslice $t_i$ and with the same statistics as the three-point functions. \par

\subsection{Renormalization} \label{subsec:renormalization}

Unlike hadron masses which are renormalization group invariants, matrix elements as given in Eq.~(\ref{eq:matrix_element}) typically require renormalization. To this end we have performed the non-perturbative renormalization for the relevant operators using the Rome-Southampton method \cite{Martinelli:1994ty} at each lattice spacing except for the finest one. The reason for this is that at lattice spacings of $a\lesssim 0.05\fm$ topological charge freezing is expected to become a severe issue, hence simulations with periodic boundary conditions as required by the Rome-Southampton method are not feasible in such a setup.
Our results are summarized in the top portion of Tables~\ref{tab:Zloc} and~\ref{tab:Ztwist2} for local and twist-2 operators, respectively. They are all given in the $\MSbar$ scheme at a scale of $\mu=2\,\gev$ and we have included results for both irreducible representations for the twist-2 operators. For the renormalization of the twist-2 matrix elements that are actually computed in our study we require only one of the irreps in each case, i.e. $Z^{\MSbar}_{v2b}$, $Z^{\MSbar}_{r2a}$ and $Z^{\MSbar}_{h1a}$ for $\avgx{-}{}$, $\avgx{-}{\Delta}$ and $\avgx{-}{\delta}$, respectively. For further details of our renormalization procedure and associated notation, we refer to appendix~\ref{app:npr}. \par

For the required values of the renormalization constants at our finest lattice spacing at $\beta=3.7$, we have resorted to extrapolations, introducing a further source of uncertainty. The numerical results from this procedure are summarized in the bottom lines of Tables~\ref{tab:Zloc} and~\ref{tab:Ztwist2} for local and twist-2 operators, respectively. The errors for the extrapolated values have been scaled by a factor of ten to account for the systematic uncertainty of this procedure. Examples of the extrapolations are shown in Fig.~\ref{fig:extrapolation_zx_b3.7} in the appendix, indicating that the final error is estimated very conservatively.\par

Nevertheless, even allowing for a generous error margin on the extrapolated $Z$-factors may not entirely disperse all doubts concerning the reliability of this procedure; however, in the case of the axial-vector matrix element we have performed a thorough cross-check using the results for $Z_A$ determined from the chirally-rotated Schr\"odinger functional \cite{DallaBrida:2018tpn}, which are available for all four values of $\beta$ used in our study. This offers the possibility for cross-checking the validity of the extrapolation that we applied in the case of $\beta=3.7$ from the perspective of the final, (combined) continuum extrapolation. Moreover, this alternative renormalization method allows for a more consistent $\mathcal{O}(a)$ improvement in the case of $g_A^{u-d}$. We will validate our extrapolation for $Z_A$ by a detailed comparison with Schr\"odinger-functional results in section~\ref{subsec:sfrimon}. \par

The values of $Z^\mathrm{SF}_{A}$ used in this study have been collected in Table~\ref{tab:ZSF} together with results for the improvement coefficient $b_A$ taken from Ref.~\cite{Korcyl:2016ugy} as well as values of $\kappa_\mathrm{crit}$ determined in Ref.~\cite{Gerardin:2018kpy}. The final, $\mathcal{O}(a)$-improved renormalization factors are then dependent on the bare coupling constant $g_0$ as well as on the quark mass $m_q=\frac{1}{2a}(\frac{1}{\kappa_q} - \frac{1}{\kappa_\mathrm{crit}})$ where $q=l,s$ and the average quark mass $\bar{m}=\frac{1}{3}(2m_l + m_s)$
\begin{equation}
 Z_A^\mathrm{imp}(g_0^2, m_q, \bar{m}) = Z_A(g_0^2) \left(1 + am_q b_A(g_0^2) + 3 a \bar{m} \tilde{b}_A(g_0^2) \right) \,.
 \label{eq:Z_A_SF_improved}
\end{equation}
The last term depends on an additional improvement coefficient $\tilde{b}_A$ for which results have not been published for all four values of $\beta$. However, it is formally of $\mathcal{O}(g_0^4)$ and hence likely to be suppressed. Moreover, it has been found in Ref.~\cite{Korcyl:2016ugy}  at the coarsest lattice spacing for ensemble H102 that the value of $\tilde{b}_A$ is indeed compatible with zero, albeit with large statistical errors. Therefore, we will drop this term from our analysis. \par

\begin{table}[!t]
 \centering
 \begin{tabular}{@{\extracolsep{\fill}}lccc}
  \hline\hline
  $\beta$ & $Z_A$ & $Z^{\MSbar}_S$ & $Z^{\MSbar}_T$ \\
  \hline\hline
  3.40 & 0.7533(18) & 0.6506(82) & 0.8336(35) \\ 
  3.46 & 0.7604(16) & 0.6290(82) & 0.8475(33) \\ 
  3.55 & 0.7706(14) & 0.6129(81) & 0.8666(33) \\ \hline 
  3.70 & 0.7879(33) & 0.575(18)  & 0.900(7)   \\ 
  \hline\hline
  \end{tabular}
  \caption{Renormalization factors corresponding to the three local operator insertions used in this study. Results are obtained from the Rome-Southampton method and given in the $\MSbar$--scheme at a scale of $\mu=2\gev$ (where applicable). Statistical and systematic errors have been added in quadrature. Values for $\beta=3.7$ are obtained by an extrapolation.}
 \label{tab:Zloc}
\end{table}

\begin{table}[!t]
 \centering
 \begin{tabular}{@{\extracolsep{\fill}}lcccccc}
  \hline\hline
  $\beta$ & $Z^{\MSbar}_{v2a}$ & $Z^{\MSbar}_{v2b}$ & $Z^{\MSbar}_{r2a}$ & $Z^{\MSbar}_{r2b}$ & $Z^{\MSbar}_{h1a}$ & $Z^{\MSbar}_{h1b}$ \\
  \hline\hline
  3.40 & 1.105(10) & 1.117(10) & 1.097(10) & 1.134(10) & 1.138(12) & 1.147(12) \\ 
  3.46 & 1.122(10) & 1.129(10) & 1.115(10) & 1.148(10) & 1.157(12) & 1.167(12) \\ 
  3.55 & 1.157(10) & 1.161(10) & 1.150(10) & 1.180(10) & 1.196(12) & 1.205(12) \\ \hline 
  3.70 & 1.209(23) & 1.204(23) & 1.203(22) & 1.224(23) & 1.253(27) & 1.262(27) \\ 
  \hline\hline
  \end{tabular}
  \caption{Renormalization factors corresponding to the twist-2 operator insertions used in this study. Results are obtained from the Rome-Southampton method and given in the $\MSbar$--scheme at a scale of $\mu=2\gev$ and values for both irreps of each operator (cf.~appendix \ref{app:npr} for notation) have been included. Statistical and systematic errors have been added in quadrature. Values for $\beta=3.7$ are obtained by an extrapolation.}
 \label{tab:Ztwist2}
\end{table}

\begin{table}[!t]
 \centering
 \begin{tabular}{@{\extracolsep{\fill}}lccc}
  \hline\hline
  $\beta$ & $Z^\mathrm{SF}_{A}$ & $b_A$ & $\kappa_\mathrm{crit}$ \\
  \hline\hline
  3.40 & 0.75485(68) & 1.71(11) & 0.1369115 \\ 
  3.46 & 0.76048(80) & 1.49(20) & 0.1370645 \\
  3.55 & 0.76900(42) & 1.38(12) & 0.1371726 \\
  3.70 & 0.78340(43) & 1.26(09) & 0.1371576 \\
  \hline\hline
  \end{tabular}
  \caption{Axial vector renormalization factors $Z_A^\mathrm{SF}$ from the Schr\"odinger functional method as given in Ref.~\cite{DallaBrida:2018tpn}, improvement coefficients $b_A$ from Ref.~\cite{Korcyl:2016ugy} and values for $\kappa_\mathrm{crit}$ as determined in Ref.~\cite{Gerardin:2018kpy}. In the notation of Ref.~\cite{DallaBrida:2018tpn} we choose $Z_\mathrm{A,sub}^l$ from the $L_1$ constant line of physics for $Z^\mathrm{SF}_{A}$. Statistical and systematic errors have been added in quadrature.}
 \label{tab:ZSF}
\end{table}

\section{Ground-state dominance} \label{sec:ground state dominance}

It is a well-established fact that nucleon structure calculations in lattice QCD are hampered by excited-state contamination \cite{Capitani:2012gj}. This is caused by a signal-to-noise problem preventing the use of sufficiently large source-sink separations in the calculation of nucleon three-point functions. Therefore, in practice it is not feasible to directly extract a reliable ground-state plateau value from lattice data for the ratio in Eq.~(\ref{eq:ratio}). We have investigated several approaches to deal with excited states and extract the final observables. \par

\subsection{Multi-state fits} \label{subsec:multi-state fits}

Our main approach to tackle excited-state contamination in nucleon structure calculations are multi-state fits to lattice data for the ratio in Eq.~(\ref{eq:ratio}). Inserting complete sets of states in the two- and three-point functions in Eqs.~(\ref{eq:2pt},\ref{eq:3pt}) their spectral representation can be parameterized as
\begin{align}
 \Ctwopt{\tsep}{q} &= \sum_{k=0}^{\infty} a_k(\vec{q}) e^{-E_k(\vec{q}) t} \,, \label{eq:2pt_exp} \\
 \Cthreept{X}{\mu_1...\mu_n}{t}{\tsep}{q} &= \sum_{k=0}^{\infty} \sum_{l=0}^{\infty} A^{X,kl}_{\mu_1 ... \mu_n}(\vec{q}) e^{-E_k(\vec{0})(\tsep-t) -E_l(\vec{q})t} \,, \label{eq:3pt_exp}
\end{align}
in terms of observable-independent energies $E_k(\vec{q})$ and observable-dependent factors $a_k(\vec{p})$, $A^{X,kl}_{\mu_1 ... \mu_n}(\vec{q})$ containing amplitudes and further kinematical expressions. The exact form of the latter will not be relevant for our purposes in this section. Moreover, suppressing all indices related to the operator insertion by introducing the shorthand $A_{kl}(\vec{q}) = A^{X,kl}_{\mu_1 ... \mu_n}(\vec{q})$ and defining
\begin{equation}
 \tilde{A}_{kl}(\vec{q})=A_{kl}(\vec{q}) / A_{\mathrm{min}(k,l) \mathrm{min}(k,l)}(\vec{q}) \,,
 \label{eq:A_tilde}
\end{equation}
the three-point function in Eq.~(\ref{eq:3pt_exp}) can be rewritten as
\begin{align}
 \Cthreept{X}{\mu_1...\mu_n}{t}{\tsep}{q} =& \sum_{k=0}^{\infty} A_{kk}(\vec{q})  e^{-E_k(\vec{0})(\tsep-t) -E_k(\vec{q})t} \times \notag \\ 
 & \ \times  \left( 1 + \sum_{l=k+1}^{\infty} \left( \tilde{A}_{kl}(\vec{q}) e^{-\Delta_{kl}(\vec{q}) t} + \tilde{A}_{lk}(\vec{q}) e^{-\Delta_{kl}(\vec{0}) (\tsep-t)}  \right) \right) \,,
 \label{eq:3pt_rewritten}
\end{align}
where we have introduced the energy gaps $\Delta_{kl}(\vec{q}) = E_{l}(\vec{q}) - E_{k}(\vec{q})$. Assuming vanishing momentum transfer $\vec{q}=0$ as required in our actual calculation and suppressing all occurrences of zero momenta in the notation the expression is further simplified to become
\begin{equation}
 \Cthreept{X}{\mu_1...\mu_n}{t}{\tsep}{0} = \sum_{k=0}^{\infty} A_{kk}  e^{-m_k \tsep} \left( 1 + \sum_{l=k+1}^{\infty} \tilde{A}_{kl} \left( e^{-\Delta_{kl} t} + e^{-\Delta_{kl} (\tsep-t)}\right) \right)  \,,
 \label{eq:3pt_zero_momentum}
\end{equation}
where we made use of the fact that $\tilde{A}_{kl}(\vec{0})=\tilde{A}_{lk}(\vec{0})$ for the current insertions we consider. Keeping only terms involving the lowest gap $\Delta = \Delta_{01}$, one arrives at the following expression for the ratio in Eq.~(\ref{eq:ratio})
\begin{equation}
 R^X_{\mu_1...\mu_n}(\vec{0},t,\tsep) = \bar{A}^{X,00}_{\mu_1...\mu_n} + \bar{A}^{X,01}_{\mu_1...\mu_n} \left( e^{-\Delta t} + e^{-\Delta (\tsep-t)} \right) + \bar{\bar{A}}^{X}_{\mu_1...\mu_n} e^{-\Delta \tsep}  \,,
 \label{eq:multi_state_fit}
\end{equation}
where we defined
\begin{equation}
 \bar{A}^{X,kl}_{\mu_1...\mu_n} = \frac{A^{X,kl}_{\mu_1 ... \mu_n}(\vec{0})}{a_0(\vec{0}) } 
\end{equation}
and
\begin{equation}
  \bar{\bar{A}}^X_{\mu_1...\mu_n} = \bar{A}^{X,11}_{\mu_1...\mu_n} - \bar{A}^{X,00}_{\mu_1...\mu_n} \cdot \frac{a_1(\vec{0})}{a_0(\vec{0})} \,.
\end{equation}
The first term on the r.h.s. is then a (linear combination of) form factor(s) at vanishing momentum transfer depending on the operator insertion $X$ and the spin-projection in the original three-point function, e.g. for $X=A$ and $\mu=3$ for our choice of projectors one finds that $\bar{A}^{A,00}_{3}$ gives the axial charge. The expression in Eq.~(\ref{eq:multi_state_fit}) represents our final fit model, which has already been applied in a previous analysis of lattice data with $N_f=2$ dynamical quark flavors in Ref.~\cite{Capitani:2017qpc}. In principle, it is possible to fit the model in Eq.~(\ref{eq:multi_state_fit}) leaving the gap as a free parameter; however, this requires very precise data and leads to rather large errors on the estimate for the corresponding observables. Still, from a theoretical point of view it is desirable to apply such fits without additional assumptions, in contrast to Ref.~\cite{Capitani:2017qpc}, where the gap was fixed to $\Delta_0=2M_\pi$ on each ensemble. Therefore, we choose a more sophisticated approach, fitting the model in Eq.~(\ref{eq:multi_state_fit}) with a single free gap $\Delta$ to all observables and for all available values of $\tsep$ simultaneously. This is possible because the gaps are only related to the interpolating operators which are chosen the same for all the nucleon matrix elements in this study. These simultaneous fits yield much more stable fits compared to fitting a free gap to a single observable only. In fact, we find that they often outperform simple two-state fits with a fixed gap with respect to the resulting error on $\bar{A}^{X,00}_{\mu_1...\mu_n}$ as correlations in the data are more thoroughly exploited. \par

Since the fit form in Eq.~(\ref{eq:multi_state_fit}) is symmetric in $t$ around $\tsep/2$ we explicitly symmetrize the data before fitting, which leaves a fit range of $t\in [t_\mathrm{fit}, \tsep/2]$ at each value of $\tsep$. Furthermore, we restrict ourselves to a consistent set of source-sink separations for each value of $\beta$ as listed in Table~\ref{tab:sim_fits}. As a result, we drop the largest available source-sink separation from the fit in a few cases. The data at these additional, largest source-sink separations are typically very noisy and do not affect the final results much within errors, and dropping them entirely can lead to more stable fits. This is especially so because the problem size is reduced, and hence the estimate of the inverse covariance matrix becomes more reliable. \par 

\begin{figure}[t]
 \centering
 \subfigure{\includegraphics[width=.48\linewidth]{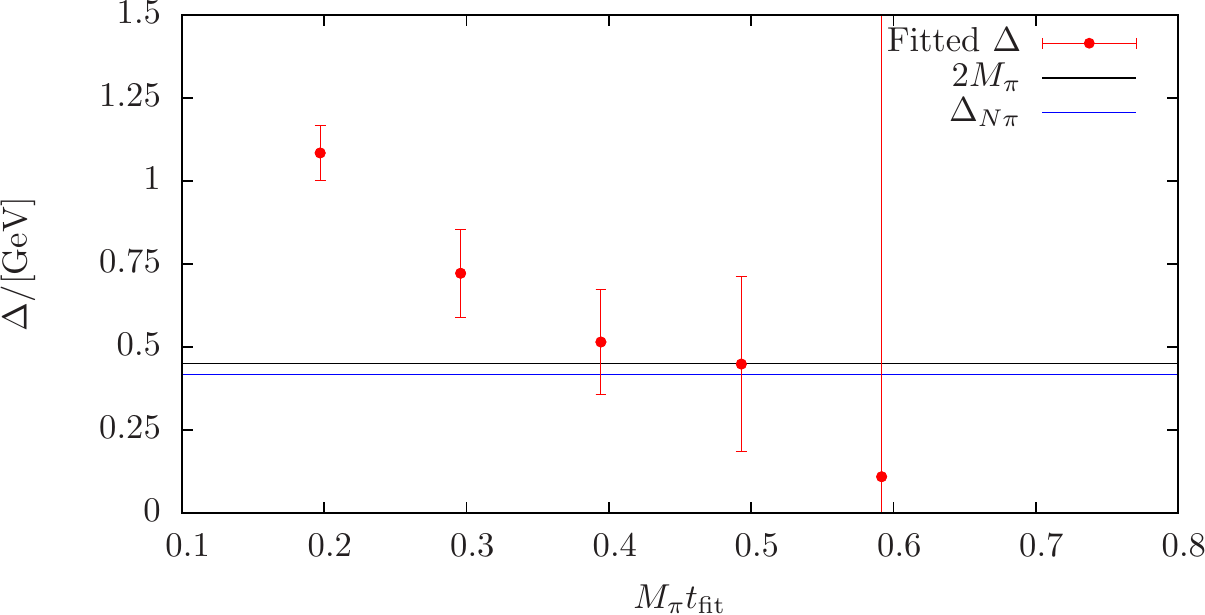}} \quad
 \subfigure{\includegraphics[width=.48\linewidth]{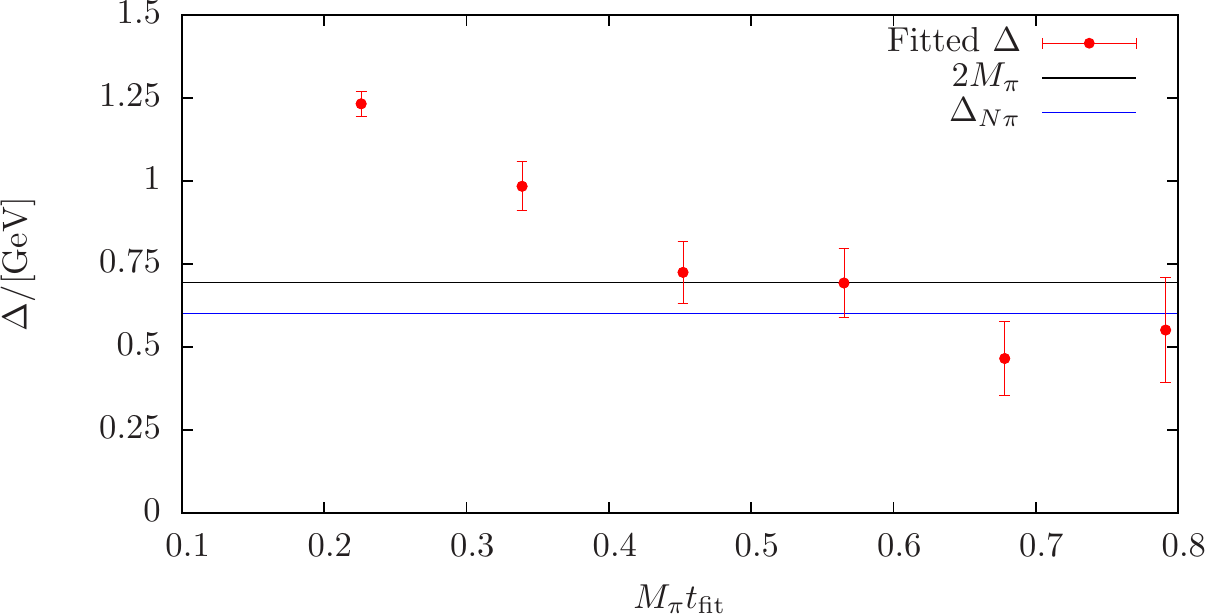}}
 \caption{Behavior of the fitted gap $\Delta$ in Eq.~(\ref{eq:multi_state_fit}) as a function of the variable $M_\pi t_\mathrm{fit}$ representing the lower bound on the fit range. Left panel: ensemble C101; right panel: ensemble N203.}
 \label{fig:gap}
\end{figure}

When selecting time intervals for the simultaneous fits, some care is required to ensure that the fitted gap is stable under variation of the fit interval, since the excitation spectrum is very dense. In the actual fits, we demand $M_\pi t_\mathrm{fit} \geq 0.4$ on all our ensembles, which we found to be a reasonable compromise between the statistical precision and the suppression of further excited states. On some of the ensembles, however, due to the high statistical precision achieved, it is necessary to be more restrictive and leave out further data points. The final choices of $t_\mathrm{fit}/a$ can be found in Table~\ref{tab:sim_fits} together with the resulting correlated $\chi^2/\mathrm{dof}$ and $p$-values, as well as the renormalized results for the individual observables on each ensemble.\footnote{Note that for the purpose of this table we have consistently applied results from the Rome-Southampton method as given in Tables~\ref{tab:Zloc} and~\ref{tab:Ztwist2}.} Demanding at least a consistent lower bound on $t_\mathrm{fit}$ in units of $M_\pi$ is motivated by the expectation that the lowest gap for our ensembles will typically be close to $2M_\pi$. On ensembles with large enough statistics it is actually possible to track the convergence of the gap as a function of $t_\mathrm{fit}$, which allows us to further corroborate the choice of $t_\mathrm{fit}$ in these cases. This is illustrated in Fig.~\ref{fig:gap} for two of our ensembles (C101, N203). Clearly, in both cases the value of $\Delta$ approaches $2 M_\pi$ within errors, which are increasing with $M_\pi t_\mathrm{fit}$. Keeping in mind that we are not actually interested in a precise determination of the value of the gap itself, we generally choose the fit range such that the gap has converged within statistical errors (at least on ensembles for which it can be sufficiently tracked) while statistical precision still allows for a stable fit and a meaningful extraction of the final observable. \par

The goal of our simultaneous multi-state fits is to suppress the residual excited-state contamination to a level which is no larger than the statistical precision. These fits can be systematically improved by
\begin{enumerate}
 \item increasing statistics while choosing more restrictive bounds on $t_\mathrm{fit}$, \\
 \item adding further terms in the fit corresponding to additional terms in Eq.~(\ref{eq:3pt_zero_momentum}), provided one has sufficient statistics to retain a stable fit, \\
 \item including further observables, which we found to stabilize the fits and reduce the resulting error. \\
\end{enumerate}
Besides, it is possible to use similar fits beyond the case of vanishing momentum transfer, by removing the assumption of symmetric plateaux. A fit model analogous to in Eq.~\ref{eq:multi_state_fit} can be derived for this case, although it will contain additional amplitudes and gaps due to the momentum transfer. Finally, we remark that from a theoretical point of view these simultaneous fits also supersede earlier attempts using a fixed gap as used in Ref.~\cite{Capitani:2017qpc} with statistically much less precise data. \par

\begin{table}[!t]
 \setlength{\tabcolsep}{4.25pt}
 \centering
  \begin{tabular}{lllllcccccc}
   \hline\hline
    ID   & $t_\mathrm{sep}/a$ used & $t_\mathrm{fit}/a$ & $\chi^2/\mathrm{dof}$ & $p$ & $g_A^{u-d}$ & $g_S^{u-d}$ & $g_T^{u-d}$ & $\avgx{-}{}$ & $\avgx{-}{\Delta}$ & $\avgx{-}{\delta} $\\
   \hline\hline
    H102 & 12,14,16    & 5 & 0.957 & 0.504 & 1.129(14) & 0.92(06) & 1.033(15) &        -- &        -- &        -- \\
    H105 & 12,14,16    & 5 & 1.271 & 0.131 & 1.101(24) & 0.70(12) & 1.005(16) & 0.223(13) & 0.255(09) & 0.270(08) \\
    C101 & 12,14,16    & 4 & 0.755 & 0.906 & 1.204(35) & 0.94(10) & 0.966(34) & 0.165(33) & 0.218(19) & 0.196(45) \\
    \hline
    S400 & 14,16,18,20 & 4 & 1.188 & 0.084 & 1.130(22) & 0.98(09) & 1.014(19) & 0.210(13) & 0.245(12) & 0.254(18) \\
    N401 & 14,16,18,20 & 5 & 1.523 & 0.001 & 1.186(22) & 0.89(07) & 1.047(14) & 0.202(13) & 0.228(11) & 0.237(16) \\
    \hline
    N203 & 16,18,20,22 & 6 & 0.894 & 0.754 & 1.195(26) & 1.27(13) & 0.976(27) & 0.161(27) & 0.193(29) & 0.195(34) \\
    S201 & 16,18,20,22 & 5 & 1.098 & 0.224 & 1.011(32) & 0.89(18) & 0.948(52) & 0.211(21) & 0.233(33) & 0.247(42) \\
    N200 & 16,18,20,22 & 5 & 0.964 & 0.592 & 1.160(16) & 1.06(07) & 0.996(16) & 0.198(10) & 0.236(08) & 0.246(11) \\
    D200 & 16,18,20,22 & 6 & 1.209 & 0.088 & 1.188(25) & 0.99(13) & 0.940(20) & 0.189(14) & 0.230(14) & 0.234(21) \\
    \hline
    N302 & 20,22,24,26 & 6 & 1.536 & 0.000 & 1.148(21) & 1.11(09) & 0.961(28) & 0.196(15) & 0.202(19) & 0.205(26) \\
    J303 & 20,22,24,26 & 8 & 0.892 & 0.757 & 1.160(19) & 0.96(09) & 1.021(18) & 0.206(09) & 0.245(10) & 0.252(14) \\
   \hline\hline
  \end{tabular}
  \caption{Parameters including correlated $\chi^2/\mathrm{dof}$ and $p$-values and renormalized results for all six observables from simultaneous fits on each ensemble. Note that the set of source-sink separations that has been used in the fits differs in a few cases from the full list of available data given in Table~\ref{tab:ensembles}; see discussion in text.}
 \label{tab:sim_fits}
\end{table}

\begin{table}[!t]
 \centering
  \begin{tabular}{llcccccc}
   \hline\hline
    ID   & $t_\mathrm{sep}/a$ used & $g_A^{u-d}$ & $g_S^{u-d}$ & $g_T^{u-d}$ & $\avgx{-}{}$ & $\avgx{-}{\Delta}$ & $\avgx{-}{\delta} $\\
   \hline\hline
    H102 & 12,14,16       & 1.166(55) &  0.70(29) & 1.006(59) & --        & --        & --        \\
    H105 & 12,14,16       & 1.23(10)  & -0.21(57) & 0.922(88) & 0.179(26) & 0.249(29) & 0.233(31) \\
    C101 & 12,14,16       & 1.173(35) &  0.91(26) & 1.019(29) & 0.200(10) & 0.201(12) & 0.259(11) \\
    \hline
    S400 & 14,16,18,20,22 & 1.151(56) &  1.13(21) & 1.048(49) & 0.220(12) & 0.246(15) & 0.258(15) \\
    N401 & 14,16,18,20,22 & 1.321(64) &  1.47(29) & 1.131(51) & 0.190(15) & 0.193(15) & 0.195(20) \\
    \hline
    N203 & 16,18,20,22,24 & 1.197(28) &  1.41(10) & 1.032(22) & 0.194(06) & 0.239(07) & 0.243(08) \\
    S201 & 16,18,20,22    & 0.97(14)  &  1.35(75) & 1.09(15)  & 0.197(29) & 0.176(35) & 0.192(41) \\
    N200 & 16,18,20,22    & 1.187(60) &  1.26(31) & 1.063(47) & 0.181(11) & 0.223(15) & 0.250(28) \\
    D200 & 16,18,20,22    & 1.193(68) &  1.46(46) & 0.929(56) & 0.127(16) & 0.200(17) & 0.196(23) \\
    \hline
    N302 & 20,22,24,26,28 & 1.039(60) &  1.27(22) & 0.934(47) & 0.179(13) & 0.193(15) & 0.217(16) \\
    J303 & 20,22,24,26    & 1.218(73) &  0.98(37) & 0.988(66) & 0.187(16) & 0.247(21) & 0.195(24) \\
   \hline\hline
  \end{tabular}
  \caption{Renormalized results from the summation method for all six observables. Here we have used data from the full set of available source-sink separations as listed in Table~\ref{tab:ensembles}.}
 \label{tab:summation_method}
\end{table}

\subsection{Summation method} \label{subsec:summation_method}

\begin{figure}[!t]
 \centering
 \subfigure{\includegraphics[totalheight=0.2\textheight]{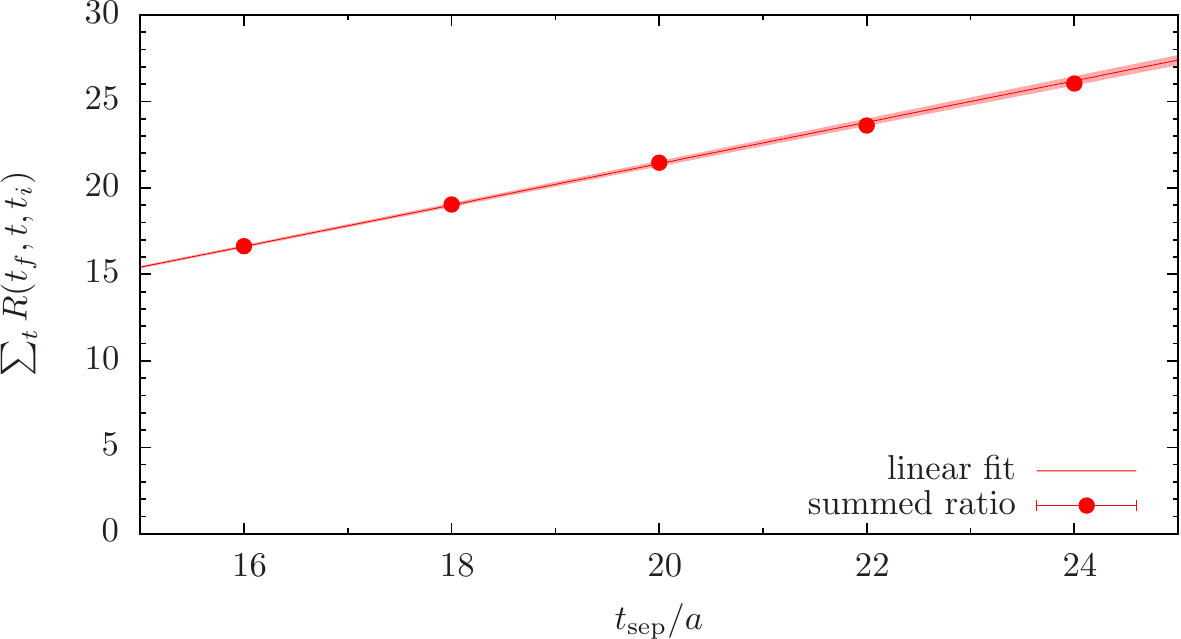}}  \quad
 \subfigure{\includegraphics[totalheight=0.2\textheight]{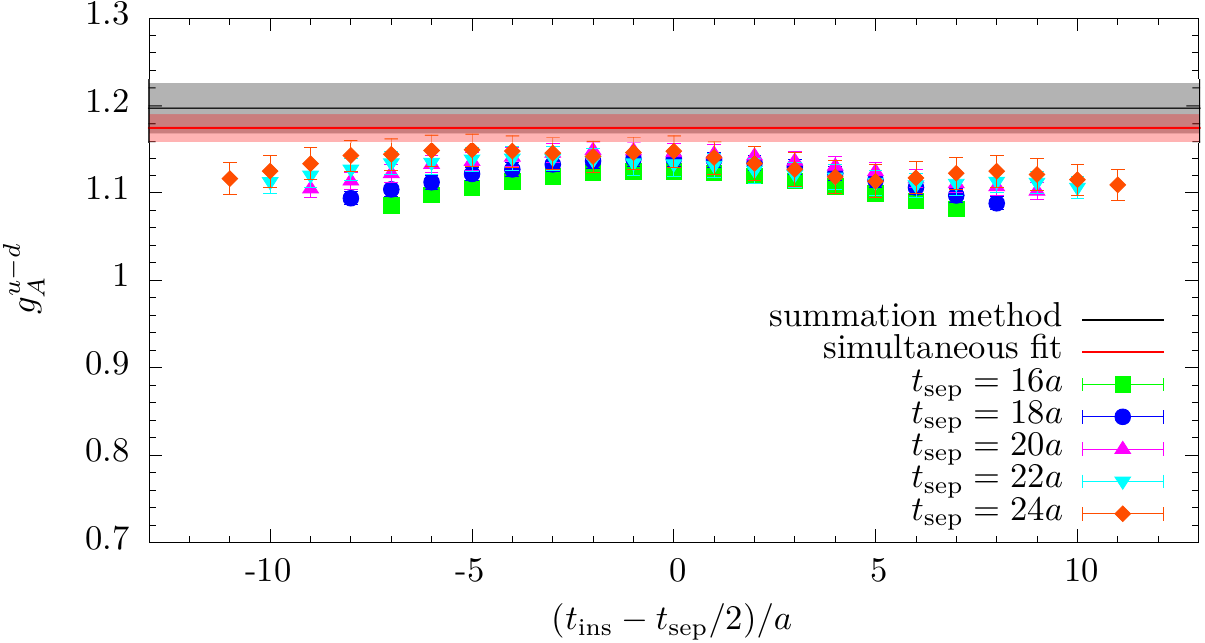}}  \\
 \subfigure{\includegraphics[totalheight=0.2\textheight]{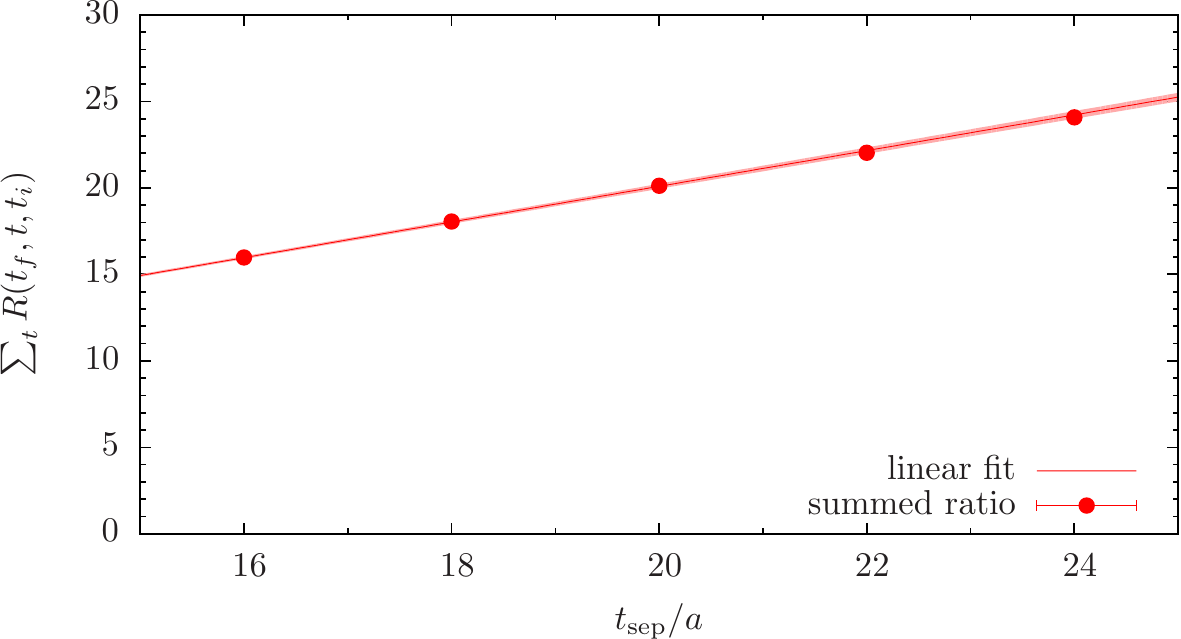}} \quad
 \subfigure{\includegraphics[totalheight=0.2\textheight]{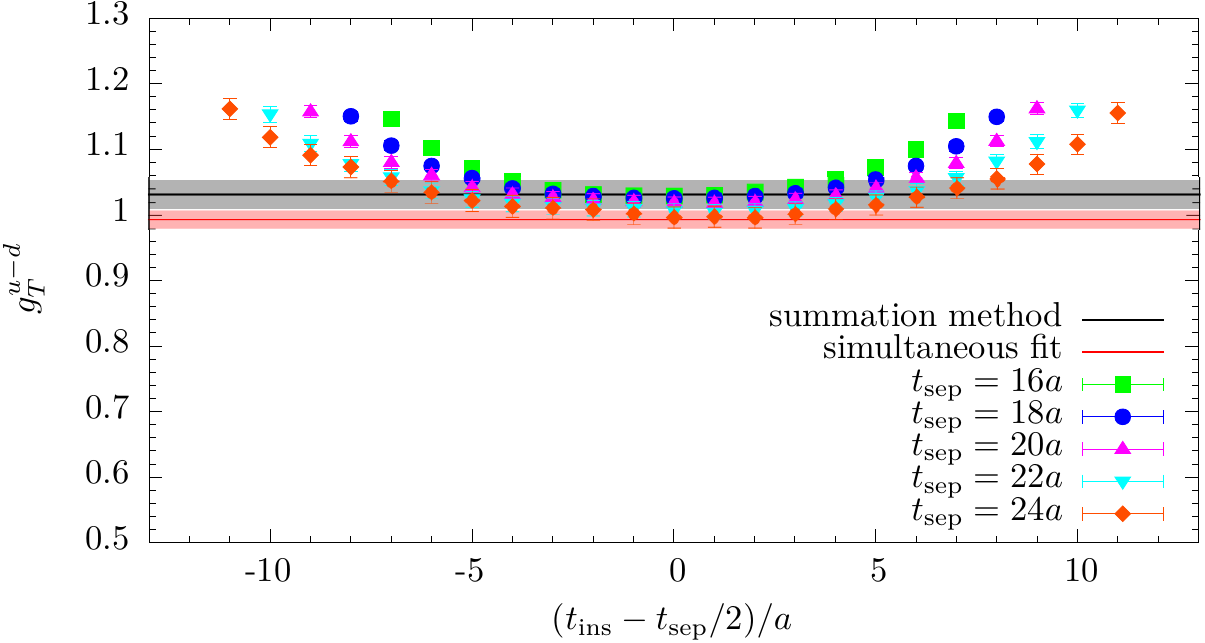}}  \\
 \subfigure{\includegraphics[totalheight=0.2\textheight]{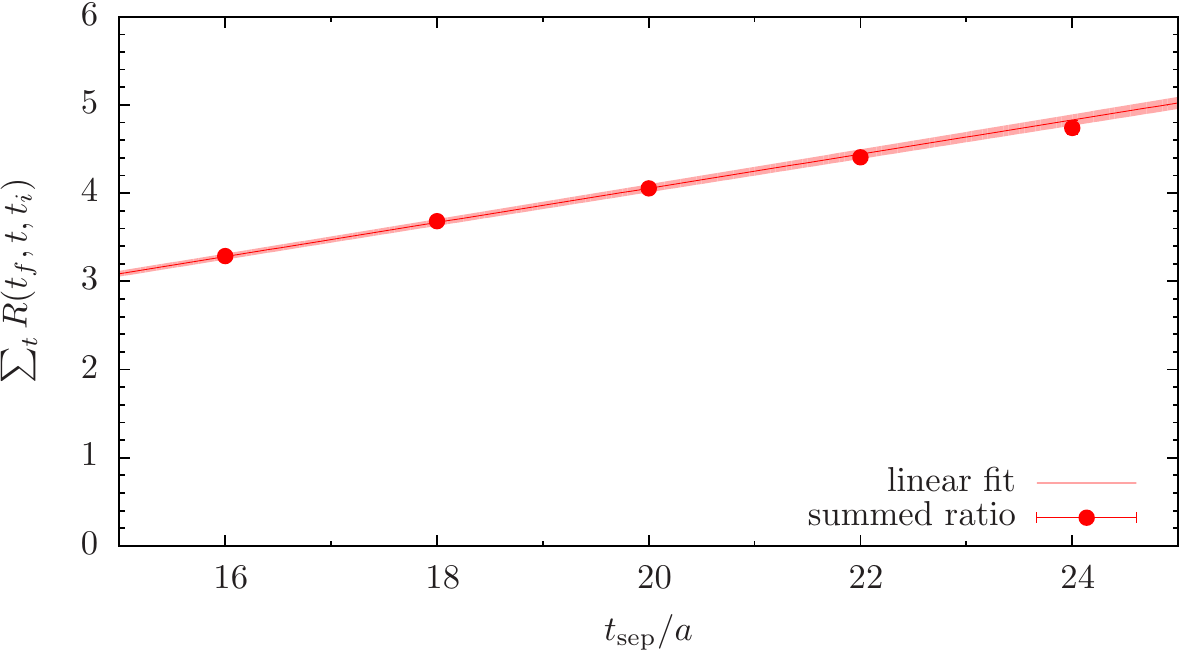}}     \quad
 \subfigure{\includegraphics[totalheight=0.2\textheight]{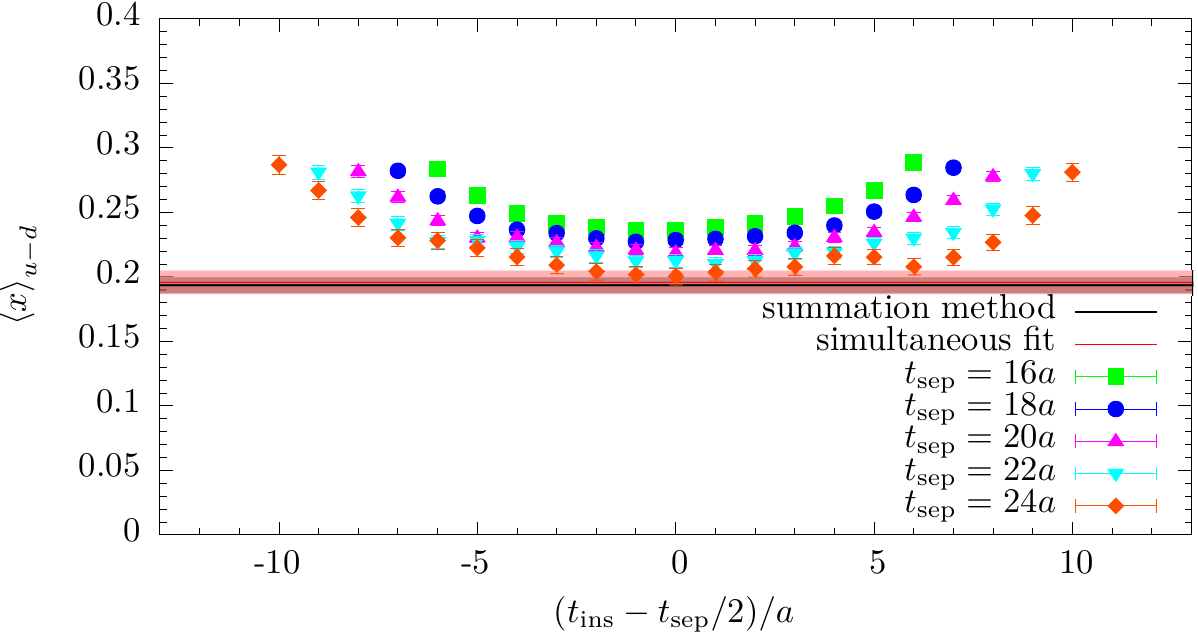}}
 \caption{Overview of results for $g_A^{u-d}$, $g_T^{u-d}$ and $\avgx{-}{}$ from summation method and simultaneous fits on ensemble N203. Left column: linear fits with error bands for summation method as given in Eq.~(\ref{eq:summation_method}). Right column: renormalized lattice data for all values of $\tsep/a$ together with results from summation method and simultaneous fits. Note that the simultaneous fits use data for all six observables, while in the case of the summation method separate fits were performed for each observable.}
 \label{fig:method_comparison}
\end{figure}

Ignoring all but the very first term for the ratio on the r.h.s of Eq.~(\ref{eq:multi_state_fit}), corresponds to a constant fit to the ratio data, which is also known as the \emph{plateau method}. In principle, one can test the convergence of the plateau method by comparing results for several increasing source-sink separations (see \cite{vonHippel:2016wid,Shintani:2018ozy}). However, due to the exponential decrease of the signal-to-noise ratio for the nucleon at increasing time separations, such a test is not feasible with our available statistics. Instead of explicitly fitting excited-state terms for the ratio in Eq.~(\ref{eq:multi_state_fit}) as discussed in the previous section, it is also possible to achieve additional suppression of excited states by appropriate summation over the operator insertion in time. This so-called \emph{summation method} was first introduced in Ref.\cite{Maiani:1987by}. Here we consider the version with explicit summation of the ratio $R^X_{\mu_1...\mu_n}(\vec{0},t,\tsep)$ over timeslices $t$ \cite{Dong:1997xr,Capitani:2012gj} which yields
\begin{equation}
 \sum\limits_{t=t_\mathrm{ex}}^{t_\mathrm{sep}-t_\mathrm{ex}} R^X_{\mu_1...\mu_n}(\vec{0},t,\tsep) = c^X_{\mu_1...\mu_n} + (\tsep - 2t_\mathrm{ex} + a) \cdot \l(\bar{A}^{X,00}_{\mu_1...\mu_n} + \bar{\bar{A}}^{X}_{\mu_1...\mu_n} e^{-\Delta \tsep}\r) + f^X_{\mu_1...\mu_n} e^{-\Delta \tsep} + ... \,.
 \label{eq:summation_method}
\end{equation}
Restricting ourselves to the terms present in Eq.~(\ref{eq:multi_state_fit}), the constant $c^X_{\mu_1...\mu_n}$ and the coefficient $f^X_{\mu_1...\mu_n}$ both receive contributions proportional to $\bar{A}^{X,01}_{\mu_1...\mu_n}$ related to transition matrix elements involving the ground state and the first excited state. In order to avoid contributions from contact terms, one (two) timeslices at both ends are excluded from the sum for local (twist-2) operators, i.e. $t_\mathrm{ex}=1$ for $X\in\{A,S,T\}$ and $t_\mathrm{ex}=2$ for $X\in\{vD,aD,tD\}$. The desired ground-state matrix element $\bar{A}^{X,00}_{\mu_1...\mu_n}$ can be obtained from a linear fit to the lattice data for the l.h.s. using several values of $t_\mathrm{sep}$. Clearly, the leading correction $\sim e^{- \Delta \tsep}$ on the r.h.s. of the above expression is then more strongly suppressed by the larger time extent $t_\mathrm{sep}$ compared to the leading correction $\sim e^{-\Delta t}$ in the case of a naive plateau fit to the data for the ratio itself. In the left column of Fig.~\ref{fig:method_comparison}, examples of fits for $g_A^{u-d}$, $g_T^{u-d}$ and $\avgx{-}{}$ are shown for the N203 ensemble. \par 

In our current setup, such summation method fits are dominated by the smallest source-sink separations, which exhibit the smallest statistical errors. Again, this is a consequence of the aforementioned signal-to-noise problem. Moreover, the efficacy of the summation method is restricted by the total number of different values of $t_\mathrm{sep}$ and the fact that data at consecutive source-sink separations tend to be strongly correlated. Typically, these issues lead to larger statistical errors for the summation method compared to the plateau method or multi-state fits. Therefore, we consider the summation method only as a cross-check rather than a stand-alone method to obtain final numbers. \par

Besides, we observe deviations from the linear behavior in Eq.~(\ref{eq:summation_method}) on ensembles with large statistics and including five values of $\tsep$. While hardly visible by eye, the values in the left column of Fig.~\ref{fig:method_comparison} exhibit non-linear curvature and lie systematically below the fitted result for $\tsep>20a$. Still, on most ensemble our data are well described by the linear fits, albeit within the rapidly increasing errors at larger values of $\tsep$. For our current setup, the summation method works best for the statistically precise axial and tensor charges. In principle, the results from the summation method might still depend on the source-sink separations used, however, it is not possible to systematically test this effect with the available number of source-sink separations and effective statistics by e.g. leaving out the smallest source-sink separation. \par 

In Table~\ref{tab:summation_method}, we have included the results from the summation method for all six observables on each ensemble. Overall we find rather good agreement with the results from the simultaneous fits in Table~\ref{tab:sim_fits}, although the errors for the summation method are significantly larger for the local operator insertions. An example for this is shown in the right column of Fig.~\ref{fig:method_comparison} where we have plotted the lattice data together with results from the summation method and a simultaneous fit for selected observables on ensemble N203. Note that for the summation method we have always used all available values of $\tsep$. For the scalar charge and the twist-2 operator insertions we observe some fluctuations when comparing the two methods. In particular, the summation method fails completely for $g_S^{u-d}$ on H105 yielding a negative value, which is clearly due to insufficient statistics. The simultaneous fits still give a reasonable result in this case as they exploit correlations between the different matrix elements. \par

In general, there is no obvious global trend in any of the observed deviations between summation method and simultaneous fits. However, it appears that there is typically a larger spread in the results from the summation method. This is still true even for the twist-2 operator insertions for which the relative statistical precision is more similar to that of the simultaneous fits than in the case of local operator insertions. However, this behavior is more or less expected because the summation method only uses data for a given observable while the two-state fits are stabilized by fitting all matrix elements simultaneously. This is another important reason why the simultaneous multi-state fits are our preferred method to deal with excited-state contamination. \par

\section{Chiral, continuum and finite-size extrapolation} \label{sec:CCF extrapolation}

Obtaining the final, physical results requires a combined chiral, continuum and finite-size extrapolation to account for unphysical quark masses and the fact that lattice simulations are performed at finite values of the lattice spacing and at finite volume. To this end we have tested several fit ans\"atze guided by chiral perturbation theory. For any given quantity $Q(M_\pi,a,L)$, the fit models used in this study are derived from the following expression,
\begin{equation}
 Q(M_\pi,a,L) = A_Q + B_Q M_\pi^2 + C_Q M_\pi^2 \log M_\pi + D_Q a^{n(Q)} + E_Q \frac{M_\pi^2}{\sqrt{M_\pi L}} e^{-M_\pi L} \,,
 \label{eq:CCF_fit_model}
\end{equation}
by an appropriate selection of non-zero fit parameters $A$, $B$, $C$, $D$ and $E$. We will label fit models by their corresponding combination of non-zero fit parameters, e.g. ``$ABD$''. The first term on the r.h.s. represents the observable in the $SU(2)_F$-chiral, continuum and infinite-volume limit, while the second and third term describe the leading chiral behavior. In the case of the axial charge, the coefficient $C_{g_A}$ of the term containing the chiral logarithm is known analytically \cite{Bernard:1992qa,Kambor:1998pi},
\begin{equation}
 C_{g_A} = \frac{- \chiral{g}_A}{(2 \pi f_\pi)^2}\l(1+2\chiral{g}_A^2\r) \,.
 \label{eq:C_g_A}
\end{equation}
The leading continuum behavior is observable dependent, i.e. by default we have $n(Q)=1$ for unimproved observables, while in case of the axial and the scalar charge we assume $n(g_A)=n(g_S)=2$ since additional counterterms at $\mathcal{O}(a)$ do not contribute to the corresponding operators at vanishing momentum transfer. The last term on the r.h.s of Eq.~(\ref{eq:CCF_fit_model}) describes the leading finite-size behavior; see Ref.~\cite{Beane:2004rf}. \par

As regards the term containing the chiral logarithm, we find that it does not describe our data at all. In the case of the axial charge, we have tested both possible choices, i.e. including the analytically known coefficient in Eq.~(\ref{eq:C_g_A}) and leaving it as a free parameter of the fit for model ABCDE. Using the analytical expression we arrive at an implausibly small value of $g_A^{u-d}=1.143\staterr{21}$. Besides, we observe a large cancellation between the chiral logarithm and the term $\sim M_\pi^2$ for which the coefficient is otherwise compatible with zero. This seems to indicate that our data are not really sensitive to the chiral logarithm. Leaving the parameter free in the fit yields a more plausible result of $g_A^{u-d}=1.275\staterr{62}$, however, with a much larger statistical error. Moreover, the fitted coefficient $C_{g_A}$ comes out with the wrong sign compared to the analytical expectation in Eq.~(\ref{eq:C_g_A}). This is similar to what has been found in an earlier, two-flavor study in Ref.~\cite{Capitani:2017qpc}. As a result we do not include this term in our final fit model. We remark that excluding data with $M_\pi>300\mev$ does not remedy any of these issues: the corresponding results $g_A^{u-d}=1.178\staterr{35}$ and $g_A^{u-d}=1.31\staterr{15}$ have larger statistical errors, but the qualitative features remain unchanged. Given that the applicability of baryonic chiral perturbation theory in the mass range studied here is by no means established, we do not necessarily expect an ansatz incorporating Eq.~(\ref{eq:C_g_A}) to be superior. \par

\begin{figure}[t]
 \centering
  \subfigure{\includegraphics[totalheight=0.223\textheight]{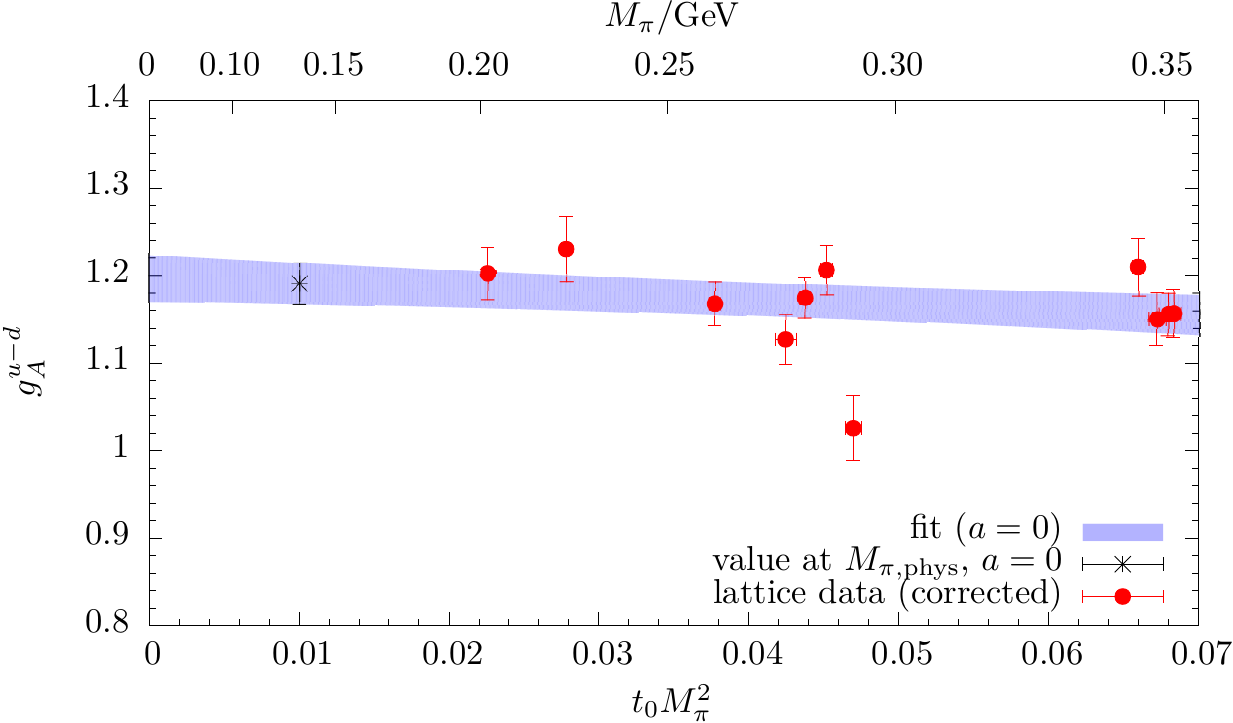}}
  \subfigure{\includegraphics[totalheight=0.223\textheight]{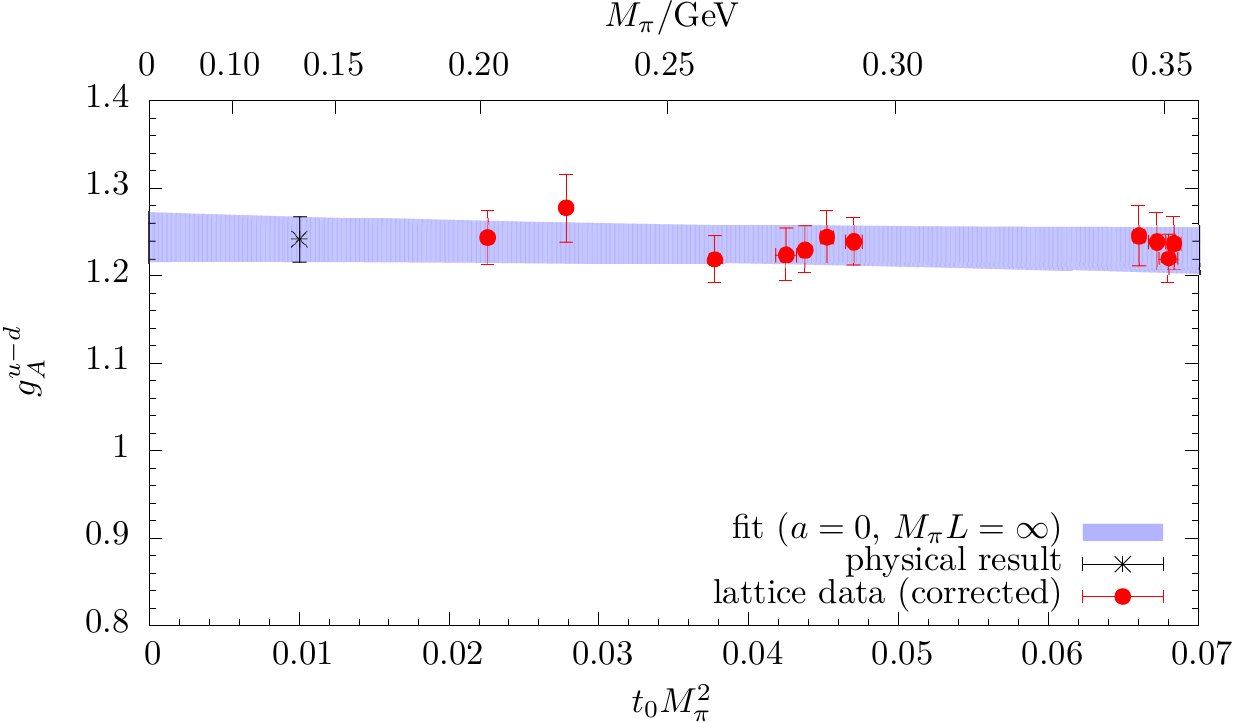}} \\
  \subfigure{\includegraphics[totalheight=0.223\textheight]{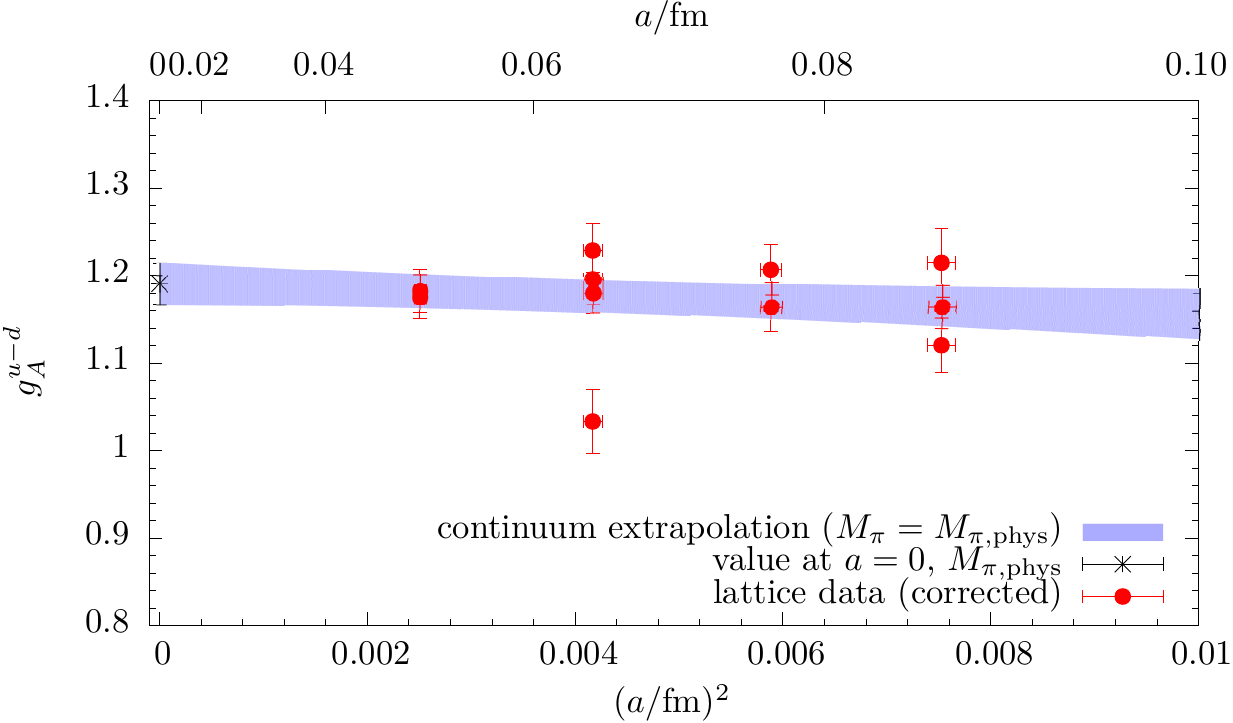}}
  \subfigure{\includegraphics[totalheight=0.223\textheight]{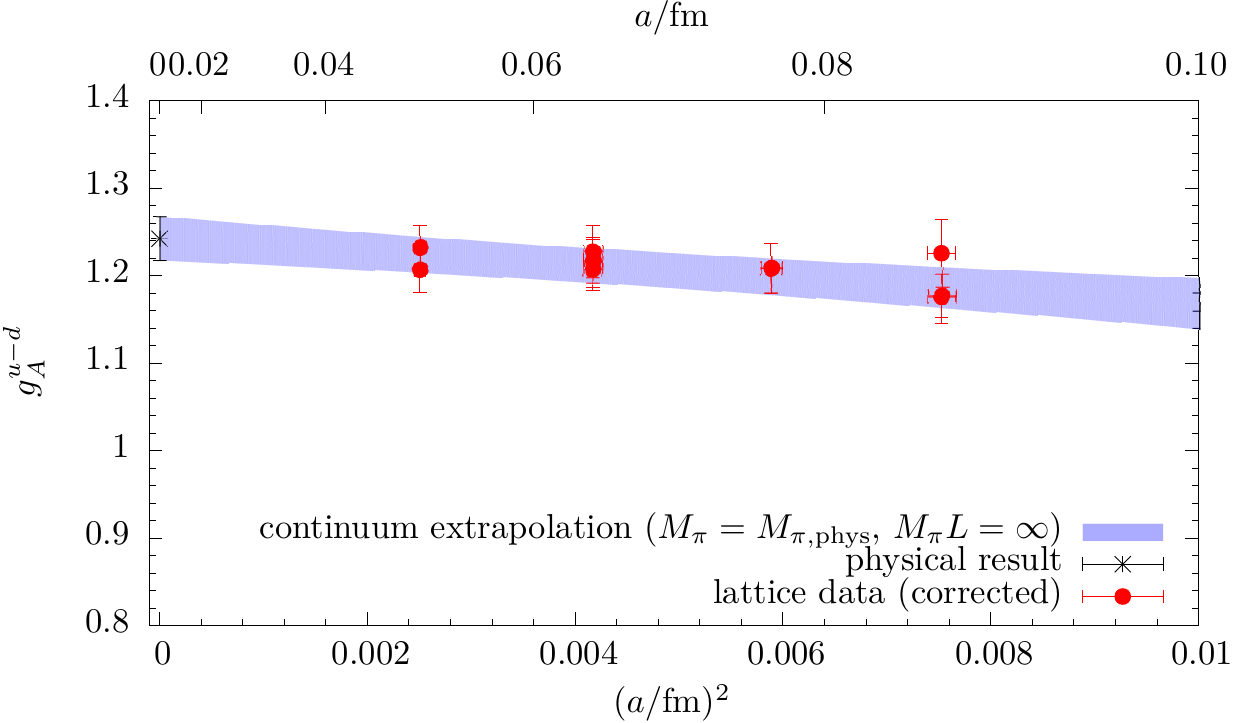}}
  \caption{Chiral behavior (upper row) and continuum behavior (lower row) for $g_A^{u-d}$. Left column: Results from CCF fit model ABD, i.e. not including finite-size corrections. Right column: Results from model ABDE including finite-size corrections. Lattice data in each panel have been corrected using parameters from the corresponding fits for all extrapolations apart from the one given by the blue band.}
  \label{fig:g_A_CC_vs_CCF}
\end{figure}

\subsection{Test of finite-size effects for \texorpdfstring{$g_A^{u-d}$}{the isovector axial charge}}

\begin{figure}[t]
 \centering
 \includegraphics[totalheight=0.225\textheight]{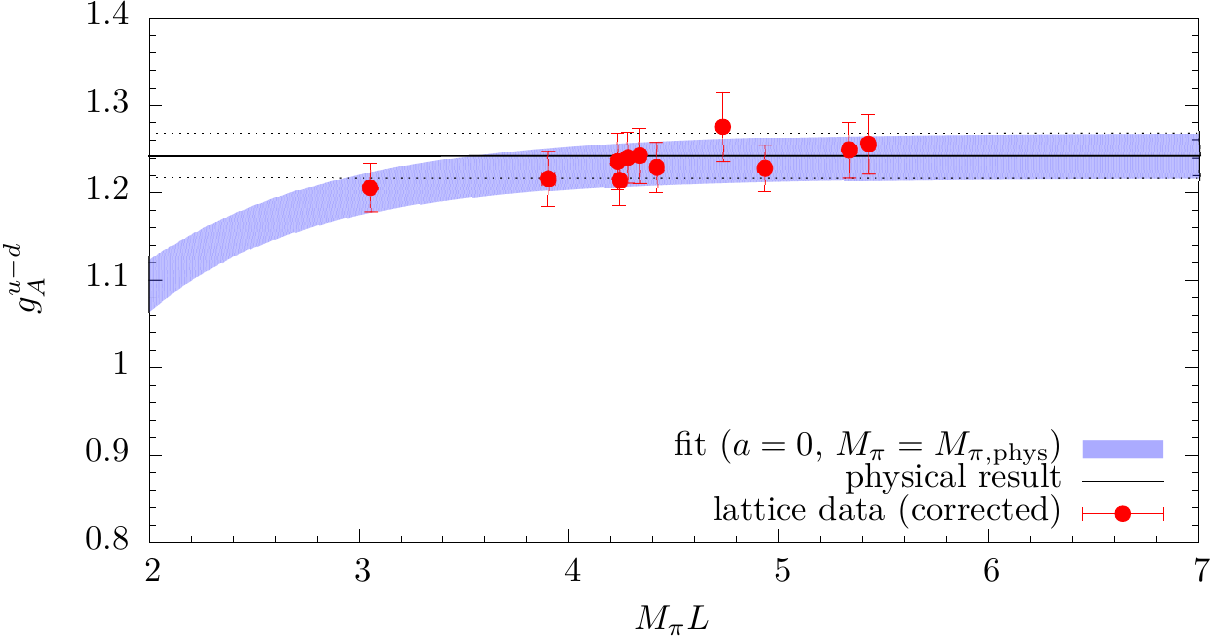}
 \caption{Finite-size extrapolation for $g_A^{u-d}$. Lattice data have been corrected to the physical value of the pion mass and the continuum limit using parameters from the fit. Therefore, the corrected data points are highly correlated.}
 \label{fig:g_A_FS_extrolation}
\end{figure}

In the left column of Figure~\ref{fig:g_A_CC_vs_CCF} we show the chiral and continuum behavior for $g_A^{u-d}$ obtained from fitting model ABD, i.e. without including a finite-size term. The lattice data in the upper and lower panel have been corrected to vanishing lattice spacing and to physical light quark mass, respectively. The resulting behavior is very flat in both $M_\pi^2$ and $a^2$. Nevertheless, a significant spread in the data remains around the blue extrapolation bands. This is reflected by a prohibitively bad value of $\chi^2/\mathrm{dof}$ for this fit, i.e. $\chi^2/\mathrm{dof}\approx4.067$. In particular, there is one outlier that lies far below all other data points. This data point belongs to ensemble S201 which is the only ensemble with $M_\pi L \approx 3$. Since it has been generated with the same input parameters as N200 apart from the spatial volume, we can perform an explicit finite-size test in this case. With respect to the continuum extrapolation shown in the lower panel we find
\begin{equation}
 g_{A,\text{S201}}^{u-d}[a{=}0.06426\fm,\,\Mpi{=}M_{\pi,\phys}]=1.033\staterr{37}
\end{equation}
and
\begin{equation}
 g_{A,\text{N200}}^{u-d}[a{=}0.06426\fm,\,\Mpi{=}M_{\pi,\phys}]=1.180\staterr{23} \,,
\end{equation}
respectively. This very significant difference can be attributed to finite-size effects. For the plots in the right column of Figure~\ref{fig:g_A_CC_vs_CCF}  we show the chiral and continuum behavior from fit model ABDE, i.e. including a finite-size term, which greatly reduces the scattering of results around the extrapolation bands. In fact, it entirely removes the spread in the values for S201 and N201 which now read
\begin{equation}
 g_{A,\text{S201}}^{u-d}[a=0.06426\fm,\, \Mpi{=}M_{\pi,\phys},\, M_\pi L {=} \infty] = 1.217\staterr{25}
\end{equation}
and 
\begin{equation}
 g_{A,\text{N200}}^{u-d}[a=0.06426\fm,\, \Mpi{=}M_{\pi,\phys},\, M_\pi L {=} \infty] = 1.207\staterr{24} \,, 
\end{equation}
respectively. We also find that the quality of the fit is greatly improved, resulting in $\chi^2/\mathrm{dof}\approx0.573$. Moreover, the introduction of the additional fit parameter barely increases the statistical error on the final results. We remark that we find finite-size effects to be relevant for all observables. However, it is only for the statistically precise axial charge that we observe such a significant improvement in the resulting value of $\chi^2/\mathrm{dof}$ when switching from model ABD to ABDE. The finite-size extrapolation for $g_A^{u-d}$ from model ABDE after taking the continuum limit and extrapolating to the physical pion mass is shown in Fig.~\ref{fig:g_A_FS_extrolation}. \par

For the final CCF extrapolation, we hence adopt model ABDE and perform the required fits using a bootstrap procedure with $N_s=10000$ samples. To this end, we apply resampling for the values of $M_\pi$, the individual results for the observables as well as for all quantities that are only $\beta$-dependent such as renormalization factors, $t_0/a^2$ and $t_{0,\phys}$. The latter enters the analysis only to fix the physical value of $M_\pi$ in units of $\mev$. For the physical pion mass, we use the FLAG value in the isospin limit $M_{\pi,\phys}=134.8(3)\mev$~\cite{Aoki:2016frl}, reflecting the fact that we impose isospin symmetry and neglect electromagnetic effects in our simulations. The bootstrap procedure allows us to propagate all individual errors and accounts also for correlations introduced in the fit by $\beta$-dependent quantities such as renormalization factors and factors of $t_0/a^2$. In fact, $t_0/a^2$ and unimproved renormalization factors are quark-mass independent and hence $100\%$ correlated at any given $\beta$. In case of the quark-mass dependent $\mathcal{O}(a)$-improved values of $Z_A^\mathrm{SF}$ the correlation of these values at fixed $\beta$ remains very large. The systematic errors on renormalization factors, $t_0/a^2$ and $t_0$ are added in quadrature to the respective statistical errors before the resampling such that they are propagated into the final error estimate as well. Therefore, the resulting errors are not purely statistical, however, the effects of these systematic uncertainties are very small compared to the actual statistical errors on the final results. \par

\subsection{Study of systematics related to renormalization} \label{subsec:sfrimon}

\begin{table}[!t]
 \centering
  \begin{tabular}{l|llll|cll}
   \hline\hline
    index & $n(g_A)$ & renormalization & $\beta$-cut & $g_A^{u-d}$ & $\chi^2/\mathrm{dof}$ & $p$ \\
   \hline\hline                                              
     1 & 2 & RIMOM   & none   & 1.242(25) & 0.537 & 0.807 \\ 
     2 & 2 & RIMOM   & $<3.7$ & 1.259(32) & 0.498 & 0.778 \\ 
     3 & 2 & SF imp. & none   & 1.231(25) & 0.532 & 0.810 \\ 
     4 & 2 & SF imp. & $<3.7$ & 1.251(32) & 0.474 & 0.796 \\ 
     5 & 2 & SF      & none   & 1.232(25) & 0.561 & 0.788 \\ 
     6 & 2 & SF      & $<3.7$ & 1.251(32) & 0.503 & 0.493 \\ 
     7 & 1 & RIMOM   & none   & 1.275(38) & 0.577 & 0.775 \\ 
     8 & 1 & SF imp. & none   & 1.258(37) & 0.574 & 0.778 \\ 
     9 & 1 & SF      & none   & 1.256(37) & 0.595 & 0.761 \\ 
   \hline\hline
  \end{tabular}
  \caption{Overview on results for $g_A^{u-d}$ from different CCF fits employing model ABDE and using data from simultaneous fits. In the column labeled renormalization the tag ``RIMOM'' refers to using renormalization factors from the Rome-Southampton method as in Table~\ref{tab:Zloc}, while ``SF imp.'' refers to using improved renormalization factors from the Schr\"odinger functional approach as defined in Eq.~(\ref{eq:Z_A_SF_improved}) and obtained from the data in Table~\ref{tab:ZSF}. ``SF'' refers to using unimproved renormalization factors from the Schr\"odinger functional approach.}
 \label{tab:CCF_g_A}
\end{table}

\begin{figure}[t]
 \centering
 \includegraphics[totalheight=0.25\textheight]{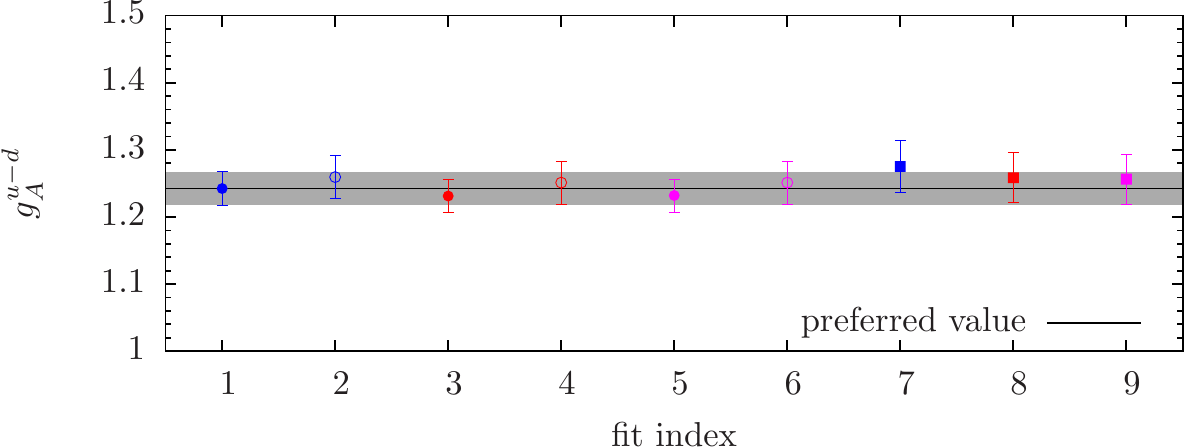}
 \caption{Overview for results for $g_A^{u-d}$ from different variations of the CCF fit model, as detailed in Table~\ref{tab:CCF_g_A}. Red symbols denote results obtained using $Z_A$ from the Rome-Southampton method. Blue and violet symbols represent data obtained using $Z_A$ from the Schr\"odinger functional with mass-dependent counterterms included and excluded, respectively. Filled symbols are used for results obtained by fitting data at all four lattice spacings, while open symbols are used for results when excluding data at $\beta=3.7$. Circles and boxes refer to fitting a lattice artifact $\mathcal{O}(a^2)$ and $\mathcal{O}(a)$, respectively.}
 \label{fig:g_A_CCF_overview}
\end{figure}

Another potential source of uncertainty concerns the renormalization factors at $\beta=3.7$ determined via the Rome-Southampton method. As discussed in Sec.~\ref{subsec:renormalization} and in appendix~\ref{app:npr}, the corresponding values have been obtained from an extrapolation. Moreover, the results for the $Z$ factors do not account for discretization effects of $\mathcal{O}(a)$ proportional to the quark mass. This may introduce residual $\mathcal{O}(a)$ artifacts for $g_A^{u-d}$ and $g_S^{u-d}$ even though no additional counterterms arise involving derivatives of quark bilinears. \par

Therefore, we have carried out additional tests to further corroborate our results for the CCF extrapolation of $g_A^{u-d}$ from fit model ABDE, as detailed in Table~\ref{tab:CCF_g_A}. Since $g_A^{u-d}$ is the statistically most precise observable, it is also expected to be the most sensitive one with respect to the aforementioned issues. Besides, for the axial vector current insertion, renormalization factors are available from the Schr\"odinger functional approach~\cite{DallaBrida:2018tpn} for all four values of $\beta$ including the mass-dependent factor in Eq.~\ref{eq:Z_A_SF_improved}. This allows us to conduct an explicit consistency check in this case. A graphical overview of the ten variations can be found in Fig.~\ref{fig:g_A_CCF_overview}. \par

The first six of these variations all assume that the leading lattice artifacts are of $\mathcal{O}(a^2)$ in the CCF fit model ABDE. They can be divided into three subgroups corresponding to the employed renormalization factors, i.e. the Rome-Southampton method and the Schr\"odinger functional, where the latter may include the mass-dependent factor or not. This allows us to test for the agreement of the two renormalization schemes and for possible deviations caused by ignoring mass-dependent counterterms in $Z_A$. Within each of these three groups, we have two variations with and without including the data at the finest lattice spacing. For results using $Z_A$ from the Rome-Southampton method this serves as a cross-check that the extrapolation required for the renormalization factors at $\beta=3.7$ is sound. With respect to the results renormalized via the Schr\"odinger functional method, we include this variation to be able to disentangle effects which arise when removing data for the finest lattice spacing from the continuum extrapolation and effects related to a potential issue with the extrapolation of $Z_A$ at $\beta=3.7$. The last three variations shown in Fig.~\ref{fig:g_A_CCF_overview} assume that the leading lattice artifact in the CCF fit is of $\mathcal{O}(a)$ instead of $\mathcal{O}(a^2)$. \par

First, we find that the results using $Z_A$ from the Rome-Southampton method and the Schr\"odinger functional are in good agreement for the extrapolations linear in $a^2$ (variations~1 to~6). Moreover, leaving out the data at $\beta=3.7$ has a very similar effect on $g_A^{u-d}$ when either the Rome-Southampton method or the Schr\"odinger functional approach is applied for the renormalization. This leads to the conclusion that a systematic effect caused by the extrapolation to $\beta=3.7$ for the Rome-Southampton method must indeed be very small. \par

A comparison of variations $\{3,\,4\}$ to $\{5,\,6\}$ reveals that the mass-dependent factor in Eq.~\ref{eq:Z_A_SF_improved} is completely negligible within the current statistical precision, i.e. both variations give practically identical results, demonstrating that residual discretization artifacts of $\mathcal{O}(a)$ are extremely small. This is also confirmed by the last three variations. While replacing the $\mathcal{O}(a^2)$ term by an $\mathcal{O}(a)$ term in the fit generally leads to somewhat larger continuum results, this behavior cannot be caused by the mass-dependent factor, since the shift is very similar in both cases, as can be inferred from variations~8 and~9. \par

\subsection{CCF-related systematics and final results}

\begin{figure}[t]
 \centering
  \subfigure{\includegraphics[totalheight=0.222\textheight]{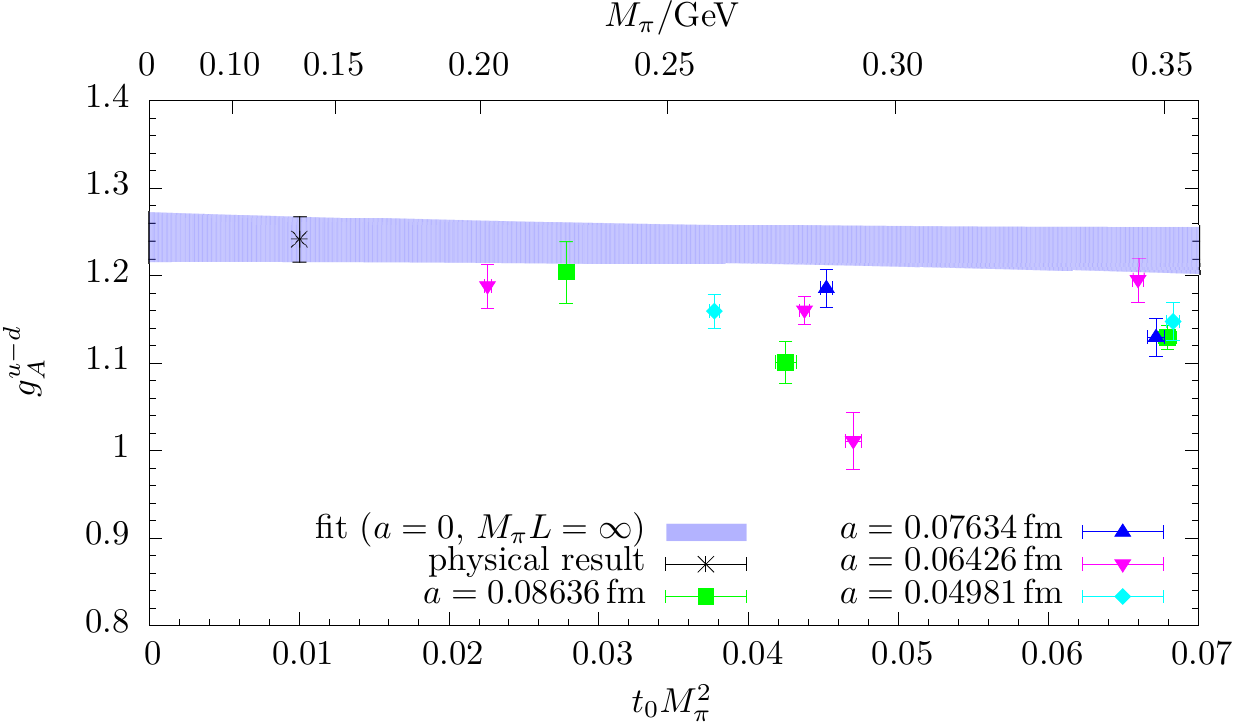}}
  \subfigure{\includegraphics[totalheight=0.222\textheight]{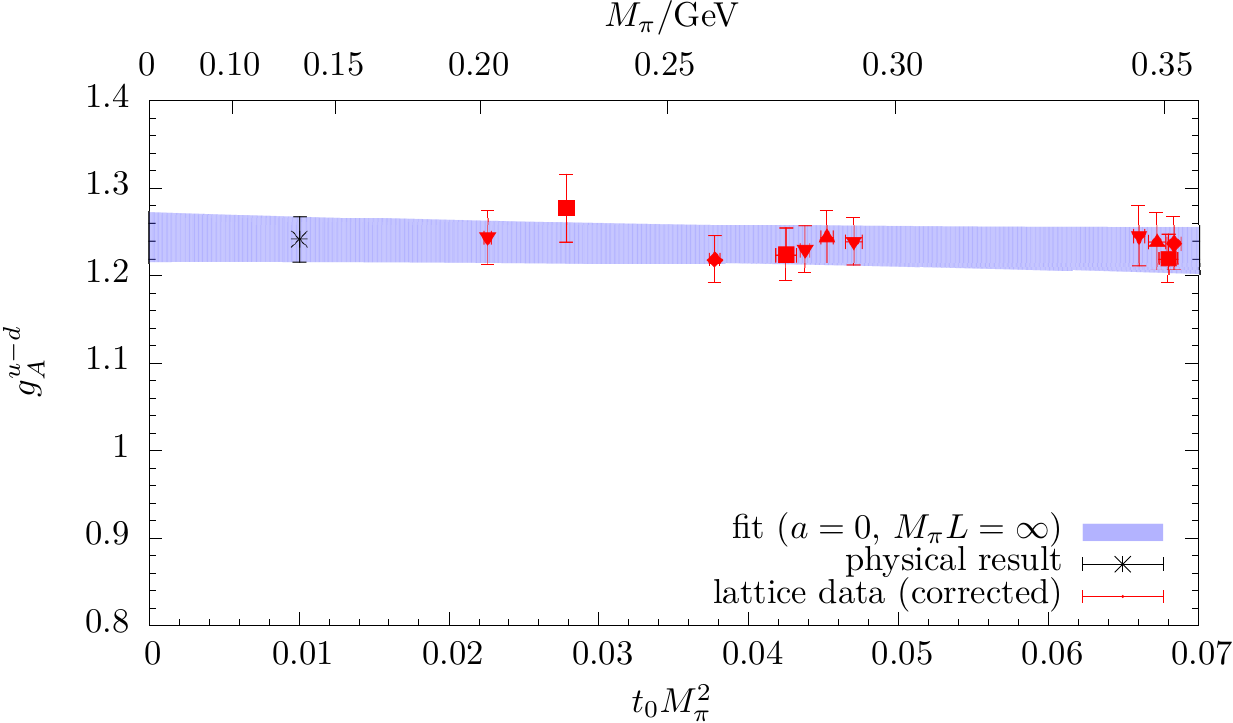}} \\
  \subfigure{\includegraphics[totalheight=0.222\textheight]{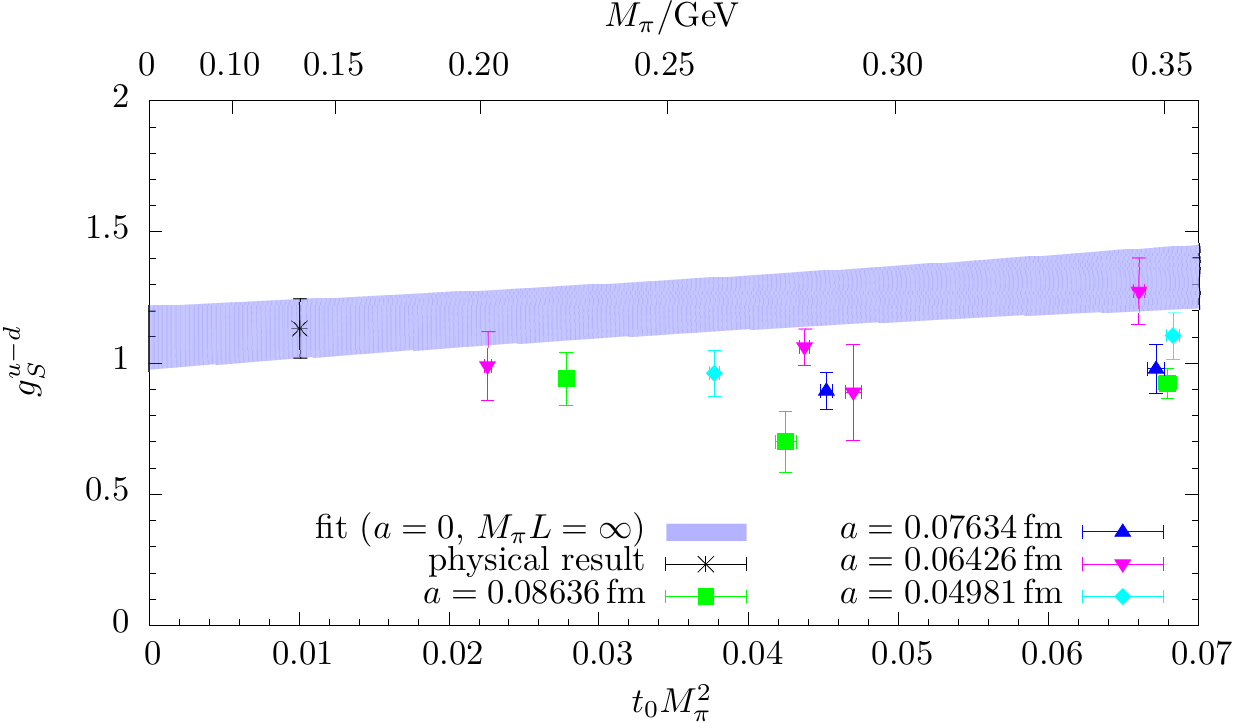}}
  \subfigure{\includegraphics[totalheight=0.222\textheight]{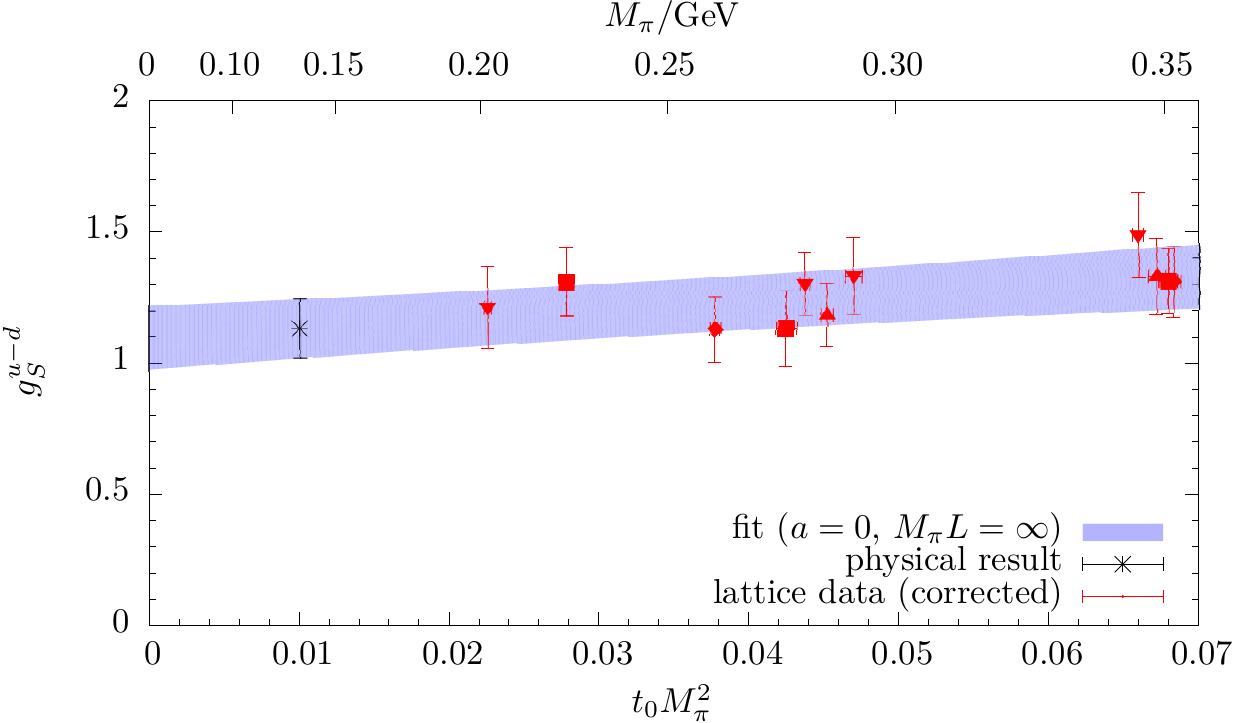}} \\
  \subfigure{\includegraphics[totalheight=0.222\textheight]{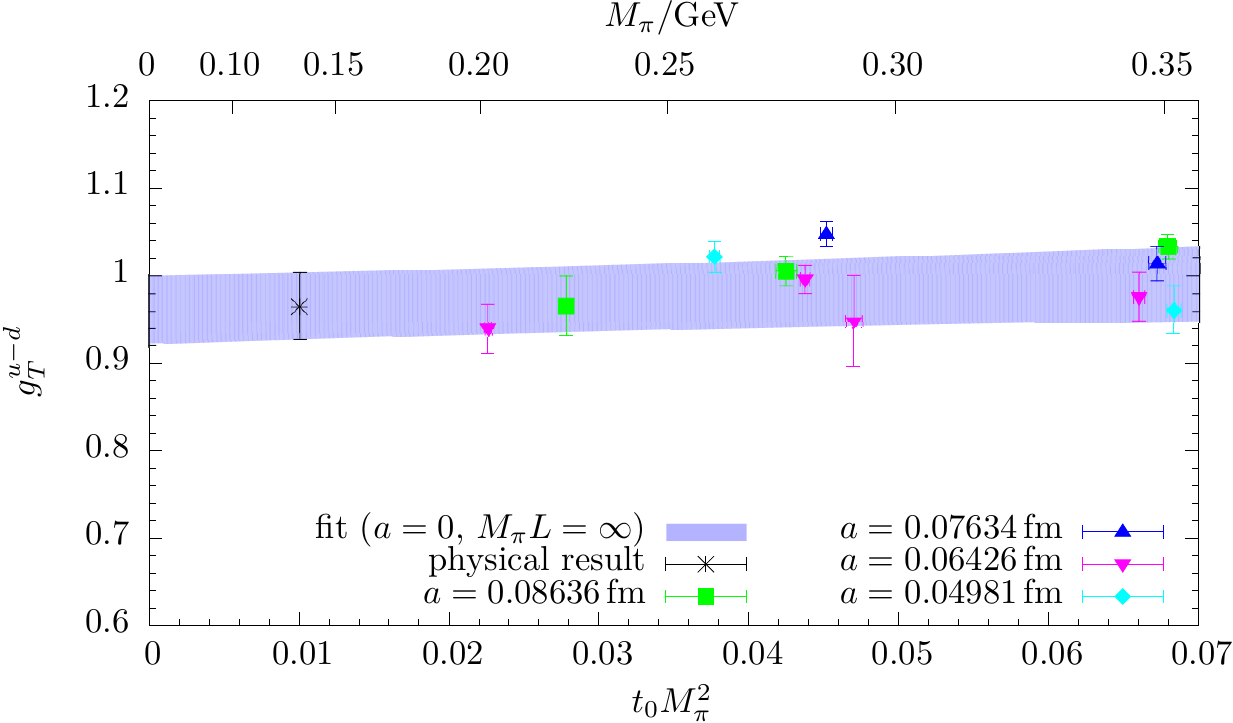}}
  \subfigure{\includegraphics[totalheight=0.222\textheight]{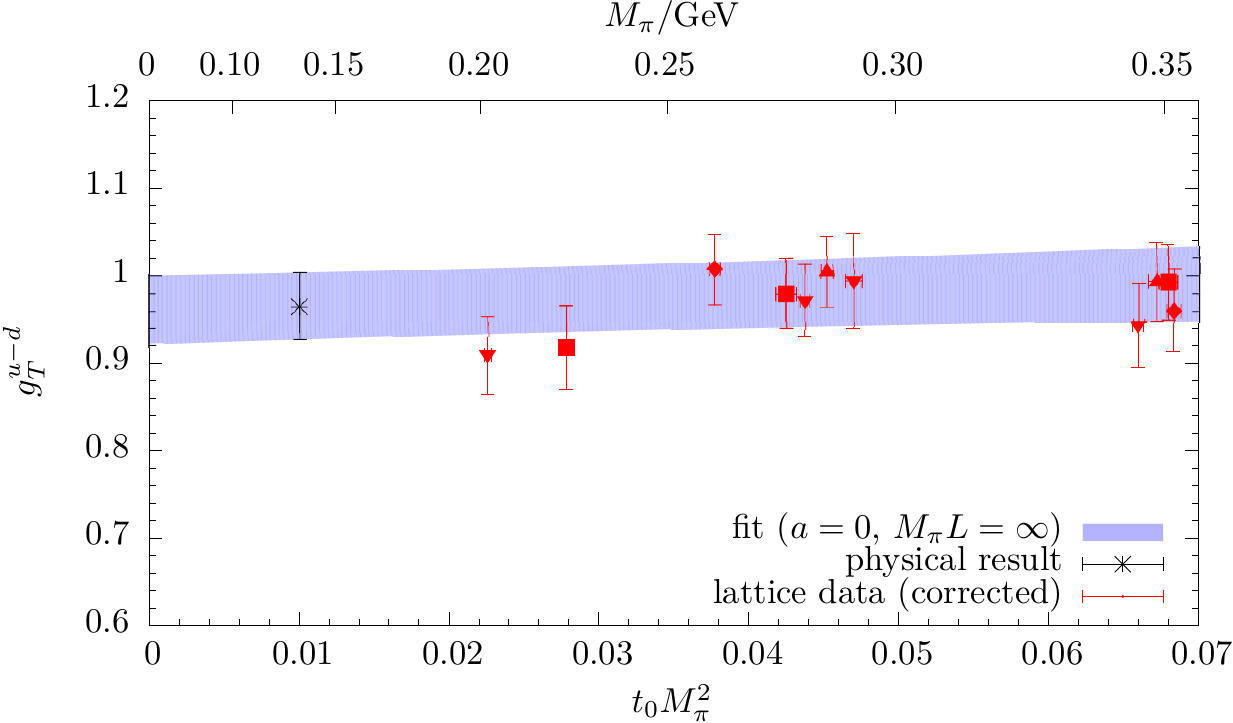}}
 \caption{Results from chiral and continuum model $ABDE$ for local charges. Data on individual ensembles have been obtained from the multi-state fit model in Eq.~(\ref{eq:multi_state_fit}). In the left column we show the chiral extrapolation together with the original data from Table~\ref{tab:sim_fits}, while in the right column the lattice data have been corrected for the continuum limit and finite-size extrapolation using the corresponding fit parameters. Therefore, the corrected data points in the right column are highly correlated within the same plot.}
 \label{fig:final_fits_local}
\end{figure}

\begin{figure}[t]
 \centering
  \subfigure{\includegraphics[totalheight=0.222\textheight]{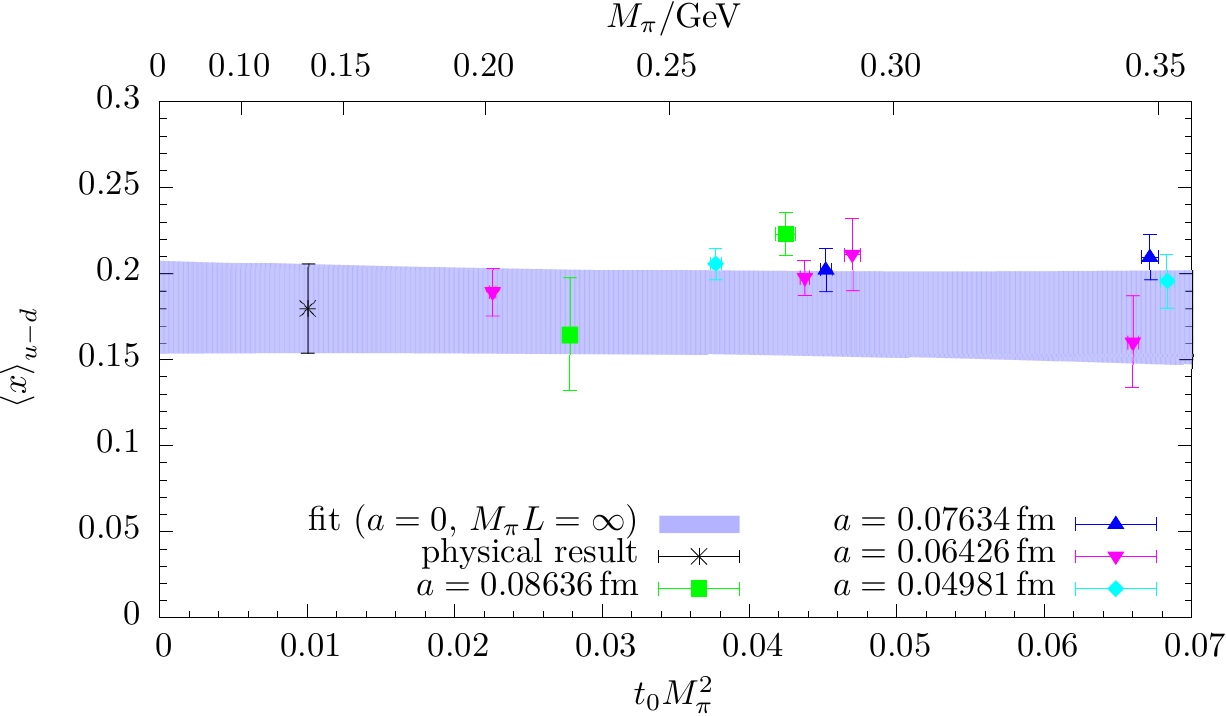}}
  \subfigure{\includegraphics[totalheight=0.222\textheight]{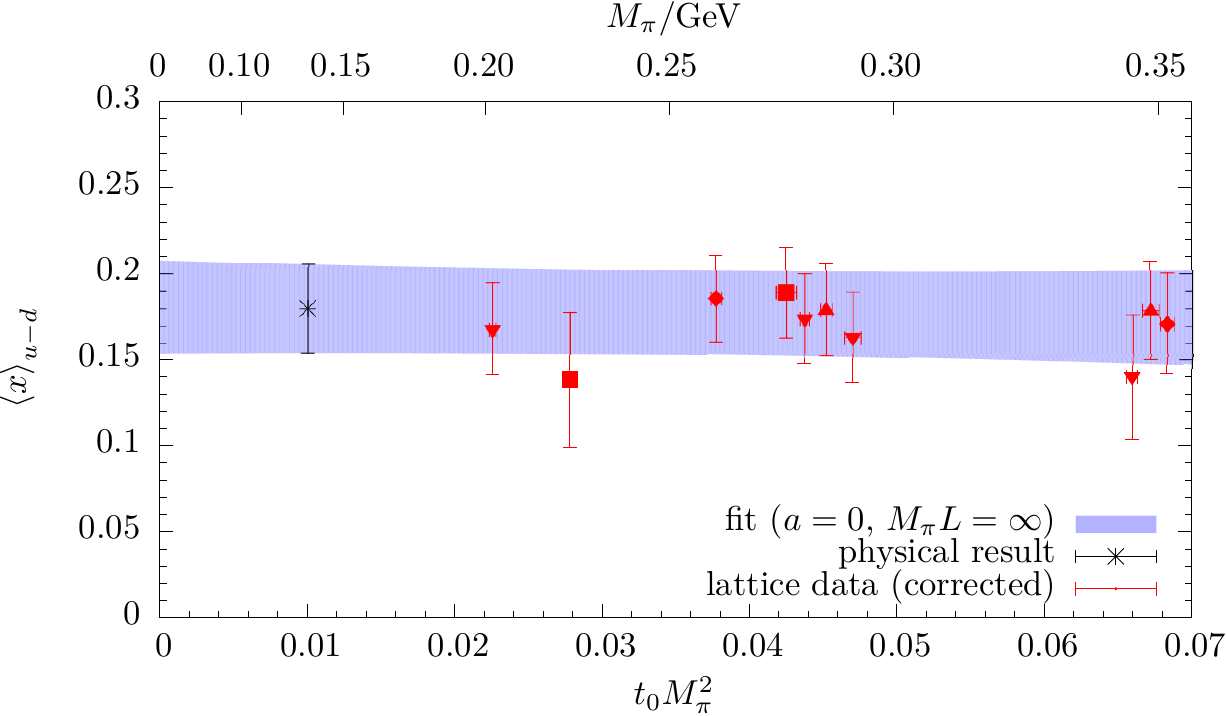}} \\
  \subfigure{\includegraphics[totalheight=0.222\textheight]{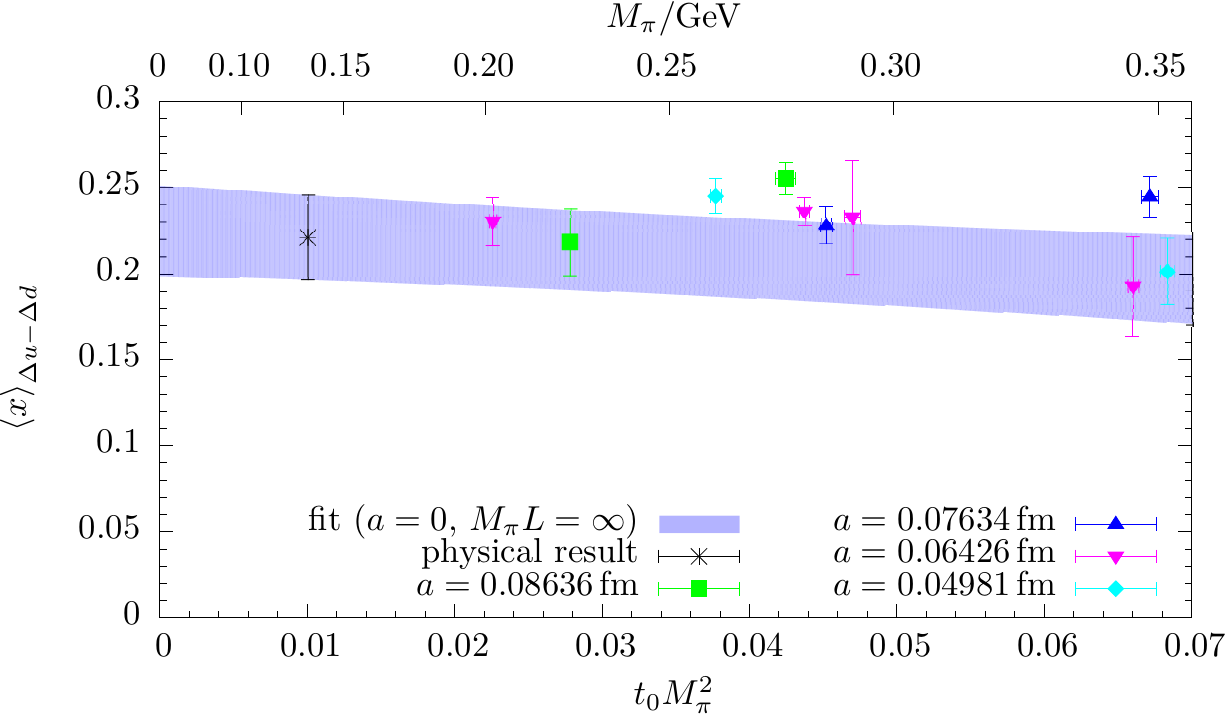}}
  \subfigure{\includegraphics[totalheight=0.222\textheight]{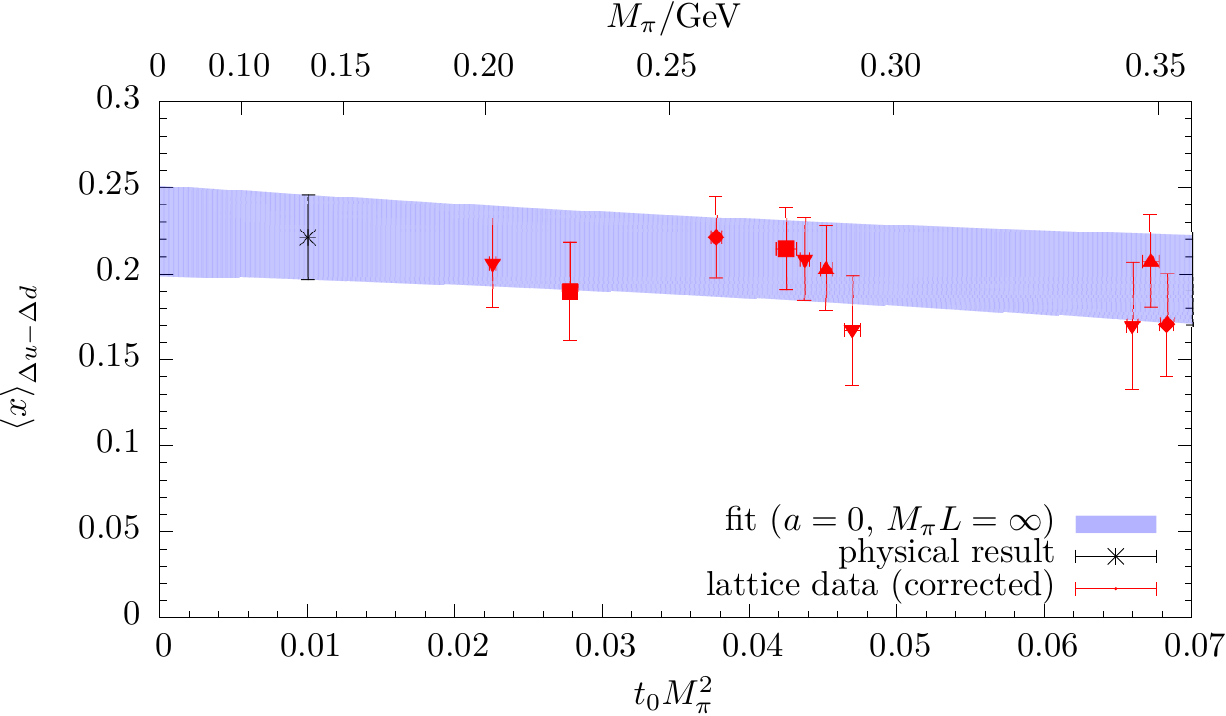}} \\
  \subfigure{\includegraphics[totalheight=0.222\textheight]{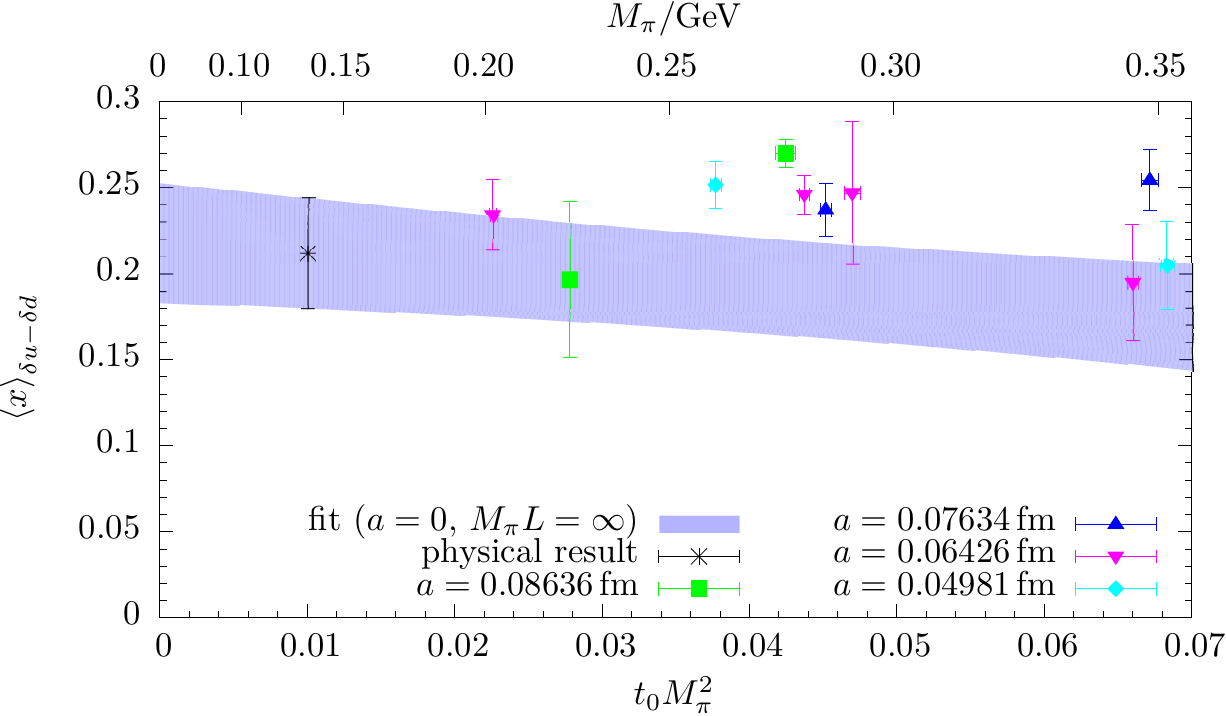}}
  \subfigure{\includegraphics[totalheight=0.222\textheight]{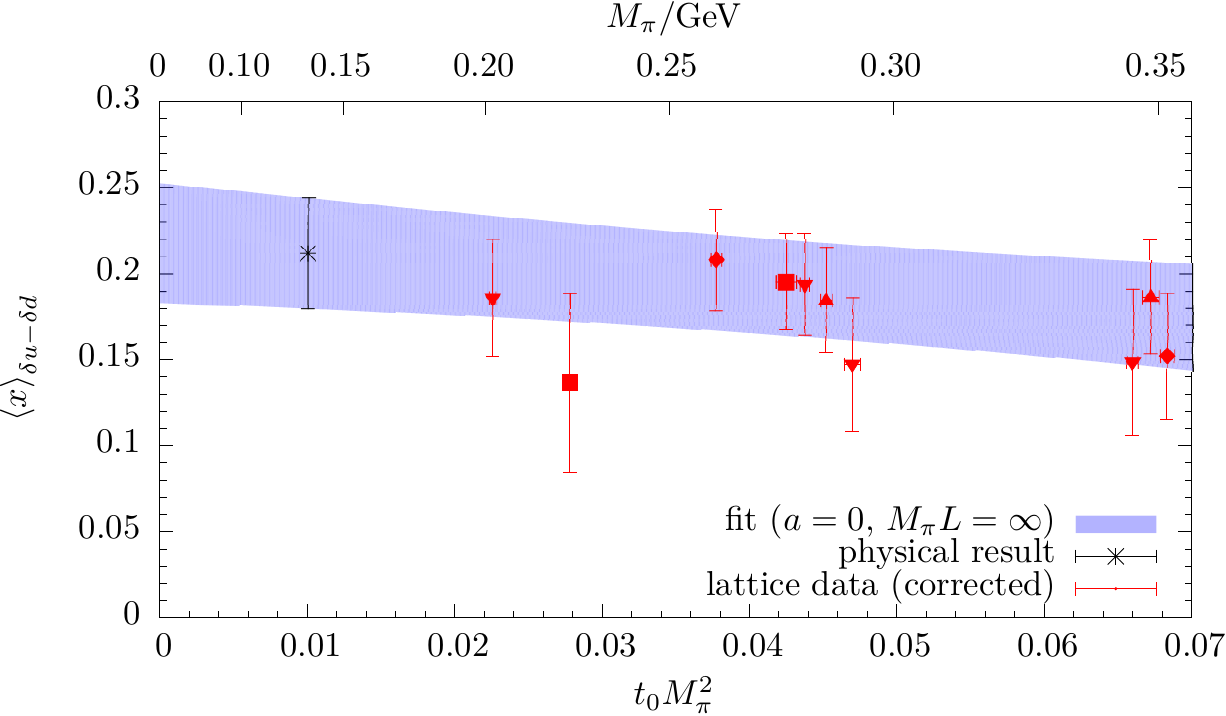}} 
 \caption{Same as Fig.~\ref{fig:final_fits_local} but for twist-2 operator insertions.}
 \label{fig:final_fits_twist2}
\end{figure}

In Fig.~\ref{fig:final_fits_local} we plot the chiral behavior for the three local isovector charges after taking the continuum limit and correcting for finite-size effects. The panels in the left column show the extrapolation band together with the original lattice data, which gives some indication for the size of continuum and finite-size corrections. For the plots in the right column the lattice data has been corrected for $a\rightarrow0$ and $\Mpi L\rightarrow\infty$ using the parameters obtained from the combined CCF fit. In general, the observed chiral behavior is very mild and the corresponding slope w.r.t. $M_\pi^2$ is often found to be compatible with zero within errors. However, the corrections for leading lattice artifacts and finite-size corrections are typically non-negligible. A qualitatively similar picture is observed for the matrix elements of the twist-2 operator insertions in Fig.~\ref{fig:final_fits_twist2}. \par

In order to estimate systematic effects in our CCF extrapolations we consider the following three, distinct variations of the fits for each observable:
\begin{enumerate}
 \item Excluding data with $M_{\pi,\mathrm{cut}}>300\mev$ to test the effect of neglecting higher order terms in the chiral extrapolation on our final results. Since the convergence properties of baryonic $\chi$PT in the regime of $M_\pi>300\mev$ are doubtful, such terms are potentially a major source of systematic errors and even more so at larger light quark masses. \\
 \item Excluding data at the coarsest lattice spacing ($\beta=3.4$) to test the convergence of the continuum extrapolation. \\
 \item Excluding data with $M_\pi L < 4$ from the CCF fits (ensembles S201 and H105) to test the stability of the finite-size extrapolation. \\
\end{enumerate}
These cuts in the data are chosen such that enough lattice data points remain for a meaningful fit in all cases. Still, at least for the first two variations they result in significantly larger errors than a fit to the full data set. For each of the three variations we assign an additional systematic error to the final results for each observable, which is given by the difference of the result from the variation and the result using the full set of data. These systematic errors for the three variations are labelled ``$\chi$'', ``$\mathrm{cont}$'' and ``$\mathrm{FS}$'', respectively. However, it should be kept in mind that these variations cannot be fully independent due to the simultaneous (and non-linear) fits. For example, removing the data at $\beta=3.4$ simultaneously removes one of the two ensembles with the smallest pion mass (C101). Therefore, this variation affects not only the continuum extrapolation as intended but in addition may potentially alter the chiral extrapolation in a rather unfavorable way, i.e. removing data at the smallest available light quark masses. This is why we believe that these estimates of systematic errors are rather conservative. Nonetheless, we find that that they are typically of similar or smaller size than the statistical errors, indicating that the final extrapolations are not dominated by systematic effects at the current level of statistical precision. \par

Our final results for the local nucleon charges read
\begin{align}
 g_A^{u-d} &= 1.242\staterr{25}\chierr{-06}\conterr{-30}\FSerr{+00} \,, \\ 
 g_S^{u-d} &= 1.13 \staterr{11}\chierr{+07}\conterr{-06}\FSerr{-01} \,, \\ 
 g_T^{u-d} &= 0.965\staterr{38}\chierr{-37}\conterr{-17}\FSerr{+13} \,,    
 \label{eq:final_results_charges}
\end{align}
while for the lowest moments of the parton distributions we obtain
\begin{align}
 \avgx{-}{}       &= 0.180\staterr{25}\chierr{-06}\conterr{+12}\FSerr{+07} \,, \\  
 \avgx{-}{\Delta} &= 0.221\staterr{25}\chierr{+01}\conterr{+10}\FSerr{+02} \,, \\  
 \avgx{-}{\delta} &= 0.212\staterr{32}\chierr{-10}\conterr{+19}\FSerr{+05} \,.     
 \label{eq:final_results_twist2}
\end{align}
The remaining fitted parameters from the final CCF fit are listed in Table~\ref{tab:CCF_final_params}. The corresponding $\chi^2/\mathrm{dof}$ and $p$-values can be found in Table~\ref{tab:CCF_chi2_p}, where we have also included the values for the three variations that have been used to assign the systematic errors. In general, we observe that our data are well described by the fit model. Only for $g_T^{u-d}$ we observe some tension, which might be related to the chiral extrapolation. This is the only case for which a cut in $\Mpi$ leads to a significant improvement of the fit. None of the other applied cuts have an effect on the fit quality, as can be seen from Table~\ref{tab:CCF_chi2_p}. However, we cannot exclude that the behavior observed for $g_T^{u-d}$ is merely a fluctuation in our data. Therefore, we prefer to quote the final result from fitting the full set of data, which is consistent with the choice for the other observables. \par

\begin{table}[!t]
 \centering
 \begin{tabular}{l|llllll}
  \hline\hline
                               & $g_A^{u-d}$ & $g_S^{u-d}$ & $g_T^{u-d}$ & $\avgx{-}{}$ & $\avgx{-}{\Delta}$ & $\avgx{-}{\delta}$ \\
  \hline\hline
   $A_Q$                       &  1.245(28)  &  1.10(12)   &  0.961(39)  &  0.181(27)   &  0.225(26)         &  0.218(35) \\
   $t_0^{-1} B_Q$              & -6(11)      &  86(47)     &  11(11)     & -2.4(8.7)    & -10.5(9.0)         & -16(13)    \\
   $t_0^{\frac{n(Q)}{2}} D_Q $ & -0.0063(31) & -0.038(13)  &  0.019(15)  &  0.008(10)   &  0.0082(93)        &  0.018(14) \\
   $t_0^{-1} E_Q$              & -398(74)    & -507(383)   & -176(91)    &  61(50)      &  96(64)            &  119(82)   \\
  \hline\hline
 \end{tabular}
 \caption{Fitted parameters for model ABDE obtained from the final CCF fits in units of $t_0$. Errors are statistical only.}
 \label{tab:CCF_final_params}
\end{table}

\begin{table}[!t]
 \centering
  \begin{tabular}{l|llllllll}
  \hline\hline
   & \multicolumn{2}{c}{final fit} & \multicolumn{2}{c}{$\Mpi<300\mev$} & \multicolumn{2}{c}{$\beta>3.4$} & \multicolumn{2}{c}{$ \Mpi L \geq 4$} \\
     observable & $\chi^2/\mathrm{dof}$ & $p$ & $\chi^2/\mathrm{dof}$ & $p$ & $\chi^2/\mathrm{dof}$ & $p$ & $\chi^2/\mathrm{dof}$ & $p$ \\
  \hline\hline
   $g_A^{u-d}$        & 0.537 & 0.807 & 0.524 & 0.666 & 0.1934 & 0.942 & 0.691 & 0.630 \\
   $g_S^{u-d}$        & 1.006 & 0.424 & 1.385 & 0.245 & 1.0567 & 0.376 & 1.149 & 0.332 \\
   $g_T^{u-d}$        & 2.539 & 0.013 & 1.611 & 0.185 & 3.4482 & 0.008 & 3.432 & 0.004 \\
   $\avgx{-}{}$       & 1.062 & 0.383 & 1.118 & 0.340 & 0.8753 & 0.478 & 1.055 & 0.377 \\
   $\avgx{-}{\Delta}$ & 1.555 & 0.156 & 1.382 & 0.246 & 1.6821 & 0.151 & 1.597 & 0.172 \\
   $\avgx{-}{\delta}$ & 1.202 & 0.301 & 1.297 & 0.273 & 1.1374 & 0.337 & 1.266 & 0.281 \\
  \hline\hline
 \end{tabular}
 \caption{$\chi^2/\mathrm{dof}$ and $p$-values from fitting CCF model ABDE for all six observables. The first two data columns contain the values for the final fit including all the data, while the remaining pairs of columns contain the values for the three variations used to estimate systematics as discussed in the text.}
 \label{tab:CCF_chi2_p}
\end{table}

\section{Summary and discussion} \label{sec:summary}

We have computed isovector nucleon axial, scalar and tensor charges as well as the isovector average quark momentum fraction, helicity and transversity moments on a set of eleven gauge ensembles using $N_f=2+1$ flavors of non-perturbatively improved Wilson fermions. The ground-state contribution has been extracted from simultaneous fits with a common, fitted energy gap. Physical results were obtained using a simultaneous extrapolation to the physical pion mass and the continuum and infinite-volume limits. \par

Adding the (directed) systematic errors in quadrature, our final results for the local charges can be summarized as
$g_A^{u-d} = 1.242\staterr{25}\syserr{\genfrac{}{}{0pt}{2}{+00}{-31}}$,
$g_S^{u-d} = 1.13 \staterr{11}\syserr{\genfrac{}{}{0pt}{2}{+07}{-06}}$, and
$g_T^{u-d} = 0.965\staterr{38}\syserr{\genfrac{}{}{0pt}{2}{+13}{-41}}$.
This is to be compared to the $N_{\rm f}=2+1$ FLAG average \cite{Aoki:2019cca} of $g_A^{u-d}=1.254(16)(30)$, and the $N_{\rm f}=2+1+1$ FLAG averages \cite{Aoki:2019cca} of $g_S^{u-d}=1.022(80)(60)$ and $g_T^{u-d}=0.989(32)(10)$.

A noticeable feature of lattice determinations of $g_A$ is that the results from most collaborations are low compared to the experimental value. Looking at our combined chiral, continuum and infinite-volume extrapolation, we find that this may potentially be explained by a conspiracy of different correction terms, all of which tend to depress the lattice value: while the chiral extrapolation is fairly flat, both the continuum and the infinite-volume extrapolation yield large positive corrections to the measured values, which come on top of the positive correction from the removal of the leading excited-state contaminations. Given that all of these effects have the same sign, even small remnants of each could considerably depress the value extracted from lattice simulations.

\begin{figure}
 \centering
  \subfigure{\includegraphics[width=0.5\textwidth]{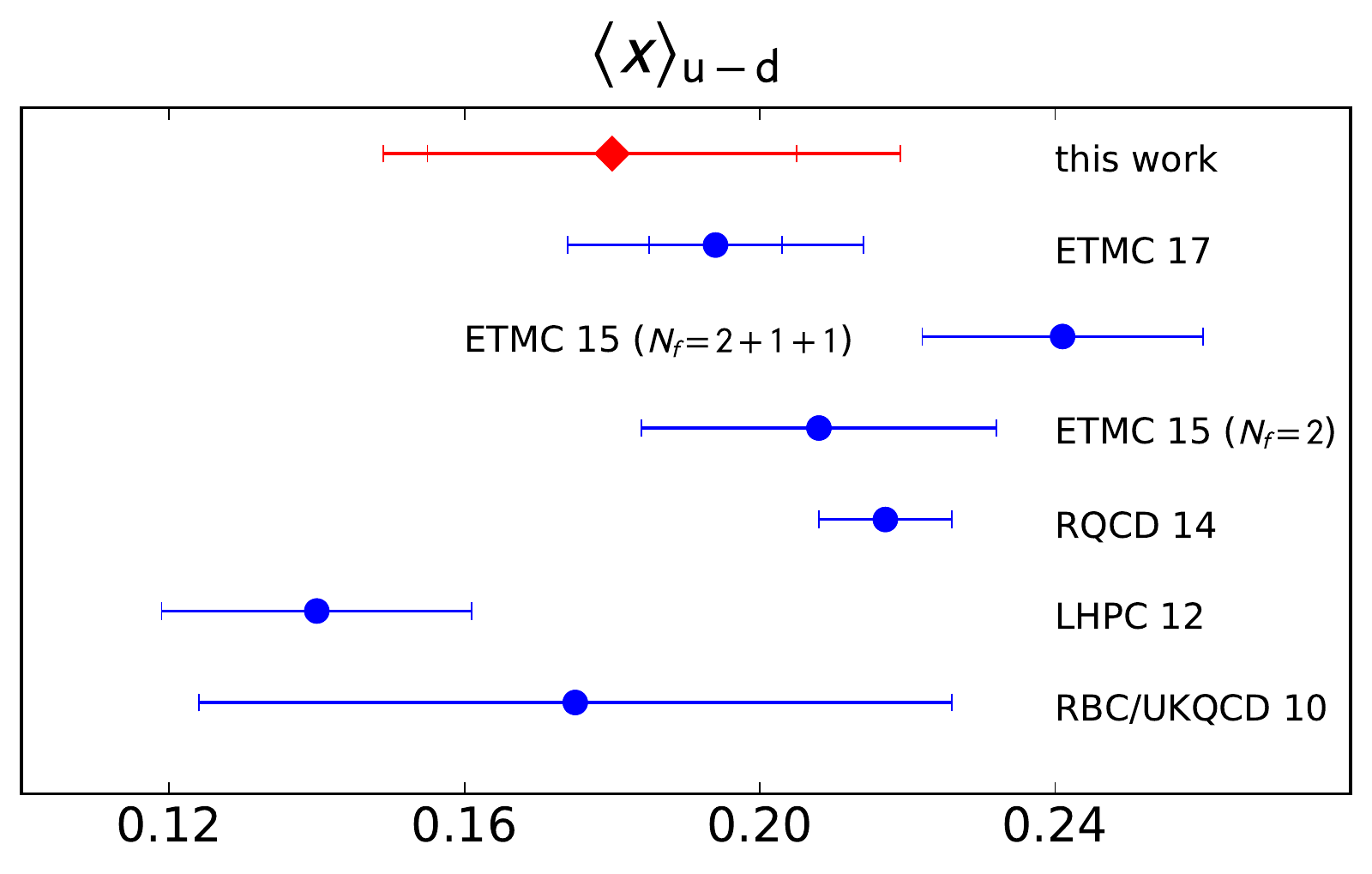}}\\
  \subfigure{\includegraphics[width=0.5\textwidth]{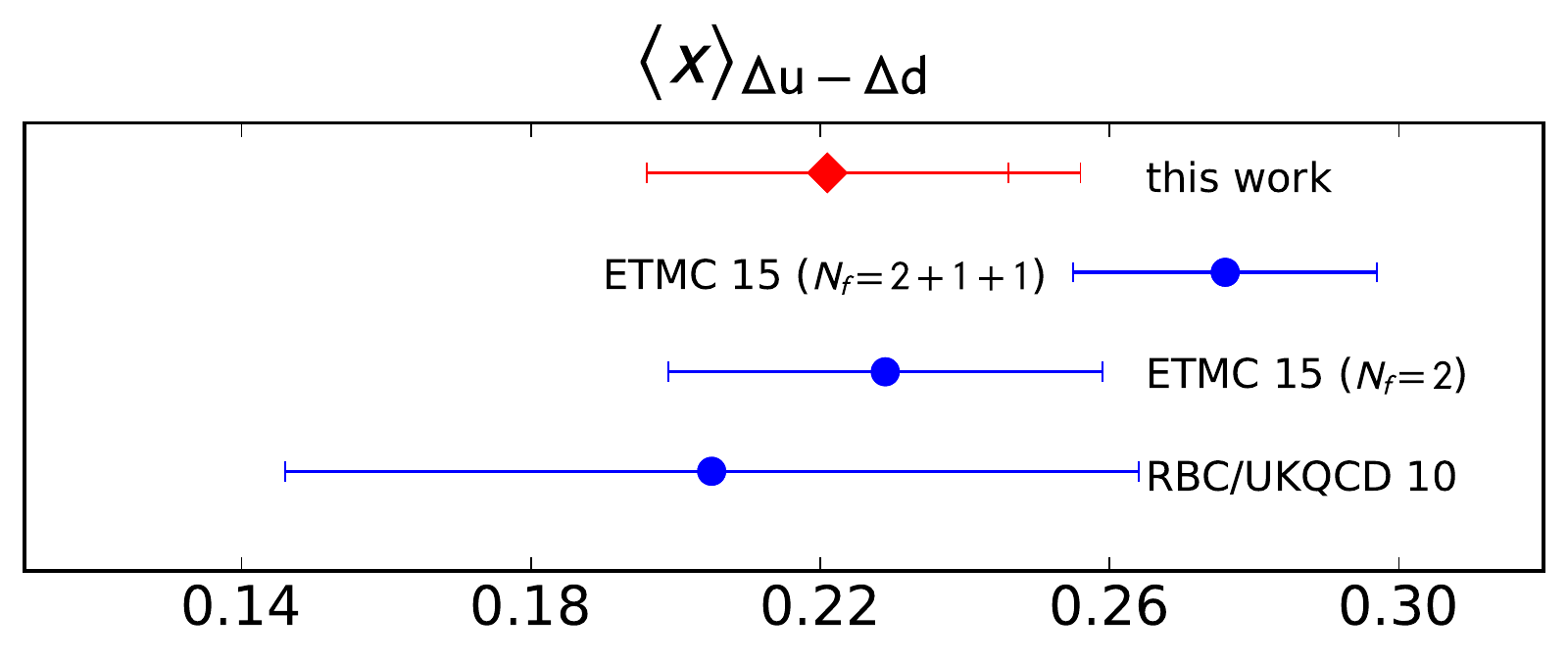}}\\
  \subfigure{\includegraphics[width=0.5\textwidth]{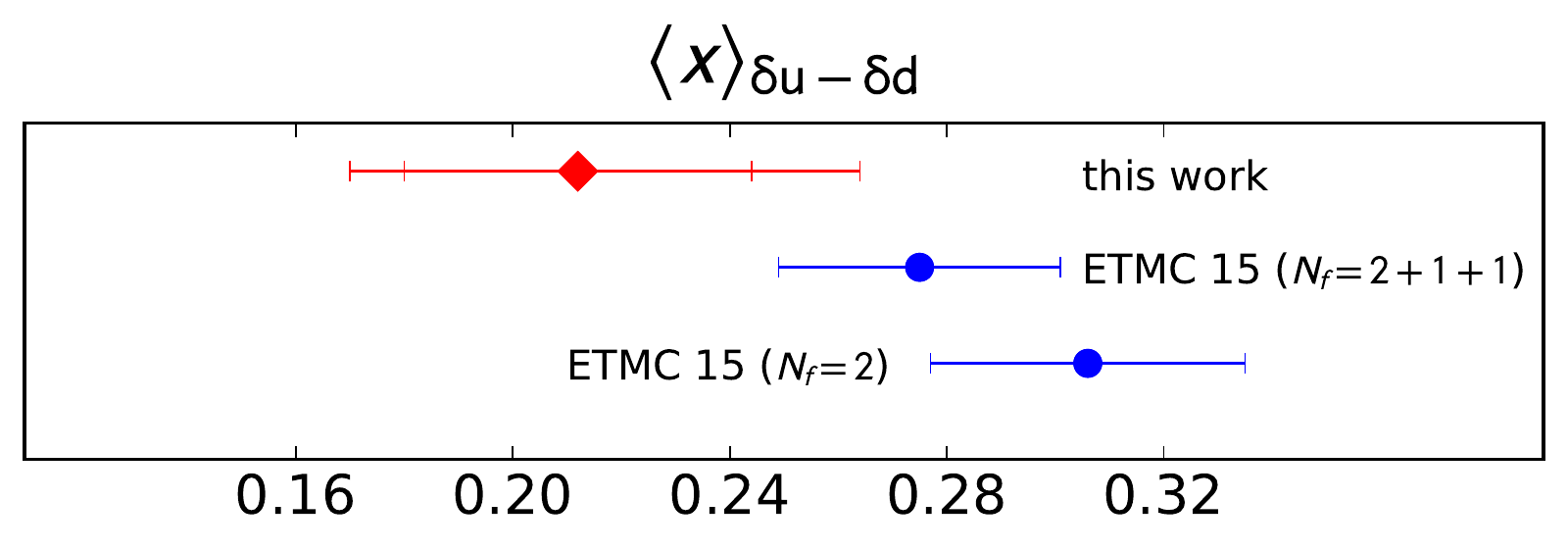}}
\caption{Comparison of our results for the twist-2 matrix elements with other recent determinations (ETMC 17 \cite{Alexandrou:2017oeh}, ETMC 15 \cite{Abdel-Rehim:2015owa}, RQCD 14 \cite{Bali:2014gha}, LPHC 12 \cite{Green:2012ud}, and RBC/UKQCD 10 \cite{Aoki:2010xg}).} 
\label{fig:compare_xumd}
\end{figure}

For the twist-2 matrix elements, our final results can be summarized as
$\avgx{-}{} = 0.180\staterr{25}\syserr{\genfrac{}{}{0pt}{2}{+14}{-06}}$,
$\avgx{-}{\Delta} = 0.221\staterr{25}\syserr{\genfrac{}{}{0pt}{2}{+10}{-00}}$ and
$\avgx{-}{\delta} = 0.212\staterr{32}\syserr{\genfrac{}{}{0pt}{2}{+20}{-10}}$. 
There are no FLAG averages to compare to for these observables, and lattice results with controlled errors are scarce, especially so for the helicity and transversity moments. In Fig.~\ref{fig:compare_xumd}, we compare our result to other recent determination of GPD moments. We note that our result is based on a full chiral and continuum extrapolation, while most of the other results were obtained at a single lattice spacing and a single pion mass. The closest comparisons for our results are $\avgx{-}{} = 0.140(21)$ (LHPC \cite{Green:2012ud}, $N_{\rm f}=2+1$), $\avgx{-}{\Delta} = 0.205(59)$ (RBC/UKQCD \cite{Aoki:2010xg}, $N_{\rm f}=2+1$), and $\avgx{-}{\delta} = 0.275(26)$ (ETMC \cite{Abdel-Rehim:2015owa}, $N_{\rm f}=2+1+1$).

There are a number of directions in which the present study can be extended:
\begin{itemize}
\item It would be highly desirable to further increase statistics on existing ensembles in a future study. Since the data at the smallest source-sink separation is already extremely precise, the most effective way to achieve this would be to include additional measurements for the larger source-sink separations such that effective statistics are comparable for each source-sink separation. We expect such an increase in statistics to greatly improve the simultaneous fits. On the one hand, it will lead to a much better determination of the excited-state-to-excited-state term in Eq.~(\ref{eq:multi_state_fit}), which will lead to even more stable fits and smaller statistical errors. On the other hand, it will allow us to further increase the value of $t_\mathrm{fit}$ and possibly even to drop the smallest source-sink separation entirely, which should lead to an additional reduction of the systematic error arising from excited-state contamination.
\item We also plan to add additional ensembles, including one with physical quark masses, in the near future. This should allow us to further reduce the uncertainty on the chiral extrapolation, and might help to remedy the issue with fitting the chiral logarithm in Eq.~(\ref{eq:CCF_fit_model}), particularly for $g_A^{u-d}$. 
\item We are also working on computing the contributions from disconnected quark loops in order to study the isoscalar counterparts of the isovector quantities considered here. This will also require the renormalization of the corresponding singlet operators, which may undergo mixing, adding a further level of complexity.
\end{itemize}
Finally, we plan to extend our analysis beyond the case of vanishing momentum transfer in order to study the isovector (and eventually the isoscalar) form factors of the nucleon. A study of the isovector electromagnetic and axial-vector form factors is currently under way.

\section*{Acknowledgments}
We thank Gunnar Bali, Sara Collins, Dalibor Djukanovic, Meinulf Göckeler, Maxwell T.~Hansen and Stefano Piemonte for useful discussions on non-perturbative renormalization.
This research is partly supported by the Deutsche Forschungsgemeinschaft (DFG, German Research Foundation) through the Collaborative Research Center SFB~1044 \emph{The low-energy frontier of the Standard Model}, under DFG grant HI~2048/1-1, and in the Cluster of Excellence \emph{Precision Physics, Fundamental Interactions and Structure of Matter} (PRISMA+ EXC~2118/1) funded by the DFG within the German Excellence strategy (Project~ID~39083149). Calculations for this project were partly performed on the HPC clusters ``Clover'' and ``HIMster2'' at the Helmholtz-Institut Mainz, and ``Mogon 2'' at Johannes-Gutenberg Universit\"at Mainz. Additional computer time has been allocated through projects HMZ21 and HMZ36 on the BlueGene supercomputer system ``JUQUEEN'' at NIC, J\"ulich. Our simulation code uses the QDP++ library \cite{Edwards:2004sx}, and the deflated SAP+GCR solver from the openQCD package \cite{openQCD}, while the contractions have been explicitly checked using QCT \cite{Djukanovic:2016spv}. We are grateful to our colleagues in the CLS initiative for sharing the gauge field configurations on which this work is based. We thank the RQCD collaboration for sharing additional ensembles that have been used in a collaborative effort for renormalization.

\bibliographystyle{h-physrev5}
\bibliography{refs} 

\clearpage
\appendix

\section{Non-Perturbative Renormalization} \label{app:npr}

\newcommand{\braket}[3]{\langle#1|#2|#3\rangle}
\def\rmd{\mathrm{d}}
\def\rme{\mathrm{e}}
\def\rmO{\mathrm{O}}
\def\calO{\mathcal{O}}
\def\trCD{\mathop{\mathrm{tr}}\nolimits_{_{CD}}}

In this appendix, we give further details of our renormalization procedure, which
follows closely that presented for the case of the $N_{\rm f}=2$ CLS ensembles
in Ref.~\cite{Hansen:2016kgh}.

\subsection{Setup}

We employ the ensembles listed in Table~\ref{tab:nprens}, which we fix to Landau gauge
by minimizing
\begin{equation}
W(U) = \sum_x\sum_\mu \tr\left[U_\mu^\dag(x)+U_\mu(x)\right]
\end{equation}
using the GLU library for Fourier-accelerated gauge fixing \cite{Hudspith:2014oja}.

\begin{table}
 \centering
 \begin{tabular}{llllllc}
  \hline\hline
  ID       & $\beta$ & $a/\mathrm{fm}$ & $T/a$ & $L/a$ & $\kappa$ & $\Mpi/\mathrm{MeV}$ \\
  \hline
  rqcd.019 & 3.40 & 0.086 & 32 & 32 & 0.1366     & 600 \\
  rqcd.016 & 3.40 & 0.086 & 32 & 32 & 0.13675962 & 420 \\
  rqcd.021 & 3.40 & 0.086 & 32 & 32 & 0.136813   & 340 \\
  rqcd.017 & 3.40 & 0.086 & 32 & 32 & 0.136865   & 230 \\
  \hline
  rqcd.029 & 3.46 & 0.076 & 64 & 32 & 0.1366     & 700 \\
  rqcd.030 & 3.46 & 0.076 & 64 & 32 & 0.1369587  & 320 \\
  X450     & 3.46 & 0.076 & 64 & 48 & 0.136994   & 250 \\
  \hline
  B250     & 3.55 & 0.064 & 64 & 32 & 0.1367     & 710 \\
  B251     & 3.55 & 0.064 & 64 & 32 & 0.137      & 420 \\
  X250     & 3.55 & 0.064 & 64 & 48 & 0.13705    & 350 \\
  X251     & 3.55 & 0.064 & 64 & 48 & 0.13710    & 270 \\
  \hline\hline
 \end{tabular}
 \caption{\label{tab:nprens}The $N_{\rm f}=3$ flavor ensembles with periodic boundary conditions used to determine the renormalization constants for this study. The ensembles labelled ``rqcd.0XX'' were made available by the RQCD collaboration as part of a joint NPR effort.}
\end{table}

\subsubsection{Renormalization scheme}

We use the RI'-MOM scheme \cite{Martinelli:1994ty,Gockeler:1998ye} in Landau gauge,
with renormalization conditions
\begin{align}
\left.\trCD\left[ S_R^{-1}(p)S_{\rm free}(p)\right]\right|_{p^2=\mu^2} &= 12,\\
\left.\trCD\left[ \braket{p}{\calO_R}{p} \braket{p}{\calO}{p}^{-1}_{\rm free}\right]\right|_{p^2=\mu^2} &= 12,
\end{align}
where $\trCD$ denotes a twelve-dimensional trace over color and Dirac indices.
Assuming multiplicative renormalization $S_R(p) = Z_q S_0(p)$, $\calO^X_R = Z_X \calO^X$,
these conditions imply that the renormalization factors are given by
\begin{align}
Z_q &= \left.\frac{1}{12}\trCD\left[S_0^{-1}(p)S_{\rm free}(p)\right]\right|_{p^2=\mu^2}, \\
Z_X &= \frac{12 Z_q}{\left.\trCD\left[\Lambda^{X}(p)\Lambda^{X,\rm free}(p)^{-1}\right]\right|_{p^2=\mu^2}}, \label{eq:zop}
\end{align}
where the bare vertex function $\Lambda^X$ is derived from the bare
Green's functions $G^X$ and $S_0$ via the amputation of its external legs,
\begin{equation}
\Lambda^X(p) = S_0^{-1}(p)G^X(p)S_0^{-1}(p)\,.
\end{equation}
The bare Green's functions are measured using momentum sources
\cite{Gockeler:1998ye} to compute position-momentum
propagators $S(y|p) = D^{-1}_{yx}\rme^{ip\cdot x}$, such that the
bare propagator is given by
\begin{align}
S_0(p) &= \left\langle\frac{1}{V}\sum_x \rme^{-ip\cdot x}S(x|p)\right\rangle,
\end{align}
the bare Green's function for a local bilinear operator
$\calO^X_{\mu_1\ldots\mu_n}(x)=\overline{u}(x)\Gamma^X_{\mu_1\ldots\mu_n} d(x)$ by
\begin{align}
G^{X}_{\mu_1\ldots\mu_n}(p) &= \left\langle\frac{1}{V}\sum_x \gamma_5S(x|p)^\dag\gamma_5\Gamma^X_{\mu_1\ldots\mu_n}S(x|p)\right\rangle,
\end{align}
and the bare Green's function of a one-link operator
$\calO^{xD}_{\mu_1\ldots\mu_n\rho}(x)=\overline{u}(x)\Gamma^X_{\mu_1\ldots\mu_n}\DBF{\rho}d(x)$
by
\begin{align}
G^{xD}_{\mu_1\ldots\mu_n\rho}(p) = \Big\langle\frac{1}{2V}\sum_x [&\gamma_5S(x|p)^\dag\gamma_5\Gamma^X_{\mu_1\ldots\mu_n}U_\rho(x)S(x+a\hat{\rho}|p)\\
&-\gamma_5S(x+a\hat{\rho}|p)^\dag\gamma_5\Gamma^X_{\mu_1\ldots\mu_n}U_\rho(x)^\dag S(x|p)]\Big\rangle. \nonumber
\end{align}

To reduce O(4) violation effects, we use only diagonal momenta of the form
$p=(\mu,\mu,\mu,\mu)$, where twisted boundary conditions
$\psi(x+L_\nu e_\nu)=\rme^{i\theta_\nu}\psi(x)$ are employed to
allow access to arbitrary momenta besides the Fourier modes.

\subsubsection{Operators and irreps}

In order to further reduce O(4) violation, we average over
the members $\ell=1,\ldots,K$ of H(4) irreps \cite{Gockeler:2010yr},
corresponding to replacing
\begin{equation}
\trCD\left[\Lambda^X(p)\Lambda^{X,\rm free}(p)^{-1}\right] \mapsto
\frac{1}{K}\sum_{\ell=1}^K\trCD\left[\Lambda^X_\ell(p)\Lambda^{X, \rm free}_\ell(p)^{-1}\right]
\end{equation}
in Eq.~(\ref{eq:zop}).

To ensure that the vector and axial vector Ward identities are respected,
we further replace \cite{Gockeler:1998ye,Green:2017keo}
\begin{equation}
\trCD\left[\Lambda^{X}(p)\Lambda^{X, \rm free}(p)^{-1}\right] \mapsto
\frac{1}{3}\sum_{\mu,\nu} \left(\delta_{\mu\nu}-\frac{p_\mu p_\nu}{p^2}\right)
\trCD\left[\Lambda^{X}_\mu(p)\Lambda^{X,\rm free}_{\nu}(p)^{-1}\right]
\end{equation}
in the case of the vector and axial currents, $X\in\{V,A\}$.

In the case of the one-link operators, there are two inequivalent H(4) irreps
in each case. For the vector and axial vector operators, there are a six- and a three-dimensional representation in each case \cite{Gockeler:1996mu},
\begin{align}
v_{2,a} ({ \tau_3^{(6)}}): &\; \{ \calO^{vD}_{\{\mu\nu\}} = \frac{1}{2}(\calO^{vD}_{\mu\nu}+\calO^{vD}_{\nu\mu}) \,|\, 1\le\mu<\nu\le 4 \}, \\
v_{2,b} ({ \tau_1^{(3)}}): &\; \{ \calO^{vD}_{11}+\calO^{vD}_{22}-\calO^{vD}_{33}-\calO^{vD}_{44},\, \calO^{vD}_{33}-\calO^{vD}_{44},\, \calO^{vD}_{11}-\calO^{vD}_{22} \}; \\
r_{2,a} ({ \tau_4^{(6)}}): &\; \{ \calO^{aD}_{\{\mu\nu\}} = \frac{1}{2}(\calO^{aD}_{\mu\nu}+\calO^{aD}_{\nu\mu}) \,|\, 1\le\mu<\nu\le 4 \}, \\
r_{2,b} ({ \tau_4^{(3)}}): &\; \{ \calO^{aD}_{11}+\calO^{aD}_{22}-\calO^{aD}_{33}-\calO^{aD}_{44},\, \calO^{aD}_{33}-\calO^{aD}_{44},\, \calO^{aD}_{11}-\calO^{aD}_{22}, \}
\end{align}
whereas for the tensor operator, there are two inequivalent eight-dimensional representations \cite{Gockeler:1996mu},
\begin{align}
h_{1,a} ({ \tau_2^{(8)}}): &\; \{ 2\calO^{tD}_{\mu\{\nu\rho\}}+\calO^{tD}_{\nu\{\mu\rho\}}, \calO^{tD}_{\nu\{\mu\rho\}} \,|\, 1\le\mu<\nu<\rho\le 4 \}, \\
h_{1,b} ({ \tau_1^{(8)}}): &\; \{ \calO^{tD}_{122}-\calO^{tD}_{133}, \calO^{tD}_{122}+\calO^{tD}_{133}-2\calO^{tD}_{144},
\calO^{tD}_{211}-\calO^{tD}_{233}, \calO^{tD}_{211}+\calO^{tD}_{233}-2\calO^{tD}_{244},\\
&\;\;\; \calO^{tD}_{311}-\calO^{tD}_{322}, \calO^{tD}_{311}+\calO^{tD}_{322}-2\calO^{tD}_{344},
\calO^{tD}_{411}-\calO^{tD}_{422}, \calO^{tD}_{411}+\calO^{tD}_{422}-2\calO^{tD}_{433} \}. \nonumber
\end{align}

\subsubsection{\texorpdfstring{Conversion to $\MSbar$ and RGI}{Conversion to MSbar and RGI}}

The measured renormalization constants in the RI'-MOM scheme at finite quark mass
are then extrapolated to the chiral limit using the ansatz
\begin{equation}
Z_X(a,\mu,M_\pi) = Z_X^{\mathrm{RI'-MOM}}(a,\mu) + c_X(a,\mu)\,(aM_\pi)^2\,.
\label{eq:rimomxe}
\end{equation}

To convert the renormalization constants obtained in the RI'-MOM scheme
to the more commonly quoted $\overline{\rm MS}$-scheme,
we use the three-loop continuum perturbation theory results of
Refs.~\cite{Chetyrkin:1997dh,Vermaseren:1997fq,vanRitbergen:1997va,Chetyrkin:1999pq,Gracey:2000am,Gracey:2003yr,Gracey:2003mr} for the conversion factors
$Z^{\overline{\rm MS}}_{\rm RI'-MOM}(\mu)$.
To check for lattice artifacts, we also determine the
Renormalization Group Invariant (RGI) values
of the renormalization factors using the three-loop $\overline{\rm MS}$
$\beta$- and $\gamma$-functions to remove the running with $\mu$,
\begin{equation}
Z_X^{\rm RGI}(a) = \Delta Z_X^{\overline{\rm MS}}(\mu) Z^{\overline{\rm MS}}_{X, \rm RI'-MOM}(\mu) Z_X^{\rm RI'-MOM}(a,\mu).
\end{equation}

\subsection{Perturbative subtraction of lattice artifacts}

The RGI renormalization factors are constructed to be independent of the
renormalization scale $\mu$. Since, however, we remove the running only at the
perturbative level, deviations are to be expected at small $\mu$,
where the running coupling becomes large and perturbation theory breaks down.
At large $\mu$, on the other hand, the running coupling is small, and perturbation
theory works well; any residual $\mu$-dependence in this regime is therefore
indicative of lattice artifacts, which in practice can be quite sizeable.

\subsubsection{General procedure}

\begin{figure}
\begin{center}
\includegraphics[width=0.7\textwidth,keepaspectratio=]{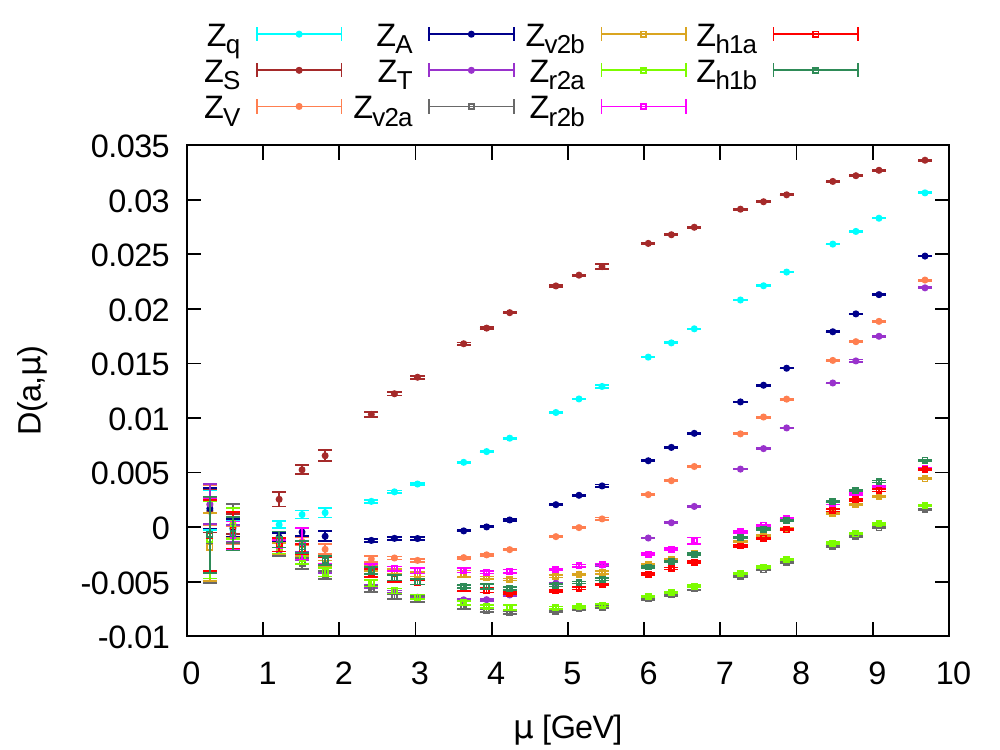}
\end{center}
\caption{\label{fig:Dmua}
The subtraction functions $D_X(\mu,a)$ for the operators $\calO^X$ considered
in this study at a lattice spacing of $a=0.086\fm$.}
\end{figure}

The use of lattice perturbation theory to reduce the size of lattice artifacts
by a perturbative subtraction has been proposed in ref.~\cite{Becirevic:2004ny},
and further explored in
refs.~\cite{Constantinou:2013ada,Simeth:2013ima,Alexandrou:2015sea}.
Here, as in ref.~\cite{Hansen:2016kgh},
we follow an approach very similar to that of ref.~\cite{Gockeler:2010yr},
subtracting all lattice artifacts at $\calO(g^2)$ by perturbatively expanding
the renormalization constants at finite lattice spacing and isolating the
lattice artifacts,
\begin{align}
Z^{\rm RI'-MOM}_X(\mu,a) &= 1 + g^2 F_X(\mu,a) +\calO(g^4)
                       = 1 + g^2 \left[\gamma^X_0\log(\mu a)+C_X+D_X(\mu,a)\right] +\calO(g^4),
\end{align}
where $\gamma^X_0$ is the analytically-known anomalous dimension, and
$D_X(\mu,a)$ is required to vanish in the continuum limit $a\to 0$.
The lattice artifacts that we wish to subtract from $Z_X^{\rm RI'-MOM}$
are then given by
\[
g^2 D_X(\mu,a) = g^2 \left[F_X(\mu,a)-\left(\gamma^X_0\log(\mu a)+C_X\right)\right],
\]
where in many cases $C_X$ is analytically known, or else can be obtained numerically
using a fit to $F_X(\mu,a)-\gamma^X_0\log(\mu a)$ in the limit $a\to0$.
Fig.~\ref{fig:Dmua} shows the subtraction functions $D_X(\mu,a)$ for the operators
$\calO^X$ considered in this study at our coarsest lattice spacing of $a=0.086\fm$;
results at the other lattice spacings are qualitatively very similar.

We can then define a subtracted renormalization constant
\begin{equation}
Z^{\rm RI'-MOM,sub.}_X(\mu,a) = Z^{\rm RI'-MOM}_X(\mu,a) - g^2 D_X(\mu,a),
\end{equation}
and we expect the corresponding RGI renormalization constant
$Z_X^{\rm RGI, sub}(a)$ to show only very mild lattice artifacts when considered
as a function of $\mu$.

\subsubsection{Automated perturbation theory}

Since the Feynman rules for lattice perturbation theory are quite complex
and do not usually allow for an analytical evaluation of Feynman integrals,
we employ the HiPPy/HPsrc packages \cite{Hart:2004bd,Hart:2009nr}, which
separate the (complicated, action-dependent) Feynman rules from the
(action-independent) Feynman diagrams: the diagrams are coded once and for
all in an operator- and action-independent fashion using the HPsrc library
of Fortran 95 modules; these generic diagrams can then be evaluated
numerically for in principle arbitrary operators and lattice actions.
The automated derivation of the action- and operator-dependent Feynman
rules is performed in a separate step using the HiPPy library of Python modules,
which takes a human-readable expression for an action or operator as input
and outputs the corresponding Feynman rules in a machine-readable format suitable
for use with HPsrc.

In this manner, we have been able to reuse much of the code written in the
context of our study of non-perturbative renormalization for the $N_{\rm f}=2$
CLS ensembles \cite{Hansen:2016kgh}, even though the gluonic action used 
is different in the two- and three-flavor cases.

\subsubsection{Choice of coupling}

\begin{figure}
\begin{center}
\includegraphics[width=0.7\textwidth,keepaspectratio=]{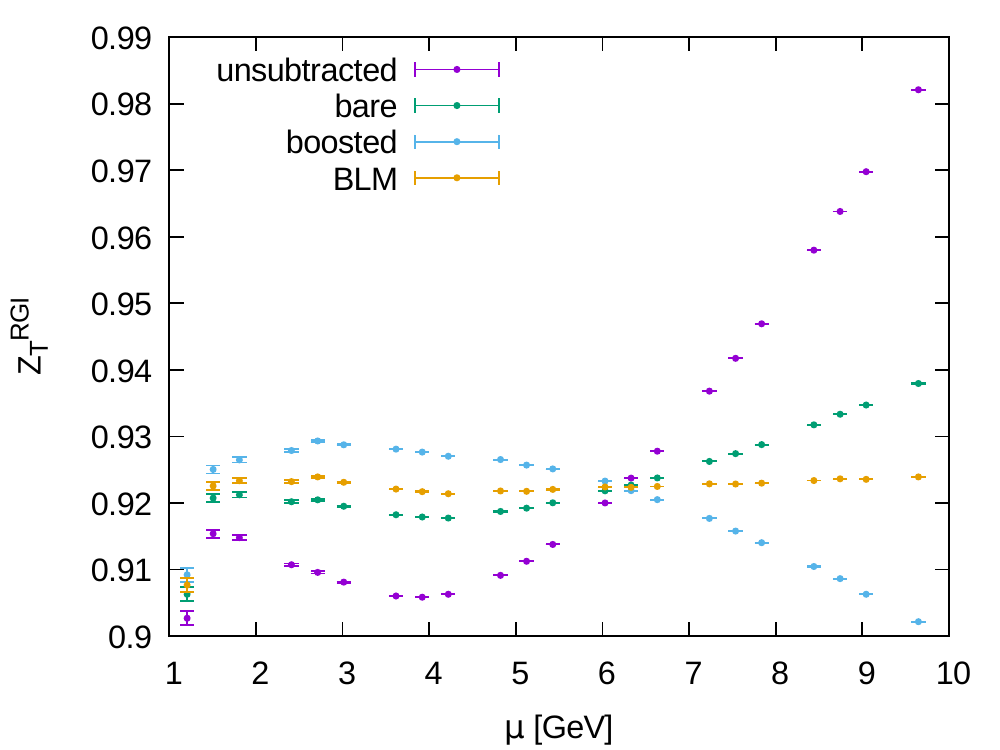}
\end{center}
\caption{\label{fig:couplings}
         Comparison of the unsubtracted and subtracted values of the RGI tensor
         renormalization constant $Z_T^{\rm RGI}$, using the bare coupling $g_0$,
         the boosted coupling $g_{\rm b}$, or the BLM coupling $g_{\rm BLM}$ for the
         perturbative subtraction. It can be seen that the BLM coupling is most
         efficient in removing the lattice artifacts, which are otherwise very large.}
\end{figure}

To combine the perturbative and non-perturbative results, we need to make a
choice for the coupling. The bare coupling $g_0^2=6/\beta$ is well-known to give
generally rather poor results. A widely-used alternative is the boosted coupling
$g_{\rm b}^2=g_0^2/\langle P(g_0)\rangle$, where $P(g_0)$ is the (non-perturbatively
determined) value of the average plaquette. Using the boosted coupling amounts to
a partial resummation of higher-order terms in the perturbative expansion. To better
control this resummation, the BLM coupling \cite{Lepage:1992xa}
$g_{\rm BLM}^2 = 4\pi\alpha_V(q_*)$ can be used, where $\alpha_V(q)$ is the coupling
in the potential scheme defined by the expression
\begin{equation}
V(q)=- \frac{4\pi C_f \alpha_V(q)}{q^2}
\end{equation}
for the static potential, and $q_*$ is a process-dependent typical momentum scale
given by
\begin{equation}
\log(q_*^2)=\frac{\int \rmd^4q\,f(q)\,\log(q^2)}{\int \rmd^4q\,f(q)}.
\end{equation}
We find that using the BLM coupling is highly efficient in removing most of the
lattice artifacts using one-loop lattice perturbation theory.
In Fig.~\ref{fig:couplings}, we show a representative example, i.e. a comparison
between the different couplings
in the case of the tensor renormalization constant $Z_T^{\rm RGI}$; it can clearly
be seen that the use of the BLM coupling leads to a nearly perfect subtraction
of the (rather large) lattice artifacts and is vastly superior in efficiency to
the use of either the bare or boosted couplings.

\subsection{Systematic uncertainties}

\begin{figure}
\begin{center}
\includegraphics[width=0.7\textwidth,keepaspectratio=]{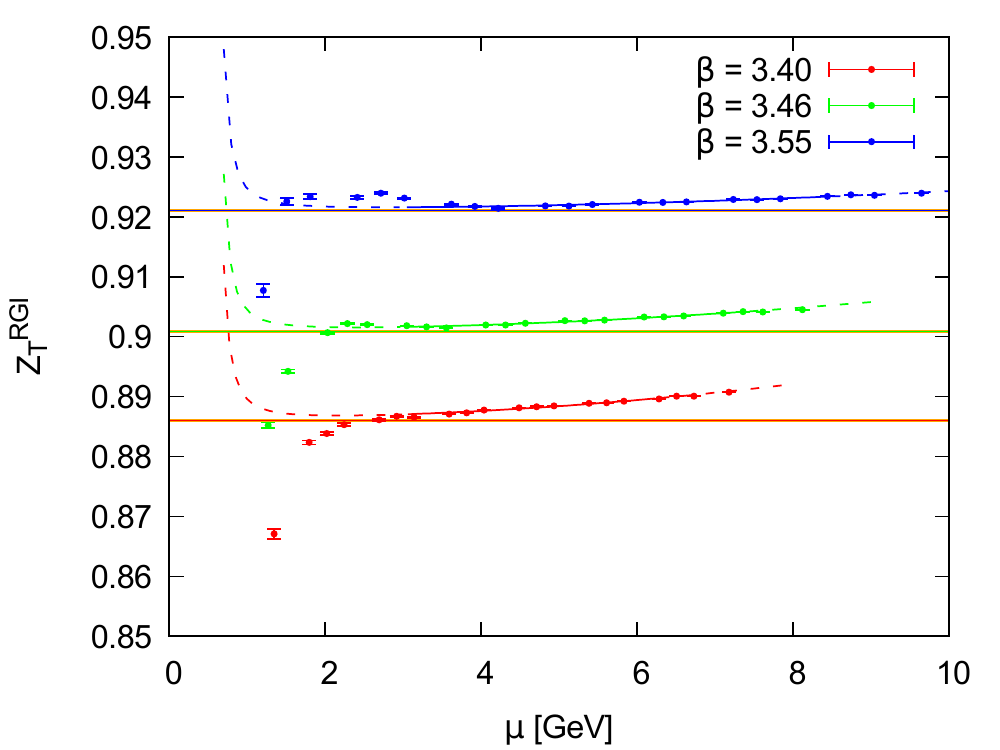}
\end{center}
\caption{\label{fig:zrgifit}
        The final fit used to extract $Z_T^{\rm RGI}$ for our three value of $\beta$.
        The solid lines denote the fit ranges, whereas the dashed lines indicate how
        the fit form of Eq.~(\ref{eq:zrgifit}) extrapolates beyond the fit range,
        while the different colors correspond to the different lattice spacings.
        Final fit results for $Z_T^{\rm RGI}(\beta)$ are shown by the horizontal bands.}
\end{figure}

\subsubsection{Final fits}

To remove the residual $\mu$-dependence of the subtracted RGI renormalization
constants, we perform the fit
\begin{equation}
Z_X^{\rm RGI, sub}(a,\mu) = Z_X^{\rm RGI}(\beta)\left\{1+d^X_1 g_{\MSbar}^8(\mu)\right\}
+d^X_2(\beta)\,(a\mu)^2 \Delta Z_X^{\overline{\rm MS}}(\mu) Z^{\overline{\rm MS}}_{X, \rm RI'-MOM}(\mu),
\label{eq:zrgifit}
\end{equation}
where the $\beta$-independent term with coefficient $d_1$ accounts for the use of
three-loop continuum perturbation theory in converting from the RI'-MOM scheme,
and the term with coefficient $d_2(\beta)$ accounts for the use of the perturbative
subtraction leaving residual discretization artifacts.

To keep both higher-order perturbative effects and lattice artifacts small,
the fit region should ideally satisfy
\begin{equation}
\Lambda^{\MSbar} \ll \mu \ll a^{-1}.
\end{equation}
Since we cannot realistically fulfil both of those inequalities at the same time,
we have chosen to take the lower end of the window at $\mu_{\rm min} = 3\,\gev$,
but allow renormalization scales as large as $\mu_{\rm max} = 2.75a^{-1}$
in the fit, because we rely on the perturbative subtraction of the leading
artifacts. An example of the resulting fits is shown in Fig.~\ref{fig:zrgifit}.

\subsubsection{Fit variants}

To explore possible sources of systematic error, we employ the following fit variants:
\begin{itemize}
\item adding either a higher-order chiral term $\tilde{c}(a,\mu)(aM_\pi)^4$ or a finite-volume term $d(a,\mu)\rme^{-M_\pi L}$ to the chiral extrapolation~(\ref{eq:rimomxe}),
\item varying the value of $a\Lambda^{\MSbar}$ within the uncertainties of
 $\Lambda^{\MSbar}$, and
\item narrowing the fit window by increasing the lower bound on the fit intervals to $\mu_{\rm min}=4\,\gev$, or by decreasing the upper bounds on the fit intervals to $\mu_{\rm max}=2.5a^{-1}$.
\end{itemize}
Our final estimate of the systematic error is obtained conservatively by adding the
spreads from all three variants in quadrature.

\subsubsection{\texorpdfstring{Extrapolation to $\beta=3.7$}{Extrapolation to beta=3.7}}

\begin{figure}[t]
 \centering
  \subfigure{\includegraphics[totalheight=0.25\textheight]{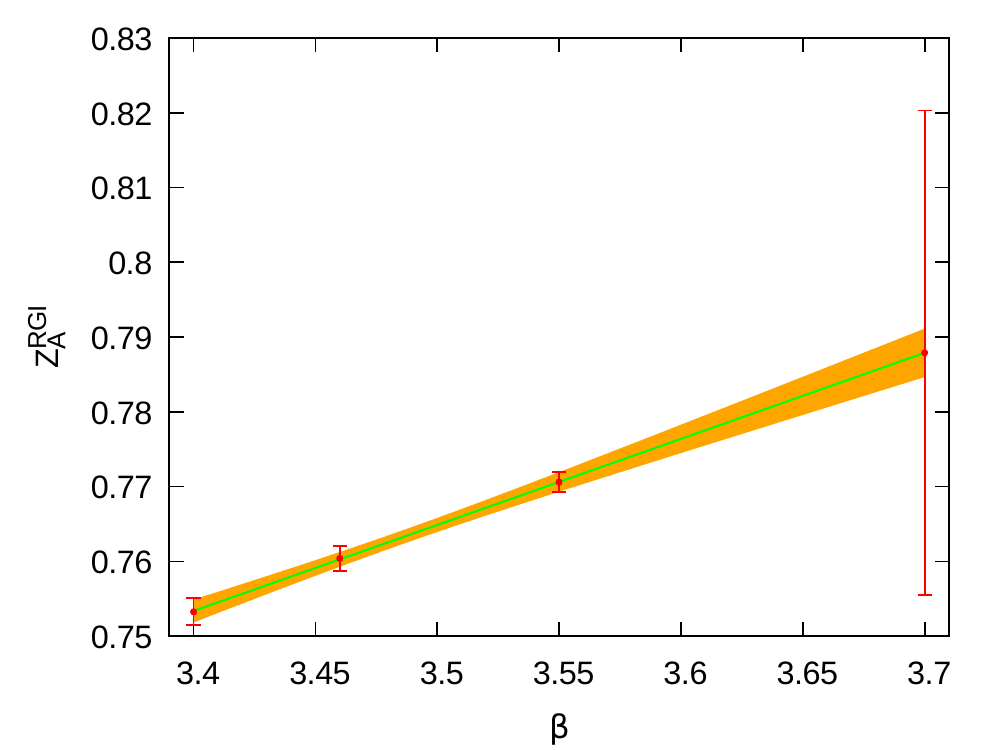}}
  \subfigure{\includegraphics[totalheight=0.25\textheight]{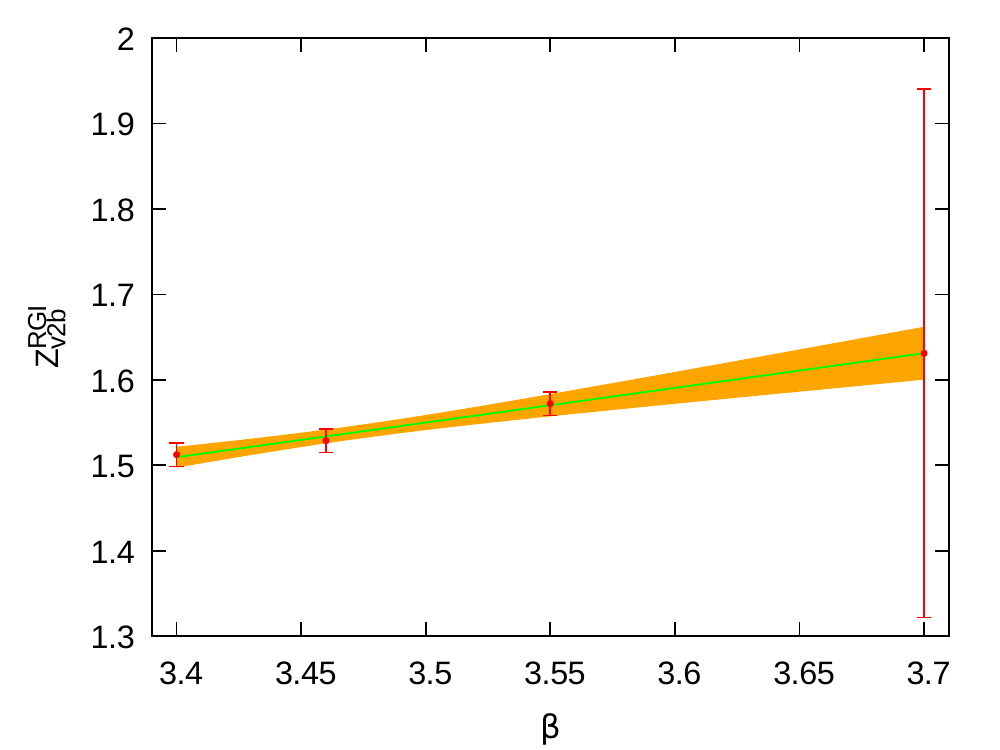}} \\
  \subfigure{\includegraphics[totalheight=0.25\textheight]{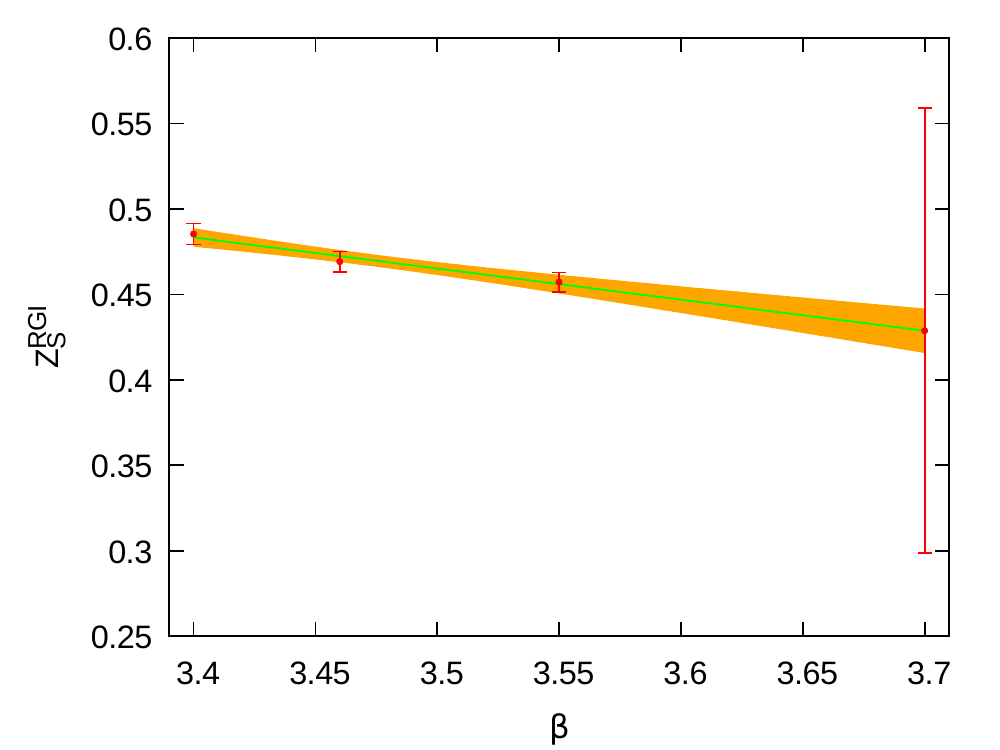}}
  \subfigure{\includegraphics[totalheight=0.25\textheight]{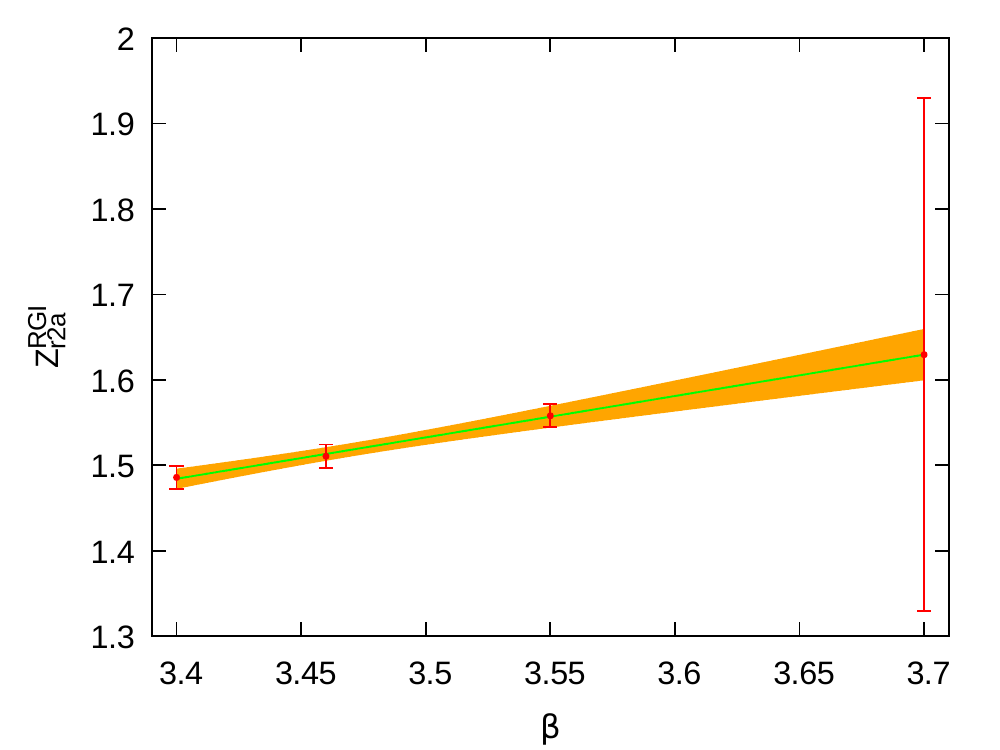}} \\
  \subfigure{\includegraphics[totalheight=0.25\textheight]{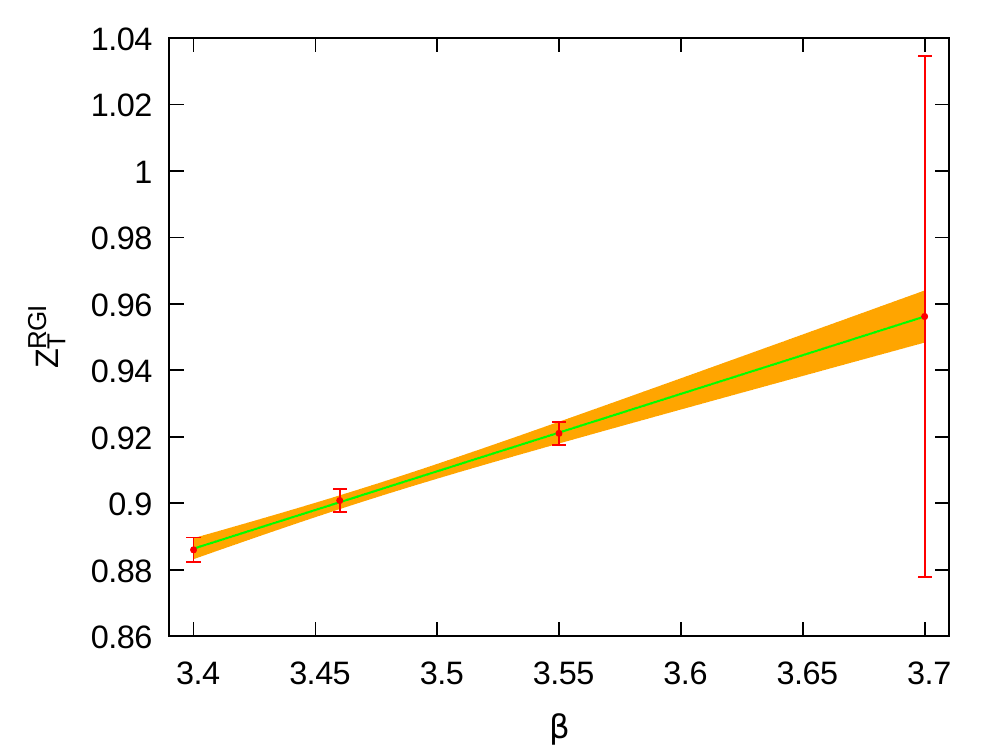}}
  \subfigure{\includegraphics[totalheight=0.25\textheight]{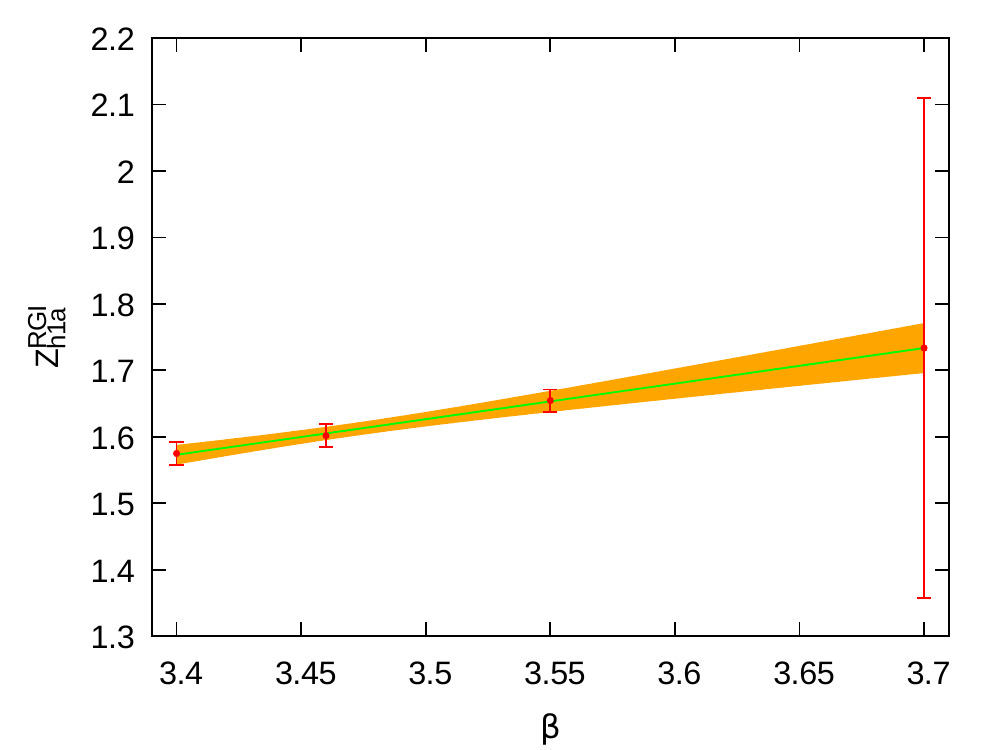}}
 \caption{Extrapolation of the renormalization constants used in the final analysis to $\beta=3.7$. Shown are the values of $Z_X$ for the local operators $X\in\{A,S,T\}$ (left column) and the one-link operators in irreps $X\in\{v2b,r2a,h1a\}$ (right column) as measured at $\beta\in\{3.4,3.46,3.55\}$, the linear fit in $\beta$ with its error band, and the extrapolated value at $\beta=3.7$ with its ten-fold inflated final error.}
 \label{fig:extrapolation_zx_b3.7}
\end{figure}

Since the RI'-MOM scheme is defined in terms of quantities at well-defined four-momenta,
it requires a four-dimensional Fourier transform and thus implicitly relies on the
gauge ensembles being generated with periodic boundary conditions in time. Due to
the extreme critical slowing-down observed in quantities related to the global topology,
the generation of sufficiently large and properly thermalized gauge ensembles
with periodic boundary conditions at $\beta=3.7$ is not feasible with currently
existing computer resources, and the existing $\beta=3.7$ ensembles with open
boundary conditions are not suitable for use with RI'-MOM. While there are some
proposals how to bypass this issue \cite{privcomm,wip}, for this study we will rely on
an extrapolation of the measured renormalization constants to $\beta=3.7$.
Given three values of $\beta$ at which we have data, we use a linear extrapolation
in $\beta$ to obtain the central value, but do not trust the errors from the fit
to account for the full uncertainty. We therefore very conservatively inflate them
by an \emph{ad hoc} factor of ten to cover the full range of uncertainty involved
in the extrapolation. The extrapolations for the renormalization constants used
in the final analysis are shown in Fig.~\ref{fig:extrapolation_zx_b3.7}; it can be
seen that the inflated error covers the whole range in which the final value could
conceivably lie. As an alternative to a linear extrapolation in $\beta$, we have also
considered a linear extrapolation in $g_0^2$, and the results of both are compatible
within their errors.

\end{document}